\documentclass[a4paper,fleqn,usenatbib]{mnras}

\usepackage[T1]{fontenc}
\usepackage{ae,aecompl}

\usepackage{graphicx}	
\usepackage{amsmath}	
\usepackage{amssymb}	
\usepackage{booktabs}
\usepackage{eso-pic}

\usepackage{ulem}       

\usepackage{times}

\newcommand{\beq}{\begin{equation}}
\newcommand{\eeq}{\end{equation}}
 
\newcommand{\htwo}{\hspace{2pt}}
\newcommand{\nhat}{\hat{\textbf{n}}}

\newcommand{\mb}{\mathbf}

\newcommand{\LCDM}{$ \Lambda $CDM~}

\newcommand{\Planck}{{\slshape Planck~}}

\newcommand{\mbl}{\ensuremath{\mathbf{l}}}

\newcommand{\kpc}{h^{-1}\mathrm{kpc}}
\newcommand{\Mpc}{\mathrm{Mpc}}

\makeatletter
\let\ftype@table\ftype@figure
\makeatother

\title[Detection of the kSZ effect with DES and SPT]{Detection of the kinematic Sunyaev-Zel'dovich effect with DES Year~1 and SPT}

\author[B. Soergel, S. Flender, K. Story et al.]{
\parbox{\textwidth}{
\Large
B.~Soergel$^{1,2}$\thanks{E-Mail: bsoergel@ast.cam.ac.uk},
S.~Flender$^{3,4}$,
K.~T.~Story$^{5,6}$,
L.~Bleem$^{3,4}$,
T.~Giannantonio$^{1,2,7}$,
G.~Efstathiou$^{1,2}$,
E.~Rykoff$^{5,8}$,
B.~A.~Benson$^{4,9,10}$,
T.~Crawford$^{4,10}$,
S.~Dodelson$^{4,9,10}$,
S.~Habib$^{3,4,11}$,
K.~Heitmann$^{3,4,11}$,
G.~Holder$^{12}$,
B.~Jain$^{13}$,
E.~Rozo$^{14}$,
A.~Saro$^{15,16}$,
J.~Weller$^{16,17,18}$,
F.~B.~Abdalla$^{19,20}$,
S.~Allam$^{9}$,
J.~Annis$^{9}$,
R.~Armstrong$^{21}$,
A.~Benoit-L{\'e}vy$^{19,22,23}$,
G.~M.~Bernstein$^{13}$,
J.~E.~Carlstrom$^{4,10,24,3}$,
A.~Carnero~Rosell$^{25,26}$,
M.~Carrasco~Kind$^{27,28}$,
F.~J.~Castander$^{29}$,
I.~Chiu$^{15,16}$,
R.~Chown$^{12}$,
M.~Crocce$^{29}$,
C.~E.~Cunha$^{5}$,
C.~B.~D'Andrea$^{30,31}$,
L.~N.~da Costa$^{26,25}$,
T.~de~Haan$^{32}$,
S.~Desai$^{15,16}$,
H.~T.~Diehl$^{9}$,
J.~P.~Dietrich$^{15,16}$,
P.~Doel$^{19}$,
J.~Estrada$^{9}$,
A.~E.~Evrard$^{33,34}$,
B.~Flaugher$^{9}$,
P.~Fosalba$^{29}$,
J.~Frieman$^{4,9}$,
E.~Gaztanaga$^{29}$,
D.~Gruen$^{5,8}$,
R.~A.~Gruendl$^{27,28}$,
W.~L.~Holzapfel$^{32}$,
K.~Honscheid$^{35,36}$,
D.~J.~James$^{37}$,
R.~Keisler$^{6}$,
K.~Kuehn$^{38}$,
N.~Kuropatkin$^{9}$,
O.~Lahav$^{19}$,
M.~Lima$^{39,25}$,
J.~L.~Marshall$^{40}$,
M.~McDonald$^{41}$,
P.~Melchior$^{21}$,
C.~J.~Miller$^{34,33}$,
R.~Miquel$^{42,43}$,
B.~Nord$^{9}$,
R.~Ogando$^{26,25}$,
Y.~Omori$^{12}$,
A.~A.~Plazas$^{44}$,
D.~Rapetti$^{15,16}$,
C.~L.~Reichardt$^{45}$,
A.~K.~Romer$^{46}$,
A.~Roodman$^{5,8}$,
B.~R.~Saliwanchik$^{47}$,
E.~Sanchez$^{48}$,
M.~Schubnell$^{33}$,
I.~Sevilla-Noarbe$^{48,27}$,
E.~Sheldon$^{49}$,
R.~C.~Smith$^{37}$,
M.~Soares-Santos$^{9}$,
F.~Sobreira$^{25}$,
A.~Stark$^{50}$,
E.~Suchyta$^{13}$,
M.~E.~C.~Swanson$^{28}$,
G.~Tarle$^{33}$,
D.~Thomas$^{31}$,
J.~D.~Vieira$^{27,51}$,
A.~R.~Walker$^{37}$,
N.~Whitehorn$^{32}$
(The DES and SPT Collaborations)
}
\vspace{0.4cm}\\
\parbox{\textwidth}{Author affiliations are listed at the end of this paper}
}

\pubyear{2016}

\begin{document}
\label{firstpage}
\pagerange{\pageref{firstpage}--\pageref{lastpage}}
\maketitle

\begin{abstract}
We detect the kinematic Sunyaev-Zel'dovich (kSZ) effect with a statistical significance of $4.2 \sigma$ by combining a cluster catalogue derived from the first year data of the Dark Energy Survey (DES) with CMB temperature maps from the  South Pole Telescope Sunyaev-Zel'dovich (SPT-SZ) Survey.
This measurement is performed with a differential statistic that isolates the \textit{pairwise} kSZ signal,
providing the first detection of the large-scale, pairwise motion of clusters using redshifts derived from photometric data.
By fitting the pairwise kSZ signal to a theoretical template we measure the average central optical depth of the cluster sample, $\bar{\tau}_e = (3.75 \pm 0.89)\cdot 10^{-3}$.
We compare the extracted signal to realistic simulations and find good agreement with respect to the signal-to-noise,
the constraint on $\bar{\tau}_e$, and the corresponding gas fraction.
High-precision measurements of the pairwise kSZ signal with future data will be able to place constraints on the baryonic physics of galaxy clusters,
and could be used to probe gravity on scales $ \gtrsim 100$ Mpc.
\end{abstract}
%

\AddToShipoutPictureBG*{%
  \AtPageUpperLeft{%
    \hspace{0.75\paperwidth}%
    \raisebox{-4\baselineskip}{%
      \makebox[0pt][l]{\textnormal{DES-2015-0154}}
}}}%

\AddToShipoutPictureBG*{%
  \AtPageUpperLeft{%
    \hspace{0.75\paperwidth}%
    \raisebox{-5\baselineskip}{%
      \makebox[0pt][l]{\textnormal{FERMILAB-PUB-16-036-AE}}
}}}%



\section{Introduction}

Galaxy clusters, as the largest gravitationally-bound structures in the Universe, are important probes of cosmology and astrophysics.
These massive systems imprint their signature on the cosmic microwave background (CMB) through both the \textit{thermal} Sunyaev-Zel'dovich (tSZ) effect
--- in which $\lesssim 1$\% of CMB photons passing through the centre of a massive cluster inverse-Compton scatter off electrons in the hot, ionized intra-cluster gas \citep{SZ1970,SZ1972,Birkinshaw1998,Carlstrom2002} ---
as well as the \textit{kinematic} SZ effect (kSZ) in which the bulk motion of clusters imparts a Doppler shift to the CMB signal  \citep{SZ1972,SZ1980}.
The kinematic and thermal SZ effects can also be thought of as first- and second-order terms of the same physical process: the scattering of photons with a Planck distribution on moving electrons. 
The first-order kSZ effect shifts but does not distort the CMB blackbody spectrum, whereas the second-order tSZ imparts spectral distortions.
Because the thermal electron velocities within the cluster are much larger than its bulk velocity, the second-order effect dominates here:  for typical cluster masses and velocities, the amplitude of the kSZ effect is an order of magnitude smaller than its thermal counterpart (e.g.~\citealt{Birkinshaw1998}).

The tSZ effect has been well characterized, both through its contribution to the CMB temperature power spectrum (see e.g.~\citealt{Das2013,George2014}), and  via measurements on individual clusters (e.g.~\citealt{plagge10,Planck12,Bonamente12,Sayers13}).
The kSZ signal, however, has proved to be more elusive, both because of its smaller amplitude and its spectrum identical to that of primary CMB temperature fluctuations.
While challenging to measure, the kSZ effect has great potential for constraining both astrophysical and cosmological models
(see e.g. \citealt{Rephaeli1991,Haehnelt1996,Diaferio2004,Bhattacharya2006,Bhattacharya2007}).
From an astrophysical point of view, the kSZ signal can be used to probe so-called `missing baryons' (e.g.~\citealt{HM2015,Schaan2015})
--- i.e.~those baryons that reside in diffuse, highly ionized intergalactic media (see e.g.~\citealt{McGaugh2008}).
Conversely, peculiar velocities estimated from the kSZ effect, together with external constraints on cluster astrophysics, provide independent measurements of the amplitude and growth rate of density perturbations.
The latter in turn can be used to test models of dark energy, modified gravity \citep{Keisler2012, MaZhao2014, Mueller2014a, Bianchini2015} and massive neutrinos \citep{Mueller2014b}.

The first detection of the kSZ signal was reported in \citet[H12 henceforth]{Hand2012}, using high-resolution CMB data from the Atacama Cosmology Telescope (ACT, \citealt{ACT}) in conjunction with the Baryon Oscillation Spectroscopic Survey (BOSS) spectroscopic catalogue \citep{2012ApJS..203...21A}.
To isolate the kSZ signal, H12 applied a differential (or pairwise) statistical approach, which we also adopt in this paper.
H12 rejected the null hypothesis of zero kSZ signal with a $p$-value of 0.002.
Subsequently, the \Planck collaboration \citep{Planck_KSZ} used the Central Galaxy Catalog derived from the Sloan Digital Sky Survey \citep{Abazajian:2008wr} to report $1.8-2.5\sigma$ evidence for the pairwise kSZ signal with a template fit.
Other recent detections (${\sim} 3\sigma$) of the kSZ signal have been obtained via cross-correlation of CMB maps with velocity fields reconstructed from galaxy density fields \citep{Planck_KSZ,Schaan2015}; see also \citet{Li2014} for a demonstration of this method using simulations.
Indirect evidence for a kSZ component in the CMB power spectrum was also seen in power spectrum measurements from the South Pole Telescope (SPT, \citealt{George2014}).
Lastly, the kSZ signal has been measured locally for one individual cluster by \cite{Sayers2013}.

In this work, we measure the pairwise kSZ signal by combining
a catalogue of galaxy clusters derived from the Dark Energy Survey (DES; \citealt{DES}, \citealt{DESnonDEoverview}) Year 1 data with a CMB temperature map from the 2,500 square degree South Pole Telescope Sunyaev-Zel'dovich (SPT-SZ) Survey.
Our paper is organised as follows: in Section~\ref{sec:theory} we briefly review the kSZ effect and the theory of pairwise velocities, and derive an analytic template for the pairwise kSZ effect. Section~\ref{sec:data} introduces the two input data sets from DES and SPT
and in Section~\ref{sec:analysis} we detail the analysis methods.
In Section~\ref{sec:sims} we briefly describe the new suite of realistic high-resolution kSZ simulations by \citet{Flender2015} and validate the pairwise kSZ template and the analysis methods on these simulations.
We proceed by showing our main results and comparing them both to analytic theory and the expectation from simulations in Section~\ref{sec:results}. The various checks and different null tests that we perform to demonstrate the robustness of our results against systematic uncertainties are described in Section~\ref{sec:systematics}. Finally, we discuss the implications of our detection for cluster astrophysics in Section~\ref{sec:interpretation}.

Unless otherwise specified, we use the \Planck 2015 TT+TE+EE+lowP cosmological parameters,
i.e. the Hubble parameter $H_0 = 67.3$ km s$^{-1}$ Mpc$^{-1}$, cold dark matter density $\Omega_c h^2 = 0.1198$, baryon density $\Omega_b h^2 = 0.02225$, current root mean square (rms) of the linear matter fluctuations on scales of $8~h^{-1}\Mpc$, $\sigma_8= 0.831$, and spectral index of the primordial scalar fluctuations $n_s=0.9645$ \citep{Planck2015params}, to compute theoretical predictions and to translate redshifts into distances.

\section{Theory}
\label{sec:theory}
\subsection{The pairwise kSZ effect}
In the non-relativistic limit and assuming only single scatterings for individual photons,
the kSZ effect produced by a galaxy cluster $i$ observed in the angular direction $\mathbf{\hat n}_i$
corresponds to a change in the CMB temperature $T_{\mathrm{CMB}}$ given by
\beq
\frac{\Delta T}{T_{\mathrm{CMB}}} (\nhat_i) = - \tau_{e,i} \frac{\hat{\mathbf{r}}_i \cdot \mathbf{v}_i}{c} \, ,
\label{eq:ksz}
\eeq
 \citep{SZ1980}.
 Here $\hat{\mathbf{r}}_i \cdot \mathbf{v}_i$ is the projection of the cluster velocity $\mathbf{v}_i$ along the line of sight $\hat{\mathbf{r}}_i$, and $c$ is the speed of light. The Thomson optical depth $\tau_{e,i}$ for CMB photons passing through a cluster $i$ is given by the line-of-sight integral of the free electron number density $n_{e,i}$,
\beq
\tau_{e,i}  = \int  \mathrm{d}l\htwo  n_{e,i}(\mathbf{r}) \sigma_T \, ,
\eeq
where $\sigma_T$~is the Thomson cross section. Therefore the kSZ effect probes the
bulk momentum of the ionized cluster gas projected onto the line of sight.

Measuring the velocities of individual clusters is currently only possible in rare exceptions (e.g.~in the detection by \citealt{Sayers2013}) since the kSZ signal has the same spectral shape as the primary CMB, and its amplitude is small compared to the tSZ amplitude.
This has motivated alternative methods of isolating the kSZ signal.
On scales smaller than the homogeneity scale, clusters will --- on average --- fall towards each other under their mutual gravitational attraction (e.g. \citealt{Bhattacharya2006, Bhattacharya2007}).
Because of the kSZ effect, this pairwise motion creates a particular pattern in the CMB, consisting of temperature increments and decrements at the locations of clusters moving towards and away from the observer, respectively (e.g. \citealt{Diaferio2000}).
The CMB pattern caused by such motion of cluster pairs is called the \textit{pairwise} kSZ signal.

Whereas the kSZ signal from one individual cluster is sensitive to the line-of-sight velocity of that cluster, the amplitude of the pairwise kSZ signal from a sample of clusters at comoving pair separation $r$ can be related to the mean relative velocity $v_{12}(r)$ of the clusters (independent of the line of sight; see equation~\ref{eq:v12}).
We write the pairwise kSZ amplitude as
\beq
T_{\mathrm{pkSZ}}(r) \equiv \bar{\tau}_e \, \frac{v_{12}(r)}{c} \, T_{\mathrm{CMB}} \, ,
\label{eq:templ}
\eeq
where $\bar{\tau}_e$
is the average optical depth of the cluster sample.\footnote{Note that with equation~\ref{eq:templ} we implicitly make the ansatz $\langle \tau_e v \rangle \simeq \langle \tau_e \rangle \langle v \rangle$. We further discuss this assumption in Section~\ref{subsec:uncert}.}
In our sign convention $T_{\mathrm{pkSZ}} < 0$ indicates that clusters are on average approaching each other ($v_{12} <0$).
We first build a model for $v_{12}(r)$ in Section~\ref{subsec:pairwiseModel}, then relate the line-of-sight velocities inferred from the data to the total signal $T_{\mathrm{pkSZ}}(r)$ in Section~\ref{sec:pkszest}.

\subsection{Modelling the pairwise velocity of clusters}
\label{subsec:pairwiseModel}
Clusters of galaxies --- or, more generally, the dark matter haloes that host them --- are located at the peaks of the cosmic density field.
The latter is described by the overdensity $\delta (\mathbf{x}) \equiv \rho(\mathbf{x}) / \bar \rho - 1$ with the matter density $\rho(\mathbf{x})$ and its mean $\bar \rho$.
Similarly, the overdensity of haloes is $\delta_h (\mathbf{x}) \equiv n(\mathbf{x}) / \bar n - 1 $, where $n(\mathbf{x})$ is the number density of haloes and $\bar n$ their mean density.
At linear level,
and under the assumption of deterministic, local bias \citep{Fry1993},
$\delta_h (\mathbf{x})$ can be related to $\delta (\mathbf{x})$ as $\delta_h (\mathbf{x}) = b_h \, \delta (\mathbf{x})$, where $b_h$
is the linear halo bias.

The total apparent velocity  $\mathbf{v} (\mathbf{x}) $ of a dark matter particle can be decomposed into the Hubble flow and a peculiar velocity $\mathbf{u} (\mathbf{x}) $ as $\mathbf{v} (\mathbf{x}) = a H \mathbf{x} + \mathbf{u} (\mathbf{x})$, where $a$ is the scale factor in the Friedmann-Lema\^{i}tre-Robertson-Walker metric, and {\it H} is the Hubble rate at scale factor $a$.
In linear perturbation theory, the velocity field is completely described by its divergence $\vartheta (\mathbf{x}) \equiv \nabla \cdot \mathbf{u} (\mathbf{x}) $. The linearised continuity equation  relates the density and velocity fields \citep{Bernardeau2002}:
\beq
 \vartheta  (\mathbf{x}) = -  \dot \delta (\mathbf{x})  =  - a H f \delta (\mathbf{x}) \, ,
\label{eq:thetadelta}
\eeq
where the dot denotes a derivative with respect to conformal time.
Furthermore, $f$ is the growth rate of density perturbations defined as \mbox{$f \equiv \mathrm{d} \ln D/ \mathrm{d} \ln a $}, where $D$ is the linear growth factor.

If the density and velocity fields are assumed to be Gaussian, their statistical properties are specified completely by their two-point statistics.
The two-point correlation function of the matter density perturbations at comoving separation $r$
is given by $\xi (r) \equiv \langle \delta (\mathbf{x}) \, \delta (\mathbf{x} + \mathbf{r})  \rangle$, while the
power spectrum $P(k)$ in Fourier space is defined as
$ \langle \delta (\mathbf{k}) \, \delta (\mathbf{k}')\rangle = (2 \pi)^3 \delta_D (\mathbf{k} + \mathbf{k}') P(k) $,
where $\delta_D$ is the Dirac delta distribution.
Using equation \ref{eq:thetadelta} and the relation between the power spectrum and the two-point correlation function, the density-velocity correlation function can be written as
\beq
\xi^{\delta v}(r,a) \equiv \langle \delta(\mb{x}) \, \hat{\mb{r}} \cdot \mb{v}(\mb{x}+\mb{r}) \rangle = -\frac{aHf}{2\pi^2} \int_0^\infty \mathrm{d}k\, kP(k,a)j_1(kr) \, ,
\label{eq:xi_deltav}
\eeq
where $j_1$ is a spherical Bessel function.

The mean pairwise velocity of haloes, $v_{12}(r)$, is a measure of the relative velocities of density peaks (e.g. \citealt{Davis1977, Peebles1980}).
Following \cite{Schmidt2010}, we now write $v_{12}(r)$ as
\begin{align}
v_{12}(r) & = \frac{\langle n(\mb{x}_1) \, n(\mb{x}_2) \, \hat{\mb{r}} \cdot (\mb{v}_2-\mb{v}_1) \rangle }{\langle n(\mb{x}_1) \, n(\mb{x}_2) \rangle} \, .
\label{eq:v12}
\end{align}
Assuming that the velocities of haloes are unbiased\footnote{This assumption will eventually become inaccurate for small pair separations ($r \lesssim 50$ Mpc), where haloes are found to have biased velocities \citep{Baldauf2014}. However, these scales do not contribute significantly to our analysis as a result of the photometric redshift uncertainties.}, equation~\ref{eq:v12} becomes
\beq
v_{12}(r) = \frac{ \langle [1 + b_h \delta(\mb{x}_1) ] \, [1 + b_h \delta(\mb{x}_2)] \, \hat{\mb{r}} \cdot (\mb{v}_2-\mb{v}_1)   \rangle }  {1+\xi_{h} (r)} \, ,
\label{eq:v12xi}
\eeq
where $\xi_{h}(r)$ is the two-point correlation function of haloes.
For comparison with observations and simulations, we are interested in the mean pairwise velocity of a cluster sample within a given mass range
(e.g.~\citealt{Bhattacharya2007,Mueller2014a}).
Here we model this by replacing $b_h$ with the mass-averaged bias
\beq
b \equiv \frac{ \int_{M_\mathrm{min}}^{M_\mathrm{max}} \mathrm{d} M \, M \, n(M) \, b_h(M)}{ \int_{M_\mathrm{min}}^{M_\mathrm{max}} \mathrm{d} M \, M \, n(M)} \, ,
\label{eq:mwbias}
\eeq
which on the scales of interest here is a good approximation for the halo bias moments of \cite{Bhattacharya2007,Mueller2014a}.
To evaluate equation~\ref{eq:mwbias}, for $n(M)$ we use the halo mass function of \citet{Tinker2008} computed with the \textsc{hmf} code \citep{Murray2013}, and for $b_h(M)$ the halo bias model by \cite{Tinker2010}.

Neglecting terms including three-point correlations and approximating \mbox{$\xi_{h} \approx b^2 \, \xi$}, equation~\ref{eq:v12xi} reduces to
\beq
v_{12} (r,a) \simeq \frac{2 \, b \, \xi^{\delta v}(r,a)}{1 + b^2 \, \xi(r,a)} \, .
\label{eq:v12_nonlin}
\eeq

As a consistency check, we note that this expression is equivalent to previous derivations of the mean pairwise velocity from the pair conservation equation (e.g.~\citealt{Peebles1980}):
Here the mean pairwise streaming velocity can be written as (\citealt{Sheth2000}; see also \citealt{Bhattacharya2007,Mueller2014a})
\beq
v_{12}(r,a) \simeq -\frac{2}{3} a \, r \, H \, f \, \frac{b \bar{\xi}(r,a)}{1+b^2\xi(r,a)} \, ,
\label{eq:v12_halo}
\eeq
where $\bar{\xi}(r,a)$ is the correlation function averaged within a sphere of comoving radius $r$ and $b$ is again an averaged bias. In linear theory, $\xi^{\delta v}(r,a) = - arHf \bar{\xi}(r,a)/3$, demonstrating the equivalence of equations~\ref{eq:v12_nonlin} and~\ref{eq:v12_halo}.

We now use equation \ref{eq:v12_nonlin} to predict the mean pairwise velocity in our theoretical template.
For the computation of $\xi$ and $\xi^{\delta v}$ we evaluate $P(k)$ using the \textsc{pycamb}\footnote{\url{https://github.com/steven-murray/pycamb}} interface to the \textsc{Camb}\footnote{\url{http://camb.info}} code \citep{CAMB}.
As we are interested only in the large-scale behaviour of the pairwise velocities, we remove small-scale fluctuations by smoothing the power spectrum with a spherical top-hat filter of radius $R = 3~h^{-1}\Mpc$.
This procedure ensures that there are no unphysical oscillations in the theory prediction for $v_{12}(r)$, which would otherwise be caused by a sharp cut-off of the highest-$k$ modes. We have checked that our results are insensitive to the exact choice of the smoothing scale.
We demonstrate with the use of mocks in Section~\ref{sec:validation} below that the model of equations~\ref{eq:templ}, \ref{eq:v12_nonlin} describes the simulations well on scales $r \gtrsim 40$ Mpc in the absence of redshift uncertainties.

\subsection{Modelling the photo-$z$ uncertainties}
\label{sec:photoz}
To account for the uncertainty in the photometric cluster redshift (photo-$z$),
we modify the pairwise kSZ template (equations~\ref{eq:templ}, \ref{eq:v12_nonlin}) to model the dilution of the signal on small scales.
The comoving distance to a cluster at redshift $z$ is given by \mbox{$d_c(z) = \int_0^z  \mathrm{d}z'\htwo c/H(z')$}, so uncertainties in cluster redshifts are converted to errors in their distances by
$\Delta {d_c} \simeq c \Delta z/H(z)$.
Using the redshift errors from the cluster catalogue described in Section~\ref{subsec:redmapper} below
and assuming they follow a roughly Gaussian distribution (see \citealt{Rykoff2016}),
we compute an rms uncertainty in the comoving distance, $\sigma_{d_c}$.
For the sample used in this work we find $\sigma_{d_c} \simeq 50~\Mpc$.
The corresponding uncertainty in the separation of the cluster pairs affects the recovered pairwise kSZ signal: redshift errors completely dilute the signal at $r \ll \sigma_{d_c}$,
 the signal is significantly reduced on scales $r  \sim \sigma_{d_c}$, and only the signal from cluster pairs with $r \gg \sigma_{d_c}$ remains unaffected.
Given that the pairwise kSZ signal is strongest at small separations ($r \lesssim 50$ Mpc), this poses one of the main challenges for our analysis.

We account for this effect heuristically by multiplying the template with a smoothing factor that models the dilution of the pairwise kSZ signal on small scales.
The final template is thus:
\beq
T_{\mathrm{pkSZ}}(r,a) =  \bar{\tau}_e \, \frac{T_{\mathrm{CMB}}}{c} \, \frac{2 \, b \, \xi^{\delta v}(r,a)}{1+b^2 \, \xi(r,a)} \times \left[ 1 - \exp\left(-\frac{r^2}{2 \, \sigma_r^2}\right) \right].
\label{eq:templ_photoz}
\eeq
The pair separation smoothing scale $\sigma_r$ --- i.e.  the rms uncertainty in cluster pair separation --- can be inferred from the redshift errors of the cluster sample; here we approximate it by setting $\sigma_r = \sqrt{2} \sigma_{d_c}$.\footnote{This is a reasonable approximation because the pairwise estimator assigns the highest weights to cluster pairs with separation along the line of sight and $\sigma_z$ is a relatively slowly growing function of $z$.}
We demonstrate in Section~\ref{sec:validation} below that the full model of equation~\ref{eq:templ_photoz}  provides a good fit to realistic simulations that include photometric redshift errors.

\subsection{Dependence on astrophysics and cosmology}
\label{subsec:cosmodependence}
Equation~\ref{eq:templ_photoz} shows that the pairwise kSZ signal constrains a combination of cluster astrophysics ($b \bar{\tau}_e$) and cosmology ($\xi^{\delta v}$, $\xi$).
While the signal shape is mostly specified by the cosmology, the amplitude is scaled by $b$ and $\bar{\tau}_e$;
by fixing the cosmological parameters and the bias prescription, a pairwise kSZ measurement can constrain the mean optical depth of the cluster sample.
This argument can also be turned around: given measurements of $b$ (e.g.~from the cluster auto-correlation) and $\bar{\tau}_e$ (e.g.~from X-ray or tSZ observations), constraints on cosmology can be derived. To illustrate the parameter dependence, we consider large scales
($r \gtrsim 60$ Mpc, see Section~\ref{sec:validation} below).
Here $b^2\xi(r) \ll 1$ so that
\beq
T_{\mathrm{pkSZ}}(r) \propto b \, \bar{\tau}_e \, \xi^{\delta v}(r) \propto b \, \bar{\tau}_e \, f \, \sigma_8^2 \,
\eeq
and hence the shape of the signal is completely specified by the cosmology via $\xi^{\delta v}(r)$.
From a cosmological perspective, the dependence on $f \sigma_8^2$ is particularly interesting. Other dynamical probes like redshift space distortions constrain primarily the combination $f \sigma_8$ (see e.g.~\citealt{Percival2008}), so a measurement of the pairwise kSZ could be used to break the degeneracy between growth and initial amplitude. The resulting constraints on $f$ could in turn be used as a probe of dark energy or modifications of gravity \citep{Keisler2012}.

\section{Data}
\label{sec:data}
\subsection{DES redMaPPer cluster catalogue}
\label{subsec:redmapper}

The Dark Energy Survey is an ongoing 5-band {\it grizY} photometric survey of $5,000$ deg$^2$ of the Southern sky that uses the Dark Energy Camera \citep{Flaugher2015} on the 4-metre Blanco Telescope at Cerro Tololo Inter-American Observatory (CTIO).
At full depth, DES will allow the extraction of cluster catalogues that are complete for clusters with $M_{500c} \gtrsim 10^{14} M_\odot$
--- where $M_{500c}$ is the mass within a spherical region with an average density of 500 times the critical density ---
out to redshifts $z < 0.9$ \citep{Rykoff2016}.

\begin{figure}
\begin{center}
	\includegraphics[width=\columnwidth]{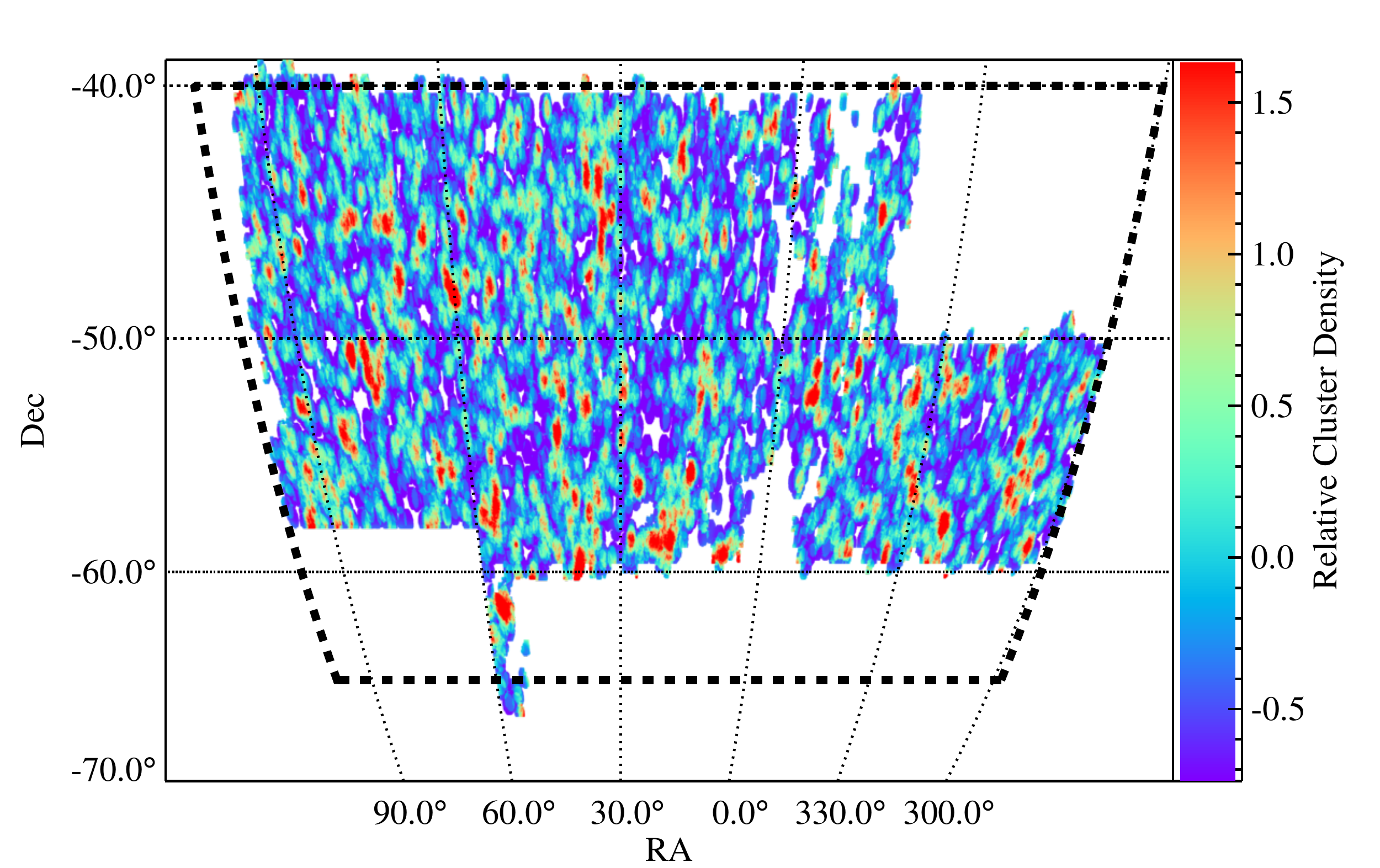}
	\caption{Relative cluster density (smoothed on a $30'$ scale for visualisation purposes) in the overlapping regions of the DES Y1 cluster catalogue and the SPT-SZ temperature map. The dashed black line marks the boundaries of the SPT-SZ survey footprint. The effective sky are for this analysis is ${\sim} 1,200$ deg$^2$.}
	\label{fig:skyplot}
\end{center}
\end{figure}

The sky footprint of DES was chosen to have almost complete overlap with the SPT-SZ survey region. This has already enabled several studies cross-correlating DES Science Verification data with SPT-SZ \citep{Giannantonio2015,Saro2015,Kirk2015,Baxter2016}.
Here we use data from the full first year of DES observations (Y1) that overlap with ${\sim} 1,400$ deg$^2$ of the SPT-SZ footprint.
The area in which clusters can be identified is reduced by $10-15\%$ because of boundary effects and masking of bright stars \citep{Rykoff2016}.
Thus the effective sky area for this analysis is ${\sim} 1,200$ deg$^2$, which we show in Fig.~\ref{fig:skyplot}.

This work uses a cluster catalogue constructed from DES data using the {\bf red}-sequence {\bf Ma}tched-filter {\bf P}robabilistic {\bf Per}colation
(redMaPPer) algorithm first described by \cite{Rykoff2013} and subsequently developed and tested against X-ray and SZ-based cluster catalogues \citep{Rozo2013,Rozo2014a}.
RedMaPPer is a photometric red-sequence based cluster finder that is trained on a sub-sample of clusters with spectroscopic redshifts to calibrate the red sequence model.
The cluster finder provides 3D cluster positions, where the angular coordinates are taken to be at the algorithm's best estimate for the central galaxy position
and the redshifts are estimated photometrically.

RedMaPPer also provides an optical richness estimate, $\lambda$, which is a low-scatter optical proxy for the cluster mass \citep{Rykoff2012, Saro2015}.
To account for partially masked clusters or member galaxies fainter than the limiting magnitude, redMaPPer applies a correction factor $s$ to the optical richness, such that $\lambda = s\tilde{\lambda}$ where $\tilde{\lambda}$ are the raw galaxy counts. Because a sample with more uniform noise properties is obtained if cuts are applied in $\tilde{\lambda}$ \citep[see][for details]{Rykoff2013,Rozo2013,Rozo2014a,Rykoff2016},
we apply richness cuts using this quantity.

The full DES Y1 redMaPPer `gold' catalogue spans the photo-$z$ range $0.1 < z < 0.95$.
As both completeness and  photo-$z$ accuracy degrade at high redshift,
for the main result of this work we use only clusters at $z < 0.8$.
This $z <  0.8$ sample is still not entirely pure (due to scatter in the richness estimate), nor is it complete to the same richness at high redshift due to depth limitations:
the effective richness threshold grows with redshift,
reaching $\lambda_{\min} \simeq 40$ at $z=0.8$.
To test for a potential bias caused by this evolution, we repeat the analysis with a more conservative redshift cut of $z < 0.65$, obtaining consistent results.

\begin{figure*}
\begin{center}
	\includegraphics[width=\columnwidth]{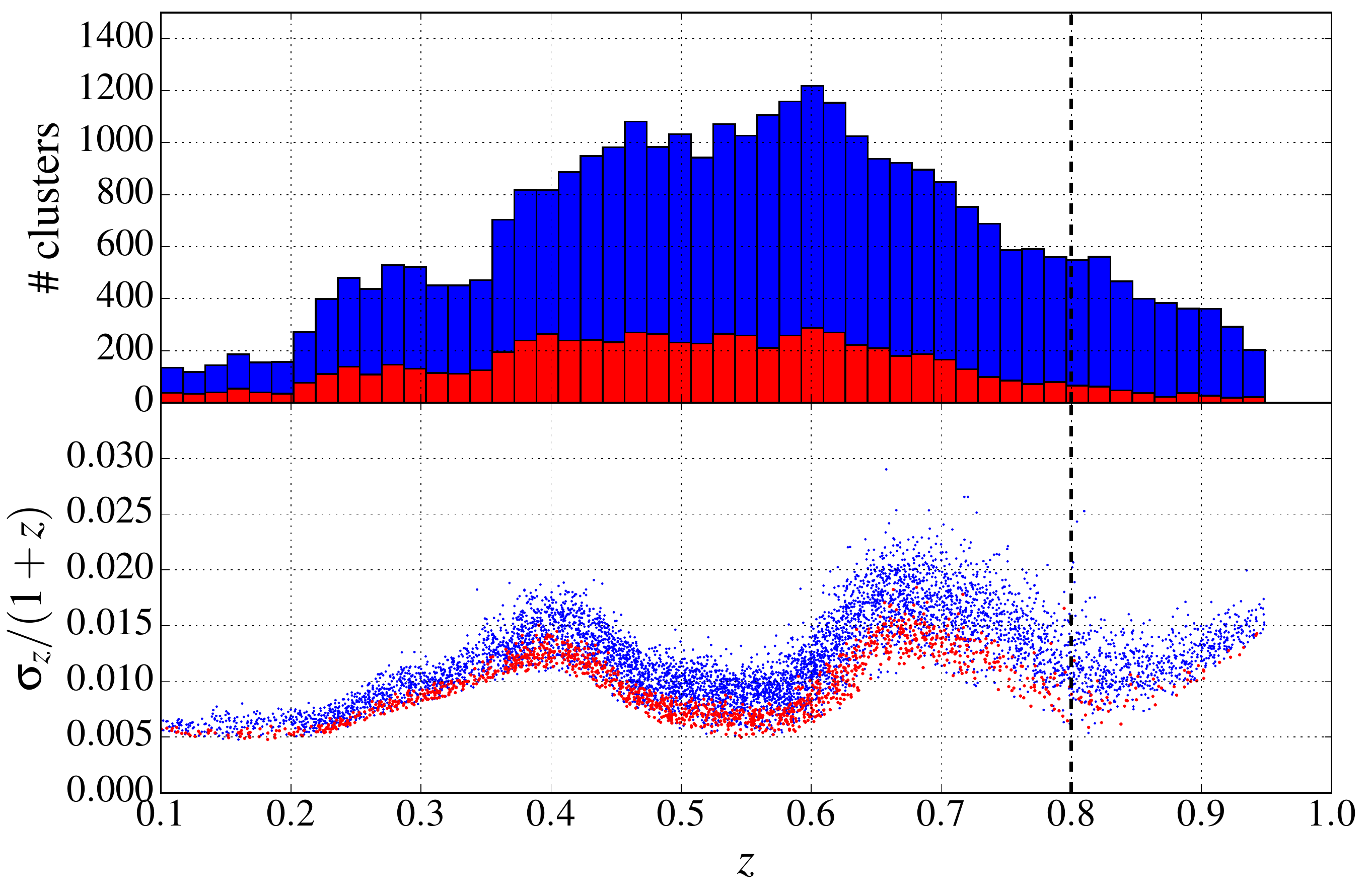}
	\includegraphics[width=\columnwidth]{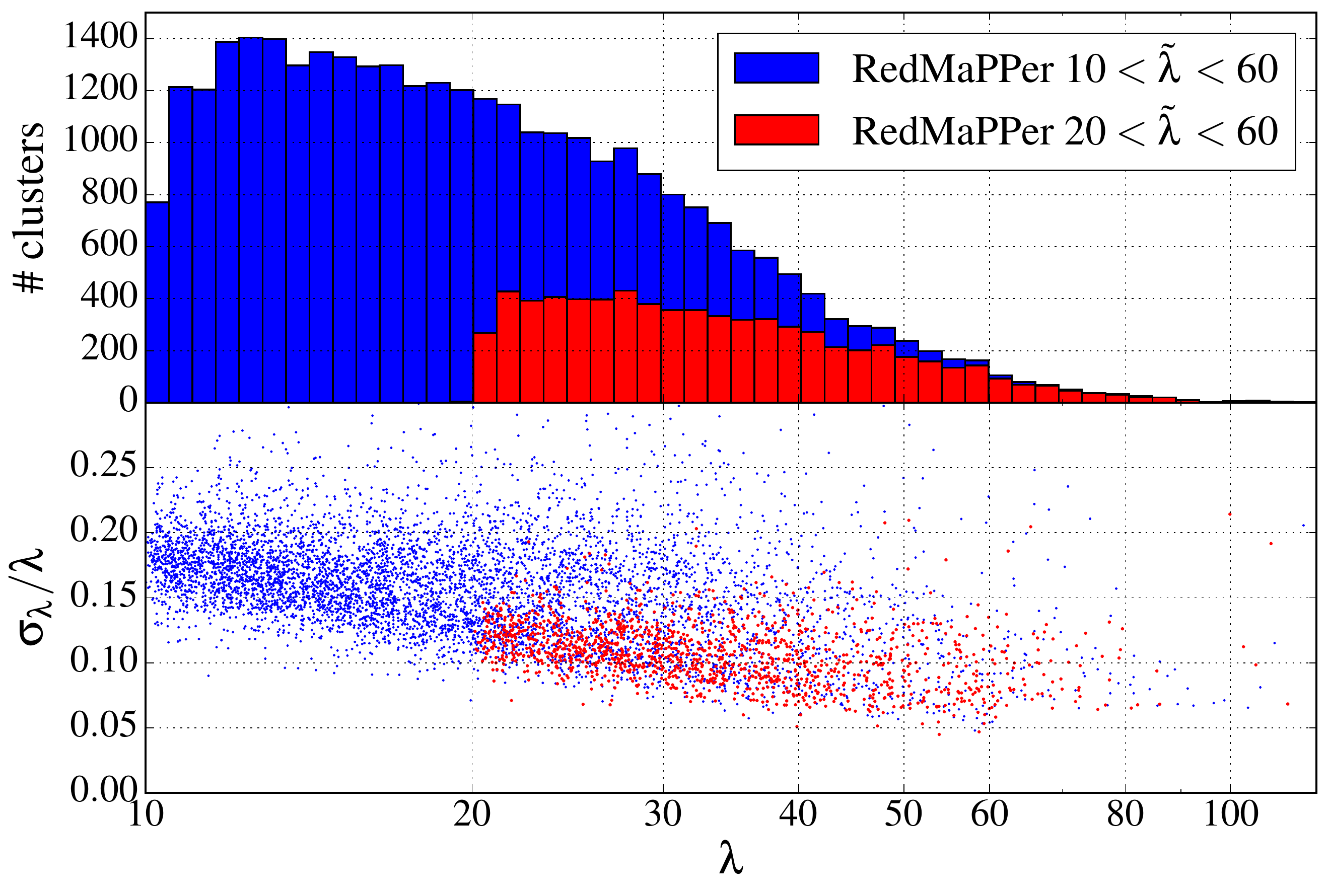}
	\caption{
		Properties of the DES Y1 redMaPPer clusters with $\tilde{\lambda} > 10$ (blue) and $\tilde{\lambda} > 20$ (red) member galaxies. For visualisation purposes we only show every fifth cluster in the bottom panels.
		\textbf{Left:}
		We show in the upper panel the photometric redshift distribution, and in the lower panel the fractional uncertainty on the photo-$z$s as a function of redshift.
		The increased level of photo-$z$ errors around $z=0.4$ ($z=0.7$) are caused by the transition of the 4000~\AA-break between the $g$ and $r$ ($r$ and $i$) bands.
		\textbf{Right:}
		We show in the upper panel the richness distribution, and in the lower panel the fractional uncertainty on the richness $\lambda$ as a function of $\lambda$.
         Since we apply cuts using the raw galaxy counts $\tilde{\lambda}$, the $\tilde{\lambda} > 20$ sample does not include all clusters with richness $\lambda > 20$.
        This (more conservative) cut removes primarily objects with higher richness uncertainties.}
	\label{fig:redmapper}
\end{center}
\end{figure*}

We use all clusters in the richness range $20 < \tilde{\lambda} < 60$
--- where $\lambda \simeq 20$ broadly corresponds to $M_{500c} \simeq 10^{14} M_\odot$ ---
for the main analysis, while also considering a larger catalogue extended to lower richness  $10 < \tilde{\lambda} < 60$ in Section~\ref{subsec:richnlimits}.
The low-$\tilde{\lambda}$ cutoff is driven by several competing factors.
As the mass function grows exponentially at the low-mass end, so does the number of clusters (and hence the number of pairs).
On the other hand, low richness clusters are more susceptible to errors in redshift and centring, which washes out the signal.
The high-$\lambda$ cutoff is set because the tSZ strongly dominates the signal in high-mass clusters.
Including these rare, heavy objects in the sample could result in an insufficient cancellation of the tSZ signal and could therefore bias the pairwise kSZ measurement.
We have found with the simulations that the upper cut chosen here maximizes the signal-to-noise, while not introducing a significant bias into the pairwise kSZ amplitude.
A potential contamination by tSZ is further investigated in Section~\ref{subsec:tszcont} below.

We finally remove clusters that fall within masked regions surrounding point sources detected in the SPT-SZ maps (see next section).
After these cuts, the catalogue contains 6,693 (28,760) clusters in the SPT-SZ footprint for the $\tilde{\lambda} > 20$ ($\tilde{\lambda} > 10$) sample; the corresponding surface density is $5.6$ deg$^{-2}$ ($24$ deg$^{-2}$).
We have tested the robustness of the results with respect to the applied cuts.

We show in Fig.~\ref{fig:redmapper} the distribution of clusters in redshift and optical richness, as well as the associated errors, for both the high- and low-richness samples.
The left panel of Fig.~\ref{fig:redmapper} shows that the photo-$z{\ss}$ errors vary for both cluster samples as a function of redshift; there are two redshift ranges where the photo-$z$ quality is significantly degraded: around $z \simeq 0.4$ and $z \simeq 0.7$.
This is due to the typical red-sequence galaxy spectral feature, the 4000 \AA-break, transitioning between the DES photometric bands.
Over the whole redshift range, we observe that the photo-$z$ errors of the high-richness sample are  $\sigma_z/(1+z) \in [0.005, 0.015]$, significantly smaller than in the low-richness case, for which $\sigma_z/(1+z) \in [0.005, 0.025]$; see also the bottom left panel of Fig.~\ref{fig:redmapper}.
As the pairwise kSZ signal is smoothed at cluster separations comparable with the photo-$z$ errors (see Section~\ref{sec:photoz}), this motivates our choice of the $\tilde{\lambda} > 20$ sample for the main analysis.

\subsection{SPT-SZ temperature maps}
The CMB data consist of temperature maps made from the SPT-SZ observations.
This analysis uses the same CMB maps as were used by \citet[hereafter B15]{bleem15};
we summarise the salient points here, and refer the reader to B15 for a complete description.

The SPT is a 10 m-diameter millimetre-wave telescope located at the National Science Foundation Amundsen-Scott South Pole Station in Antarctica.
Between 2008 and 2011, the SPT was used to observe a contiguous ${\sim} 2,500$ deg$^2$ region of sky at 90 GHz, 150 GHz, and 220 GHz at arcminute resolution
to approximate map depths of 40, 18, and 70 $\mu$K-arcmin, respectively.
The survey covers a region from 20$^h$ to 7$^h$ in right ascension (R.A.) and $-65 ^\circ$ to $-40 ^\circ$
in declination (see e.g. \citealt{story13}).
This analysis uses the 150 GHz data in the region that overlaps with the DES Y1 cluster catalogue (${\sim} 1,200$ deg$^2$, see Fig.~\ref{fig:skyplot}).
In principle, it would be possible to combine the three SPT frequencies to reduce sensitivity to noise and tSZ signal in the CMB maps; however, the 90 and 220 GHz data are significantly noisier than the 150 GHz band so that combining them would introduce additional complexity without a substantial gain.
\footnote{See, for example, the recent analysis of \citet{baxter15} which combined data at all three frequencies to construct maps free of tSZ-signal.  The resulting maps had a noise level of 55 $\mu$K-arcmin, much too high for a significant detection in this analysis.}

The survey was observed in 19 sub-patches, or \textit{fields}.
Each field was observed in at least 200 individual \textit{observations}, and maps from individual observations are coadded into a single map for each field.
The detector time-ordered-data are filtered, then projected using the telescope pointing model into maps with the Sanson-Flamsteed projection \citep{calabretta02}.
In an update from B15, absolute calibration of the maps was derived using the 2015 release of the {\it Planck} 143 GHz data \citep{Planck15Like}.
Emissive point sources are masked as follows:
a 4\arcmin\, (10\arcmin) radius mask is applied to all sources detected at $\ge 5 \sigma$ ($\ge 20 \sigma$) in the 150 GHz maps (see, e.g. \citealt{Mocanu13}),
removing ${\sim} 7\%$ ($\ll 1\%$) of the initial 2,500 deg$^2$ survey area.
Any clusters within these masks are removed from the kSZ analysis.

\section{Analysis methods}
\label{sec:analysis}

\subsection{Matched filtering and temperature estimates}
\label{sec:filter}

Since we have prior knowledge about the shape of the expected SZ signal from clusters, we can improve the signal-to-noise by filtering the CMB maps to suppress noisy modes and enhance modes with high expected signal.
We use a matched filter technique (see e.g. \citealt{Haehnelt1996,Melin2006}) identical to that used in B15.
The observed temperature in a direction $\nhat$ can be written as
\begin{equation}
  T(\nhat) = B(\nhat) \ast \left[ T_{\mathrm{clust}}(\nhat) + n_{\mathrm{astro}}(\nhat)\right]  + n_{\mathrm{noise}}(\nhat) \,.
\end{equation}
Here $B(\nhat)$ characterises the effect of the instrumental beam and data filtering
and the asterisk denotes a convolution.
The SZ signal from clusters $T_{\mathrm{clust}}(\nhat)$ is comprised of tSZ and kSZ components and
$n_{\mathrm{astro}}(\nhat)$ characterises the noise contribution from astrophysical sources that include the lensed primary CMB, emission from dusty extragalactic sources, as well as the kinematic and thermal SZ background;
all components are treated as Gaussian noise and modelled from previous SPT power spectrum measurements (\citealt{keisler11}, \citealt{shirokoff11}).
Emission from cluster members and the effect of lensing of dusty background galaxies by the cluster is removed on average by the pairwise estimator (see Section~\ref{sec:pkszest} below);
radio sources below the SPT detection threshold contribute negligibly to the maps, and are ignored.
Finally, noise from the instrument and the atmosphere is given by $n_{\mathrm{noise}}(\nhat)$.

For the purposes of filtering, we model the SZ signal from a single cluster as an amplitude $T_0$ times
an azimuthally symmetric real-space profile $\rho(\theta)$, where $\theta$ is the angular distance from the cluster centre.
For the cluster profile template $\rho(\theta)$, we use a projected isothermal $\beta$-model \citep{cavaliere76} with $\beta=1$ given by
\begin{equation}
\label{eq:beta_prof}
T(\theta) = T_0 \, \left( 1 + \theta^2 / \theta^2_c \right)^{-1} \, ,
\end{equation}
where $T_0$ is the normalisation, and the core radius $\theta_c$ controls the size of the expected signal.
This choice of profile does not affect the detection significance of the analysis, see Section~\ref{subsec:filteringscale}.
Although the tSZ and kSZ components are expected to follow slightly different profile shapes, here we assume for simplicity this single model for both components.
Averaging over multiple pairs with the pairwise estimator (see Section~\ref{sec:pkszest} below), the tSZ signal will average to zero (ignoring redshift dependence), allowing the kSZ signal to be singled out.
The kSZ contribution to the profile normalisation, $T_0^\mathrm{kSZ}$, is related to the optical depth through the centre of the cluster via equation \ref{eq:ksz} as
$T_0^\mathrm{kSZ} = -\tau_0 \frac{\hat{\mathbf{r}} \cdot \mathbf{v}}{c} T_\mathrm{CMB} $.

We now build a filter $\Psi(\nhat)$ that returns an estimate $\hat{T}_0$ of $T_0$ when centred on the cluster at $\nhat_0$:
\begin{equation}
  \hat{T}_0 = \int{ d^2\nhat\, \Psi(\nhat - \nhat_0) \, T(\nhat) } \,.
\end{equation}
The filter is constructed in Fourier space and has the form \citep{Haehnelt1996, Melin2006}
\begin{equation}
  \Psi(\mbl) = \sigma_{\Psi}^{2} \mb{N}^{-1}(\mbl)\, S_{\text{filt}}(\mbl) \,.
\end{equation}
Here $\sigma_{\Psi}^{2}$ is the predicted variance in the filtered map, defined as
\begin{equation}
  \sigma_{\Psi}^2 \equiv \left[ \int{ d^2\mbl \,S_{\text{filt}}(\mbl)^{\dagger}\, \mb{N}^{-1}(\mbl)\, S_{\text{filt}}(\mbl)} \right]^{-1} \,.
\end{equation}
The Fourier-domain noise covariance $\mb{N}(\mbl)$ includes the contributions from $n_{\text{astro}}(\nhat)$ and $n_{\text{noise}}(\nhat)$.
The instrument and residual atmosphere noise $n_{\text{noise}}(\nhat)$ is calculated by differencing pairs of observations, then coadding the resulting difference-maps in each field of sky.
The expected signal $S_{\text{filt}}(\mbl)$ is calculated as the product of the Fourier-domain cluster profile template $\rho({\mbl})$ with $B(\mbl)$.

We explore filter sizes in the range $\theta_c \in [0.25', 10']$.
For the main analysis we adopt $\theta_c = 0.5'$;
the temperature estimates obtained with this filter are shown in Fig.~\ref{fig:temp-evol}.
This choice is well matched to the average cluster extent (see Sections~\ref{subsec:physical_tau} and~\ref{subsec:filteringscale} below).
Furthermore, we have found it to yield the maximum signal-to-noise ratio when extracting the pairwise kSZ signal
from the simulated CMB maps described in Section~\ref{sec:sims}.
The significance of our main result is relatively insensitive to the exact details of the theoretical cluster profile:
the filter shape is dominated by CMB confusion at large scales and the ${\sim} 1'$ instrumental beam at small scales,
thus the cluster profile impacts $\Psi(\nhat)$ over a relatively small range of scales.
We demonstrate the robustness of our detection to the profile size and shape in Section~\ref{subsec:filteringscale} below.

\subsection{Pairwise kSZ estimator}
\label{sec:pkszest}
\citet{Ferreira1998} showed that the mean pairwise velocity of a sample of objects such as clusters, $v_{12}(r)$, can be estimated from their individual line-of-sight velocities $\hat{\mathbf{r}}_i \cdot \mathbf{v}_i$ with the estimator
\beq
 \hat{v}_{12}(r) = \frac{\sum_{i<j,r} (\hat{\mathbf{r}}_i \cdot \mathbf{v}_i -\hat{\mathbf{r}}_j \cdot \mathbf{v}_j ) \, c_{ij}} {\sum_{i<j,r} c_{ij}^2} \, , \qquad c_{ij} = \hat{\mb{r}}_{ij} \cdot \frac{\hat{\mb{r}}_i + \hat{\mb{r}}_j}{2} \, .
\label{eq:ferreira}
\eeq
Here the geometrical factor $c_{ij}$ accounts for the projection of the pair separation $\mb{r}_{ij} \equiv \mb{r}_i - \mb{r}_j$ onto the line of sight, and the sum is taken over all cluster pairs with $i<j$ and distances $|\mb{r}_{ij}| = r$.
As a reminder, $v_{12}(r) < 0$ for clusters moving towards each other.

As the kSZ effect correlates the line-of-sight velocity of a cluster with the CMB temperature $T(\nhat)$ at its angular position $\nhat$,
we can combine equations~\ref{eq:ksz}~and~\ref{eq:ferreira} to form the pairwise kSZ estimator (H12):
\beq
\hat{T}_{\mathrm{pkSZ}}(r) = - \frac{\sum_{i<j, r} \left[ T(\nhat_i)-T(\nhat_j) \right ] c_{ij}}{\sum_{i<j,r} c_{ij}^2} \,.
\label{eq:pkszest}
\eeq
Residuals of the primary CMB, foreground and noise fluctuations are uncorrelated with cluster positions and hence add noise, but average out in the pairwise measurement.
The tSZ signal, as well as cosmic infrared background (CIB) emission correlated with the clusters, are also removed on average, thus adding noise but not bias for clusters in a narrow redshift range.

\begin{figure}
	\begin{center}
		\includegraphics[width=\columnwidth]{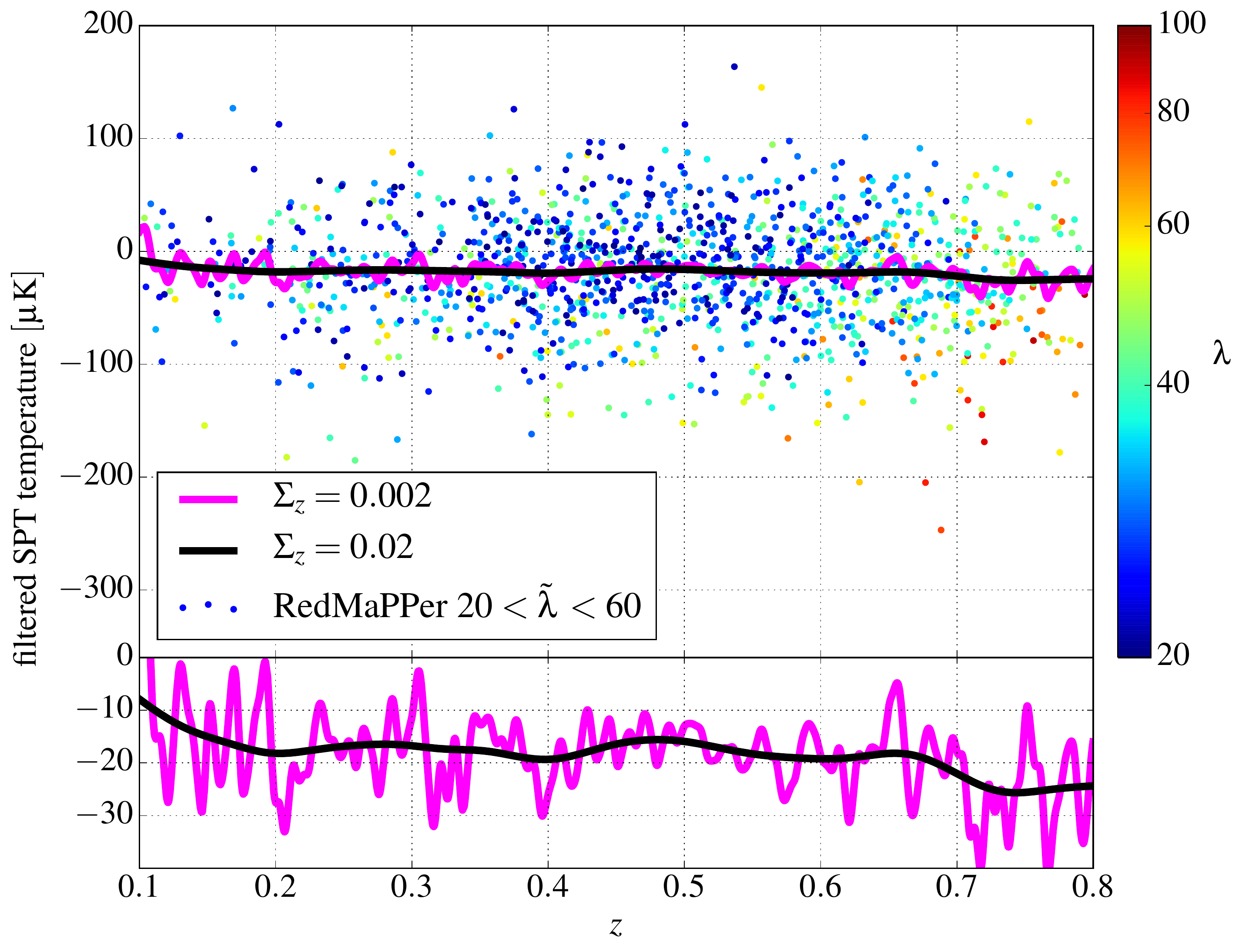}
		\caption{
			Filtered temperatures at cluster positions and correction for redshift dependence:
			\textbf{Top:} We show the CMB temperature deviations at cluster positions in the SPT-SZ map after match-filtering with a $\beta$-profile with $\theta_c = 0.5'$.
			The individual clusters are shown as a function of their redshift and colour-coded by their richness (for visualisation purposes we only show every fifth cluster).
			The solid overplotted lines are the smoothed mean temperature,
			i.e.~the second term on the rhs of equation~\ref{eq:T_zevolcorr},
			for $\Sigma_z = 0.02$ (used in the main analysis) and a smaller smoothing width of $\Sigma_z = 0.002$.
			\textbf{Bottom:} We show here the same mean temperature as in the top panel, but with a narrower temperature range, thus revealing an evolution of ${\sim} 15~\mu$K over the full redshift range. }
		\label{fig:temp-evol}
	\end{center}
\end{figure}

Over a larger redshift range, however, any evolution of these contributions with redshift would result in a bias.
Indeed, we have several known redshift-dependent components in our sample and analysis: as discussed in Sections~\ref{subsec:redmapper} and \ref{subsec:redshiftdep}, the mass-selection threshold of our sample evolves with redshift.
Furthermore, the adopted constant filter scale cannot match the average angular scale of a cluster at all redshifts and so, even in the absence of a change in the average cluster mass with $z$, the recovered temperature signal at the cluster positions will depend on redshift.
These redshift-dependent effects need to be subtracted to obtain an unbiased estimate of the separation-dependent pairwise kSZ signal (H12, \citealt{Planck_KSZ}).
We estimate and remove this bias by calculating the mean measured temperature as a function of redshift and subtracting it from the matched-filtered temperature values $\hat{T}_0(\nhat_i)$, as
\beq
\label{eq:T_zevolcorr}
T(\nhat_i) = \hat{T}_0(\nhat_i) - \frac{ \sum_j \hat{T}_0(\nhat_j) \, G(z_i,z_j,\Sigma_z)}{\sum_j G(z_i,z_j,\Sigma_z)} \, .
\eeq
The smoothed temperature at $z_i$ is calculated from the weighted sum of contributions of clusters at redshift $z_j$ using a Gaussian kernel \mbox{$G(z_i,z_j,\Sigma_z) = \exp \left [-(z_i - z_j)^2/ \left(2 \Sigma_z^2 \right)  \right]$}.
Here, we choose $\Sigma_z=0.02$, which results in smooth temperature evolution;
we have checked that our results are insensitive to this choice.
We show in Fig.~\ref{fig:temp-evol} the smoothed temperature as a function of redshift, i.e. the second term on the right-hand side of equation~\ref{eq:T_zevolcorr}.

For the main sample analysed in this work, we find only a weak trend with redshift in the range $0.2 \lesssim z \lesssim 0.7$;
this is mainly an effect of two competing processes:
at low redshift, our sample is nearly volume-limited, and as such it contains a higher number of more massive clusters due to the progress of structure formation;
at higher redshift, our sample becomes flux-limited, and by effect of sample selection, more massive clusters are more likely to be included in the sample.
However, the amplitude of the kSZ signal we are measuring is of the order of a few $\mu$K --- much smaller than the range of temperatures shown in Fig.~\ref{fig:temp-evol}.
We show for clarity in the bottom panel of Fig.~\ref{fig:temp-evol} that the change in the mean temperature with $z$ appears more considerable (${\sim} 15~\mu$K) when drawn on a temperature range that is closer to that of the kSZ measurements;
even such a weak trend with redshift could bias the results, if not appropriately subtracted.
The smoothed temperature is negative at all redshifts due to thermal SZ, but removing an overall negative offset does not affect the pairwise estimator.

We finally note that even if the filtered temperatures only consisted of CMB and/or noise residuals, the smoothed temperature would still change with redshift as the number of objects effectively contributing to the average evolves with $z$.
In that case, one would expect the smoothed temperature to fluctuate around zero, with the amplitude depending on the number density of objects as a function of $z$ (see Fig.~1 and corresponding discussion in \citealt{Planck_KSZ}).
The fact that --- unlike the \Planck analysis --- we observe a negative smoothed temperature at all redshifts, indicates that thermal SZ contributes significantly to the filtered temperatures.
However, it is important to note that the correction of equation~\ref{eq:T_zevolcorr} removes \textit{any} mean redshift evolution, regardless of its origin.

After correcting for these redshift dependent effects as described above, we measure the pairwise kSZ signal in 15 bins of comoving pair separation linearly spaced between 0 and 300 Mpc.

\subsection{Covariance estimates}
\label{sec:covariance}
In principle,
the covariance of the pairwise kSZ measurement can be estimated in a number of different ways. Analytic approaches for the covariance of pairwise cluster velocities were presented by \cite{Bhattacharya2007} and \cite{Mueller2014a}, but further work is necessary to realistically include contamination by primary CMB anisotropies and the tSZ signal;
the same holds true for the pairwise velocity covariance that \citet{Ma2015} compute from simulations.
As an alternative, the covariance could be estimated from random simulated realisations of the mm-sky with the correct power spectrum.
This procedure is insufficient, however, 
because it ignores the spatial correlation between the kSZ and tSZ signals, which causes the latter to be the largest source of uncertainty in our measurement (see Appendix~\ref{sec:decompos}).
Therefore a Monte Carlo covariance is not a good choice in our case.

On the other hand, one could generate the tSZ mock realisations from simulations in the same way as the one realisation we use for comparison with the real data (see Section~\ref{sec:sims} below).
This method becomes computationally expensive, however, for surveys that sample large volumes:
when preserving the survey geometry, only four independent projections of the DES-Y1$\times$SPT-SZ footprint can be generated from one full-sky kSZ simulation. Hence an unfeasibly large number of simulated kSZ skies (and therefore $N$-body simulations) would be required to obtain a reliable covariance estimation.

For these reasons, we adopt resampling techniques as our baseline choice for computing the covariance matrix; this approach also has the advantage of being model-independent.
We create jack-knife (JK) resamples of the pairwise kSZ measurement by splitting the cluster catalogue into $N_\mathrm{JK}$ subsamples, removing one of them, and recomputing the pairwise kSZ amplitude from the union of the remaining $N_\mathrm{JK}-1$ subsamples.
This process is repeated until every subsample has been removed from the measurement exactly once.
From the $N_\mathrm{JK}$ resamples we then estimate the covariance matrix as
\beq
\hat{C}_{ij}^\mathrm{JK} = \frac{N_\mathrm{JK}-1}{N_\mathrm{JK}} \, \sum_{\alpha = 1}^{N_\mathrm{JK}} (\hat{T}_i^\alpha - \bar{T}_i) \, (\hat{T}_j^\alpha - \bar{T}_j) \htwo ,
\label{eq:covmat}
\eeq
where $ \hat{T}_i^\alpha $ is the pairwise kSZ signal in separation bin $i$ and JK realisation $\alpha$, of mean $\bar T_i$.
For our main analysis, we use $N_\mathrm{JK} = 120$ samples.
In Appendix~\ref{sec:resamplingstab} we show that the error estimate is stable against changes in $N_\mathrm{JK}$.
Additionally, we also test and discuss alternative resampling schemes there, in particular a JK approach using sky patches and bootstrap resampling from the cluster catalogue and sky patches. We find that all four techniques give comparable results.

For the inverse of the covariance, we use the estimator
\beq
\hat{C}^{-1} = \frac{N-N_\mathrm{bins}-2}{N-1} (\hat{C}^\mathrm{JK})^{-1} \, ,
\eeq
where $N$ is the number of jack-knife or bootstrap samples used to compute the covariance, and $N_{\text{bins}}$ is the number of bins used in the measurement. The correction factor is necessary because $(\hat{C}^\mathrm{JK})^{-1}$ is a biased estimator of $C^{-1}$ \citep{Hartlap2006}.

\subsection{Amplitude fits and statistical significance}
\label{sec:significance}

We fit the pairwise kSZ signal measured with the estimator of equation~\ref{eq:pkszest} with a one-parameter template given by equation \ref{eq:templ_photoz}; the average optical depth of the cluster sample is fit as a free parameter. We then compute the statistical significance of our measurement in two different ways.

For the main results, we determine the best-fitting average optical depth of the cluster sample, $\bar{\tau}_e$, and its uncertainty by minimizing
\beq
\chi^2 (\bar{\tau}_e) = \left[ \hat{T}_\mathrm{pkSZ}-T_\mathrm{pkSZ}(\bar{\tau}_e) \right] ^\dagger \, \hat{C}^{-1} \, \left[ \hat{T}_\mathrm{pkSZ}-T_\mathrm{pkSZ}(\bar{\tau}_e) \right] \, .
\eeq
The statistical significance of the template fit is then computed as $S/N = \bar{\tau}_e / \sigma_{\bar{\tau}_e}$,
where $\sigma_{\bar{\tau}_e}$ is given by $\chi^2 (\bar{\tau}_e \pm \sigma_{\bar{\tau}_e}) - \chi^2_\mathrm{min} = 1$.
In most of this paper, we treat the optical depth $\bar{\tau}_e$ as the \textit{effective} amplitude of the extracted pairwise kSZ signal.
We show in Section~\ref{subsec:physical_tau} below that at our fiducial filtering scale it can however be interpreted as the \textit{physical} optical depth along a line of sight through the cluster centre.

Secondly, we also assess the signal significance by calculating the $\chi^2$ with respect to the no-signal hypothesis:
\beq
\chi^2_0 = \hat{T}_{\mathrm{pkSZ}}^\dagger \, \hat{C}^{-1} \, \hat{T}_{\mathrm{pkSZ}} \, .
\eeq
From the cumulative distribution function of the $\chi^2$ distribution with the same number of degrees of freedom ($d.o.f.$) as our signal, we infer the probability to exceed (PTE) the measured $\chi^2_0$ with a purely random signal.
Assuming Gaussian uncertainties, we then translate the PTE into the significance of the rejection of the no-signal hypothesis.

We expect the template fit to yield a higher statistical significance than the $\chi^2_0$ procedure:
the template fit includes the additional information of our analytic template, whereas the $\chi^2_0$ procedure makes no assumptions about the expected signal shape.
As there is a clear theoretical expectation for the pairwise kSZ signal, we adopt the template fit as our baseline choice, but also report the PTE and significance from the $\chi^2_0$ test.
\footnote{We note that our approach to compute the $S/N$ ratio of the measurement is different from the one adopted by \cite{Keisler2012}, who define the significance as $S/N = \sqrt{\chi^2}$. The latter is only a good approximation in the limit of very large $\chi^2$ per degree of freedom and significantly overestimates the $S/N$ ratio if this assumption is not fulfilled.}

\section{Simulations}
\label{sec:sims}
\subsection{kSZ simulations}
\label{sec:simdescrip}

We use realistic mock data in order to demonstrate the accuracy of our pairwise kSZ model and to estimate the impact of systematic effects such as redshift errors and mis-centring.
For these purposes, we use the simulated tSZ and kSZ maps by \citet[hereafter F16]{Flender2015}.
In the following we will briefly summarise how these maps were generated, and we refer the reader to F16 for details.
We will further describe the post-processing steps that lead from the full-sky maps and cluster catalogue to the realistic mock data used in this analysis.

The CMB maps and cluster catalogue in F16 were generated using the output from an $N$-body simulation that was run using the $N$-body code framework HACC (Hardware/Hybrid Accelerated Cosmology Code; \citealt{Habib:2014uxa}).
This $N$-body simulation is part of a suite of ${\sim} 100$ simulations that are being carried out under the Mira-Titan Universe project~\citep{Heitmann:2015xma}.
The initial conditions in this particular run adopt
the cosmological parameters $\Omega_c=0.22$, $\Omega_{\mathrm{b}}h^2=0.02258$, $\sigma_8=0.8$, and $h=0.71$, which are consistent with the best-fitting \LCDM cosmology from WMAP7 \citep{WMAP7params}.
When analysing the pairwise kSZ signal from the simulations, we use these parameters to avoid systematic errors due to an incorrect cosmological model.

F16 presented several models for the kSZ signal, all of which were generated via post-processing of the $N$-body simulation output, based on different assumptions about the intra-cluster gas.
Here, we use their ``Model\,III'', which they consider to be the most realistic model.
It is created as follows:
the cluster component of the kSZ signal is generated by adding a gas component to each halo, following the semi-analytic model of {\cite{Shaw2010}}, which takes into account star formation, feedback effects, as well as non-thermal pressure.
A diffuse gas component is added using the positions and velocities of all particles outside haloes, assuming that baryons trace the dark matter.
The tSZ signal is modelled in a similar way as the cluster component of the kSZ signal, using the same semi-analytic model.
We note that whereas in principle the SZ maps could also be generated from a full hydrodynamical simulation (e.g. \citealt{Dolag2013,Dolag2015}), currently available hydro-simulations do not provide the box size and resolution required for our purposes.

In addition to the SZ maps, we generate a random realisation of the primary CMB anisotropies based on their angular power spectrum computed with \textsc{Camb} (using the same cosmological parameters as those used in the $N$-body simulation).
We further model Poisson noise from radio galaxies and dusty star-forming galaxies, as well as the clustered component of the cosmic infrared background as Gaussian random realisations, using the best-fitting model for their power spectra presented by \cite{George2014}.

In order to generate realistic mock data we apply the following post-processing steps to the full-sky maps from F16:
\begin{itemize}
\item We project the full-sky CMB map (consisting of the SZ signal, primary CMB, and foregrounds) onto the SPT fields.
\item We convolve each field with the corresponding SPT beam and filter transfer function (note that the SPT beam and filtering depends on the field and the observation year).
\item Finally, we add to each field a random realisation of the instrumental noise in that field.
  The noise realisations for each field are calculated directly from the data by randomly pairing all observations, subtracting one observation from the other in each pair, then coadding the resulting difference maps.
\end{itemize}

We generate a mock cluster catalogue by applying the DES mask to the full-sky cluster catalogue from F16. We further apply the same point-source mask as applied to the SPT-SZ data.
For choosing the appropriate mass range to match the redMaPPer catalogue, we use the mass-richness relation that \cite{Saro2015} infer from clusters found in both DES-SV and SPT:
We compute $P(M_i | \lambda_i, z_i)$ for every cluster $i$ in the redMaPPer catalogue. Based on the average mass distribution $ \langle P(M_i | \lambda_i, z_i) \rangle_i$, we
then choose a mass range of $0.9 < M_{500c} / 10^{14} M_\odot < 4$;
this results in a mock cluster catalogue with $6,015$ clusters in the redshift range $0.1 < z < 0.8$.
The latter is comparable to the number found in the DES data with the corresponding richness range (6,693).
The number of clusters in the simulation and data catalogues are not expected to match exactly both because of potential differences between the true and simulated cosmology and Poisson noise, as well as some idealities we assume in our simulated cluster catalogue:
in our main analysis we apply a cut in mass in the simulation catalogue rather than explicitly modelling the purity and completeness of the optically-selected cluster catalogue, and we ignore the scatter in the mass-richness relation.
We note that a substantial scattering of low-mass haloes into the sample could potentially bias our results towards low optical depth.
Therefore we test the impact of mass scatter on our results by selecting an alternative sample where we model this effect explicitly; see further discussion in Section~\ref{subsec:massscatter} below.

\subsection{kSZ model validation with simulations}
\label{sec:validation}

\begin{figure}
	\begin{center}
		\includegraphics[width=\columnwidth]{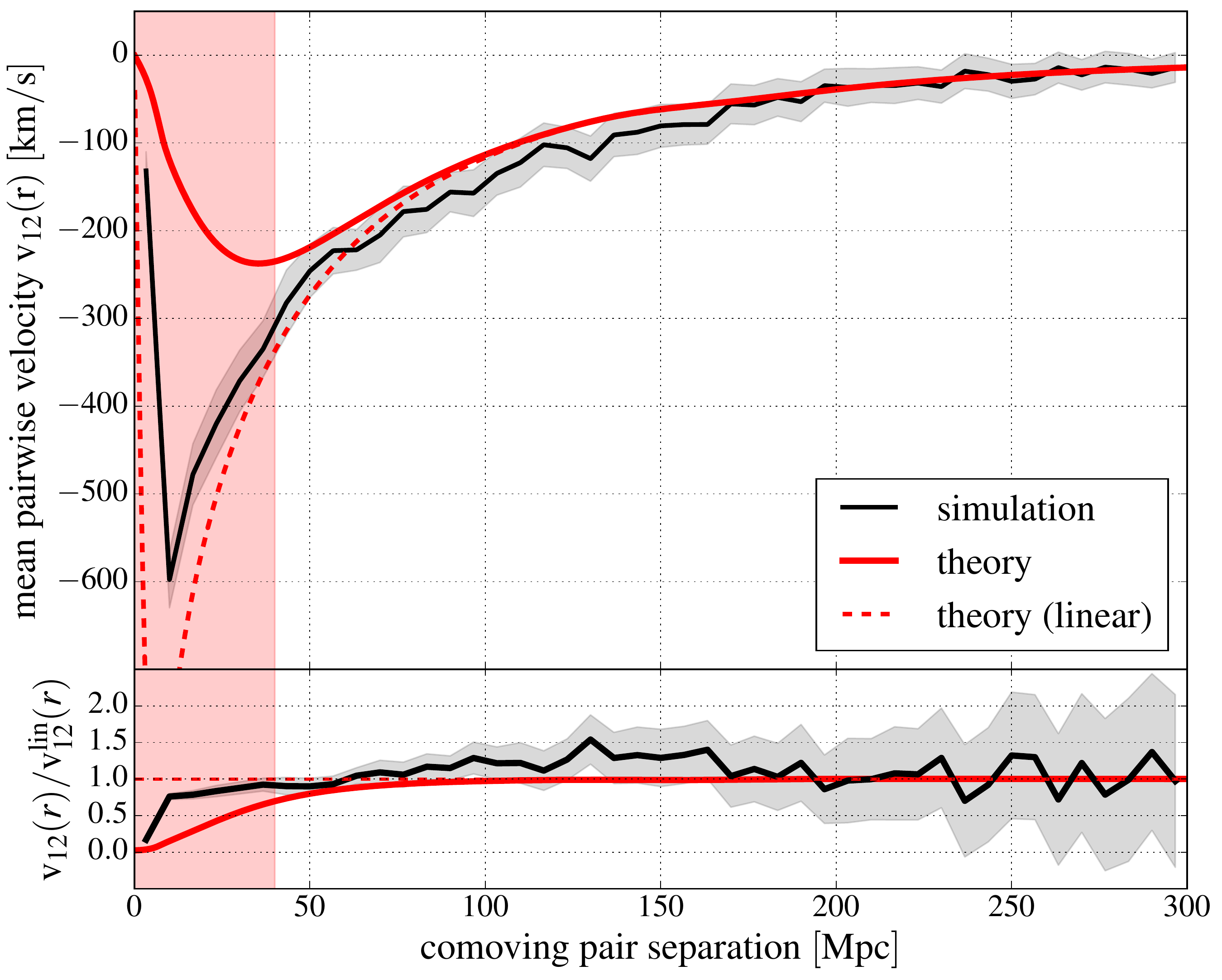}
		\caption{
			Mean pairwise velocity $v_{12}(r)$ from simulations: \textbf{Top:} We show in \textit{black} the measurement from the clusters in our mock catalogue,
			where the shaded regions indicate the $1\sigma$ uncertainties.
			The \textit{solid red line} shows the mean pairwise velocity model of equation~\ref{eq:v12_nonlin} evaluated at the median redshift of $z_m \simeq 0.5$,
			whereas the \textit{dashed red line} represents
			the leading-order term
			 (the numerator of equation~\ref{eq:v12_nonlin}).
			\textbf{Bottom:} We show here the residuals of the upper panel with respect to linear theory. In both panels, the red shaded region ($r < 40$ Mpc) indicates scales that we exclude from our analysis, as the simulations deviate by more than $2 \sigma$ from the theoretical models.}
		\label{fig:veltheory_sim}
	\end{center}
\end{figure}

Before proceeding to measure the pairwise kSZ on the data, we apply the estimators of equations~\ref{eq:ferreira},~\ref{eq:pkszest} to the kSZ simulations introduced in Section~\ref{sec:simdescrip}, in order to validate the theoretical modelling introduced in Section~\ref{sec:theory}.

We first show in Fig.~\ref{fig:veltheory_sim} the mean pairwise velocity computed from the cluster line-of-sight velocities with the estimator of equation~\ref{eq:ferreira},
compared with the mean pairwise velocity model $v_{12}(r)$ (equation \ref{eq:v12_nonlin}) evaluated at the median redshift of $z_m \simeq 0.5$.
The individual $r$-bins of the simulation result are significantly correlated, especially for $r \gtrsim 80~\Mpc$,
where the elements of the correlation matrix $R_{ij} \equiv C_{ij}/\sqrt{C_{ii} C_{jj}}$ are $\gtrsim 0.7$ for bins separated by  $\lesssim 70~\Mpc$.
We find good agreement between our model and simulations for large and intermediate scales, i.e.  for pair separations $r \gtrsim 40~\Mpc$ (linear and mildly non-linear scales).
\footnote{The agreement between simulations and linear theory on large scales also matches with the findings of \cite{HM2015}, who used Gaussian simulations of the linear matter density field and the linearised continuity equation to generate a theory prediction for the pairwise kSZ signal.}
Although the template tends to marginally (${\sim}10\%$) underestimate the pairwise velocities on intermediate scales ($r \sim 100~\Mpc$),
it is consistent with the simulation result computed from the true line-of-sight velocities within the statistical uncertainties.
As presented below, the uncertainties in the pairwise kSZ measured from current data are several times larger, so that this level of accuracy does not introduce a significant bias into our results.

We note that at scales $r \lesssim 60~\Mpc$ linear theory (the numerator of equation~\ref{eq:v12_nonlin}) starts deviating from the full model, which provides a marginally better match to the shape of the signal measured from simulations down to scales  $r \sim 40~\Mpc$.
At even smaller scales, the model does not describe the simulation result accurately, as also shown by \cite{Bhattacharya2007}. This is not surprising: we are probing the velocities of the highest peaks of the density field, so that perturbation theory cannot be expected to provide accurate answers at small scales.
We account for this by excluding the separation bins $r < 40~\Mpc$, where our template deviates from the simulation result by more than $2\sigma$, from the remainder of this analysis.
This exclusion is mainly relevant for the simulation results that use the true redshifts, but has only small impact when adding redshift uncertainties to the simulations or when analysing the real data.
The redshift errors in the latter completely erase the signal at scales $r \lesssim 50~\Mpc$ (see section \ref{sec:photoz} and below), so that on the scales of interest for our analysis equation \ref{eq:v12_nonlin} provides a good representation of the pairwise velocities.

\begin{figure}
	\begin{center}
		\includegraphics[width=\columnwidth]{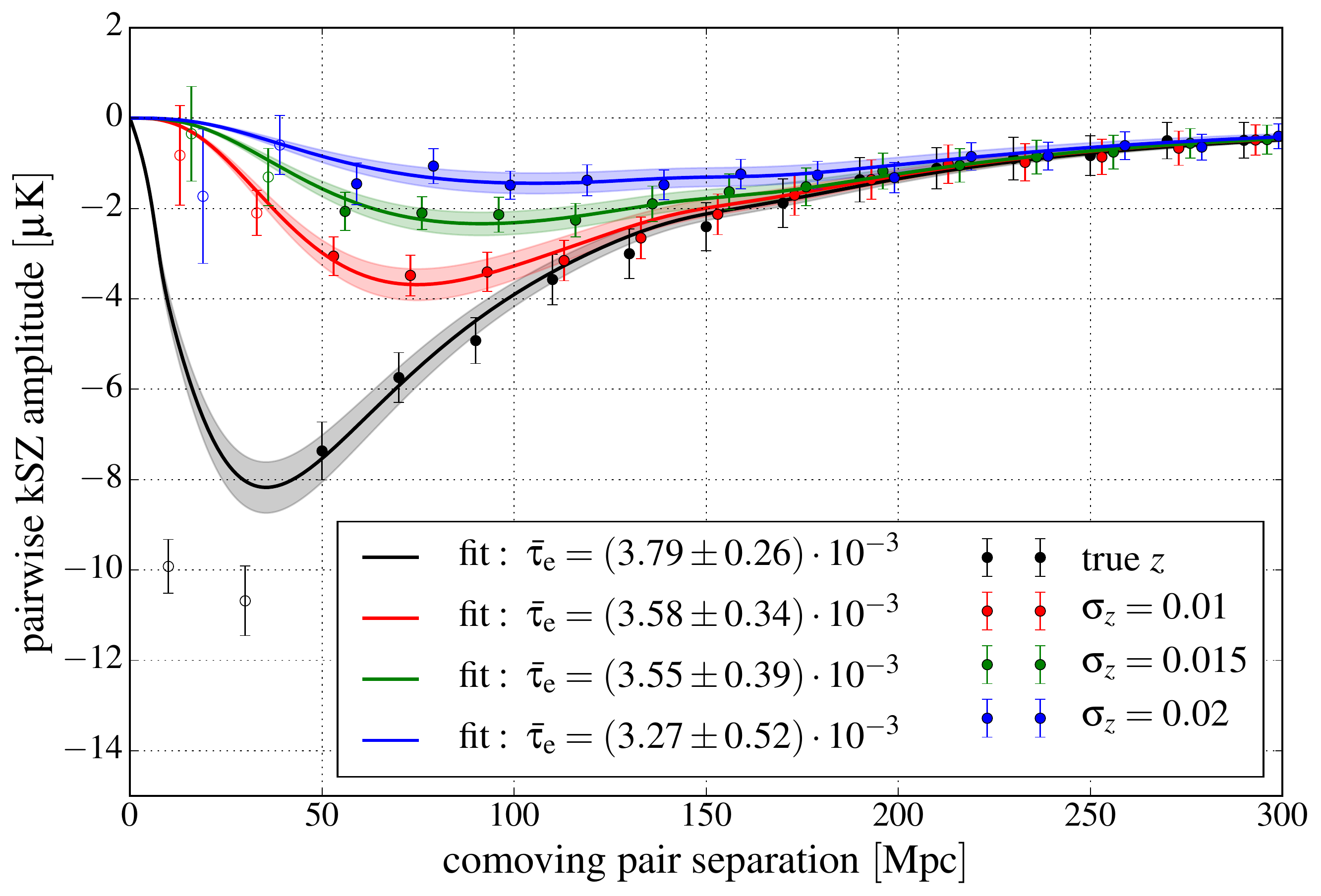}
		\caption{The effect of photo-$z$ uncertainties on simulations:
			We show the pairwise kSZ amplitude measured from \textit{kSZ-only} simulations and the template fits including the photo-$z$ model.
			The recovered pairwise kSZ signal is shown for several levels of redshift errors; the corresponding solid line and shaded region are the template fit and its uncertainty.
			In all cases the two lowest-separation points are excluded from the fit as on these scales perturbation theory is not valid (denoted by the open symbols in this and all subsequent figures).
			The results for the optical depth in the `photo-$z$' cases are consistent with the `true-$z$' results, demonstrating that redshift uncertainties reduce the significance but do not bias our results.}
		\label{fig:sim_kszonly_zerrtest}
	\end{center}
\end{figure}

We proceed by measuring the pairwise kSZ amplitude on the simulations, applying the estimator of equations~\ref{eq:pkszest} and \ref{eq:T_zevolcorr}.
In all cases we discuss below, the error bars of the pairwise kSZ measurement are estimated by applying jack-knife resampling (see Section~\ref{sec:covariance} for details) to the respective simulated data set.
We then fit the pairwise kSZ signal measured from the simulations with our template of equation \ref{eq:templ_photoz}.
To disentangle the effect of photo-$z$s from other sources of uncertainty, we first demonstrate this procedure using kSZ-only simulations and show the results in Fig.~\ref{fig:sim_kszonly_zerrtest}.
Without added redshift errors --- i.e. ~using the exact pair separation ---
our template provides an excellent fit to the measured signal; the best-fitting amplitude is given by an optical depth of $\bar{\tau}_e = (3.79 \pm 0.26)\cdot 10^{-3}$.

We next add photometric redshift uncertainties with rms $\sigma_z = \{0.01,0.015,0.02\}$.
This corresponds to an rms error in the comoving distance of $\sigma_{d_c} \simeq \{35,50,65\}$ Mpc; the central value is comparable to the distance uncertainty of the typical DES cluster, for which $\sigma_{d_c} \simeq 50$ Mpc.
Adapting the pair separation smoothing scale accordingly (see Section~\ref{sec:photoz}), we repeat the kSZ template fit; these results are also displayed in Fig.~\ref{fig:sim_kszonly_zerrtest}.
In all three cases, the template, and hence our simple model for the effect of redshift uncertainties (equation~\ref{eq:templ_photoz}), provides a good fit to the results computed from simulations.
Increasing the redshift errors significantly suppresses the signal on small scales (see Section~\ref{sec:photoz}),
but with the adapted template our estimates of $\bar{\tau}_e$ remain consistent with the `true-$z$' result within the errors.

\begin{figure}
	\begin{center}
		\includegraphics[width=\columnwidth]{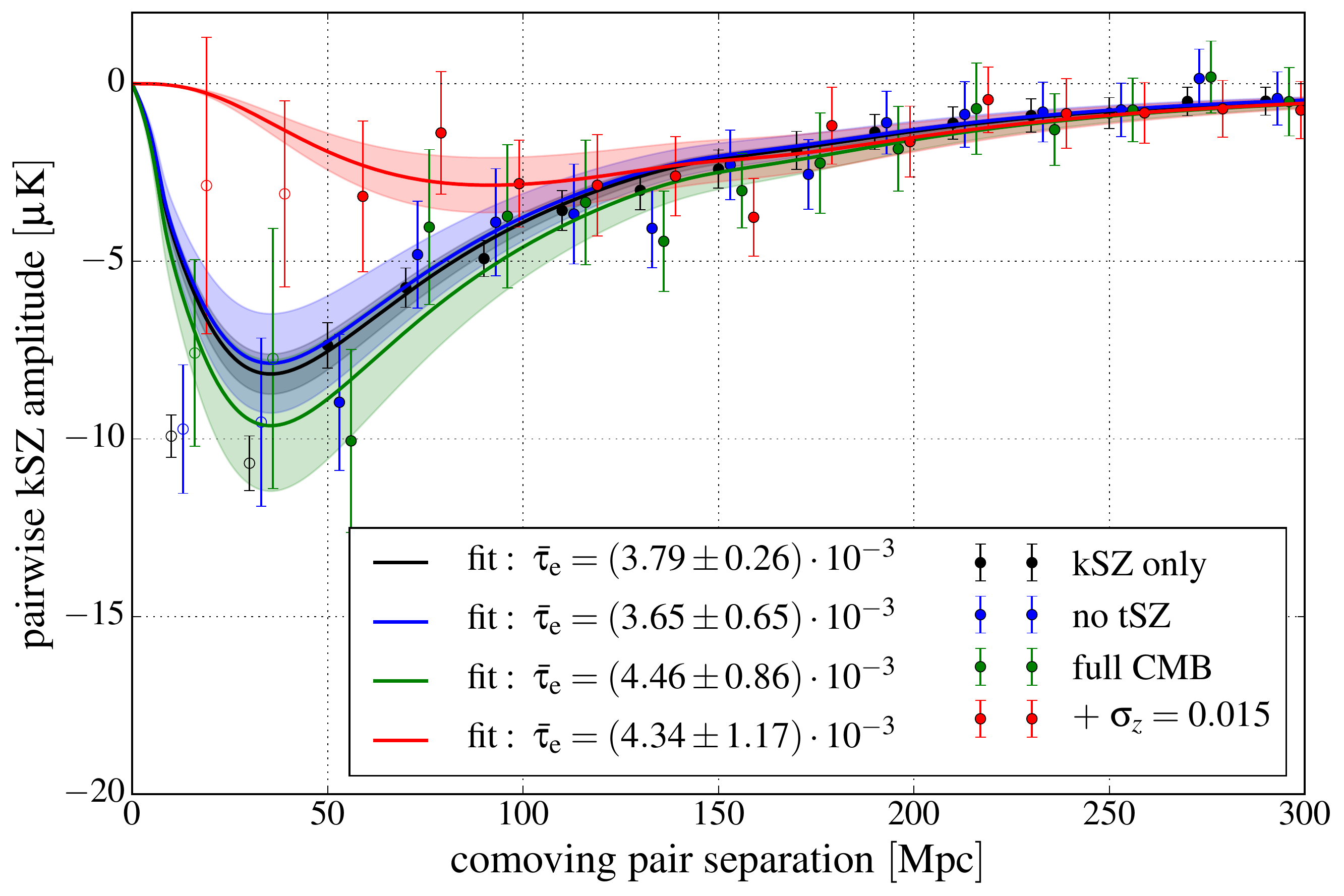}
		\caption{Pairwise kSZ measurements on realistic simulations: We show the pairwise kSZ amplitude for mock clusters with $9 \cdot 10^{13} M_\odot < M_{500c} < 4 \cdot 10^{14} M_\odot$, roughly corresponding to the DES $20 < \tilde{\lambda} < 60$ sample. The black points are from a kSZ-only simulation, the blue colour shows the results with added primary CMB, foregrounds and noise (but no tSZ) and green refers to the full simulation. In the red case we additionally add redshift errors at a similar level as in the real data.
			In all cases, the solid line in the respective colour shows the theory template scaled with the best-fitting optical depth; the shaded regions display the $1\sigma$ uncertainties of the template fit.
			Once again the best-fitting optical depth values remain consistent within the uncertainties, demonstrating that our estimate is unbiased.}
		\label{fig:sim_full_zerr}
	\end{center}
\end{figure}

We finally show in Fig.~\ref{fig:sim_full_zerr} the pairwise kSZ results from the full simulations including noise, primary CMB, and tSZ.
The individual measurements and theory curves refer to, in order: the kSZ-only simulation,
a simulation with added primary CMB, noise, and foregrounds (but no thermal SZ\footnote{We further discuss tSZ contamination as a potential contaminant in Section~\ref{subsec:tszcont} below.}),
our full mock catalogue (including tSZ), and the full mock with added redshift uncertainties. This comparison serves two purposes: first, it shows that
within the current measurement uncertainties
we recover an unbiased estimate of the kSZ amplitude in the presence of potential contaminants.
Secondly, by adding a random realisation of the redshift uncertainties, we obtain a realistic mock analogue of our measurement on real data.
From the latter, we obtain a best-fitting optical depth of $\bar{\tau}_e = (4.34 \pm 1.17)\cdot 10^{-3}$, corresponding to a $3.7 \sigma$ detection from the mock catalogue. As we discuss in Section~\ref{subsec:results} below, the pairwise kSZ amplitude measured from the mocks is therefore consistent with our main result obtained from the real data, although with slightly higher uncertainties.
This minor difference can be explained by the lower number of clusters compared to the real data, in addition to other effects such as the difference between a richness cut and a mass cut.

~
\subsection{Physical interpretation of $\bar{\tau}_e$}
\label{subsec:physical_tau}
Throughout this paper, $\bar{\tau}_e$ is primarily used as an effective parameter to characterize the amplitude of the measured signal and its detection significance.
However, if the matched filter profile is indeed a good match to the average profile of the clusters in the sample, then $\bar{\tau}_e$ can potentially be interpreted as a physically meaningful quantity: the average central optical depth.
In Section~\ref{sec:filter} we found that a $\beta$-filter scale of $\theta_c=0.5'$ maximizes the detection S/N in the simulations;
thus this scale should roughly match the actual scale in our cluster sample.
Therefore, the matched filter with this scale should return a reasonable estimate of the peak kSZ temperature $T_0$, which can be translated into the central optical depth.
We test this hypothesis by comparing the `true' central optical depth, $\bar{\tau}_e^\mathrm{true}$ to the match-filtered central optical depth, $\bar{\tau}_e^\mathrm{filt}$, as explained below.

We calculate $\bar{\tau}_e^\mathrm{true}$ from equation~\ref{eq:ksz}.
For every cluster in the mock catalogue, we use the true amplitude $T_0$ of the kSZ temperature profile computed as described in section~2.2 of F16, and the proper line-of-sight velocity from the underlying $N$-body simulation;
$\bar{\tau}_e^\mathrm{true}$ is then the slope of the best-fitting linear scaling relation between the two quantities.
This approach is similar to the technique proposed by \cite{Li2014} and used by \cite{Schaan2015};
however, instead of reconstructing velocities from the observed density field, we use the actual velocities from the simulations, and furthermore the true kSZ signal instead of a filtered CMB map.
We obtain
\beq
\bar{\tau}_e^\mathrm{true} = (3.39 \pm 0.02) \cdot 10^{-3} \, .
\eeq

Repeating the above procedure with the temperatures measured from the full, match-filtered CMB map (including all potential contaminants), we find
\beq
\bar{\tau}_e^\mathrm{filt}  = (3.13 \pm 0.20) \cdot 10^{-3} \,
\eeq
with a correlation coefficient (Pearson $r$) of $r = 0.2$.
The agreement between $\bar{\tau}_e^\mathrm{filt}$ and $\bar{\tau}_e^\mathrm{true}$ at the ${\lesssim} 10\%$ level indicates that indeed our fiducial filtering scale of $0.5'$ is reasonably well matched to the actual cluster scale.

Finally, the optical depths estimated directly from the temperatures at cluster positions in the simulations
with or without match-filtering, $\bar{\tau}_e^\mathrm{true}$ and $\bar{\tau}_e^\mathrm{filt}$, should also be consistent with the pairwise values recovered from the kSZ-only simulations, \mbox{$\bar{\tau}_e^{^\mathrm{sim, kSZ-only}} = (3.79 \pm 0.26)\cdot 10^{-3}$}, as well as with the result from the full realistic simulations derived in Section~\ref{sec:validation} above.
Averaging over multiple realisations of the photo-$z$ errors, we obtain \mbox{$\bar{\tau}_e^{^\mathrm{sim, full}} = (4.07 \pm 1.26)\cdot 10^{-3}$}.
We find that the $\bar{\tau}_e$ values recovered with these different methods are
all consistent within the realistic statistical uncertainty of our measurement, as the differences are always $ \Delta \bar{\tau}_e < \, \sigma_{\bar{\tau}_e}^{^\mathrm{sim, full}} $.
Therefore --- at the precision level of the current measurement --- we can interpret the measured value of $\bar{\tau}_e$ at the matched filter scale as a physically meaningful parameter, i.e.  the average central optical depth of the clusters in the sample.
We discuss the implications of our results for cluster astrophysics in Section~\ref{sec:interpretation} below.

We note that the changes in $ \bar{\tau}_e $ appear instead more significant when compared with the small statistical uncertainty of the true-redshift, kSZ-only simulations:
this indicates that future high-significance measurements will require further work to improve the pairwise kSZ modelling beyond the present accuracy level.
A further discussion of potential systematic uncertainties is given in Section~\ref{sec:systematics}.

\section{Results}
\label{sec:results}

\subsection{Pairwise kSZ signal with DES and SPT}
\label{subsec:results}

\begin{figure*}
	\begin{center}
		\includegraphics[width=1.3\columnwidth]{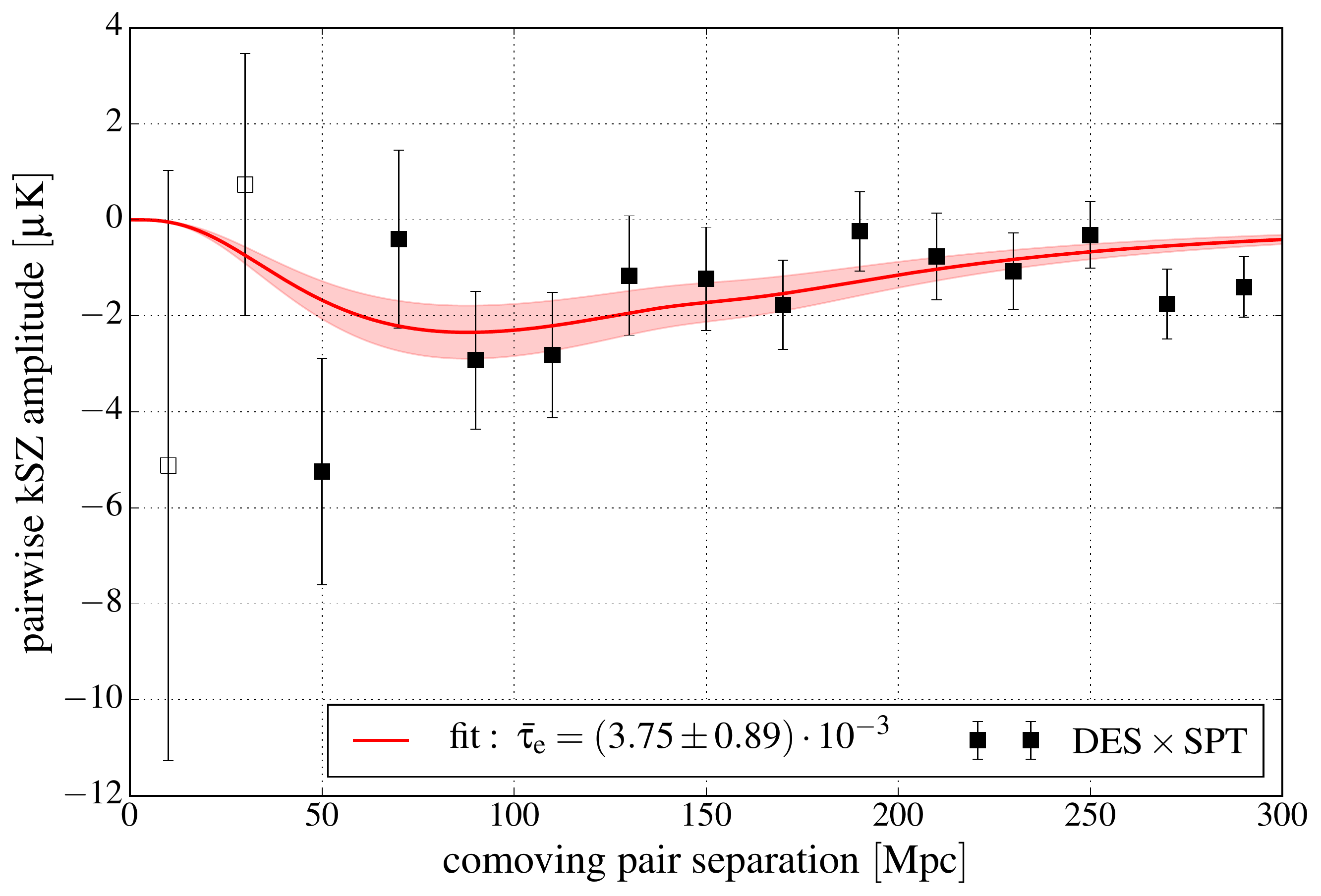}
		\caption{Pairwise kSZ amplitude measured from the DES Y1 redMaPPer catalogue and the SPT-SZ temperature maps, using the baseline sample of clusters with $20 < \tilde{\lambda} < 60$. The solid red line shows the analytic pairwise velocity template (equation~\ref{eq:templ_photoz}) scaled with the best-fitting optical depth $\bar{\tau}_e$; the shaded regions are the corresponding $1\sigma$ uncertainties. As before, the two lowest-separation points shown with empty symbols are excluded from the fit, as on these scales perturbation theory is not valid.}
		\label{fig:pksz_result}
	\end{center}
\end{figure*}
\begin{figure}
	\begin{center}
		\includegraphics[width=0.8\columnwidth]{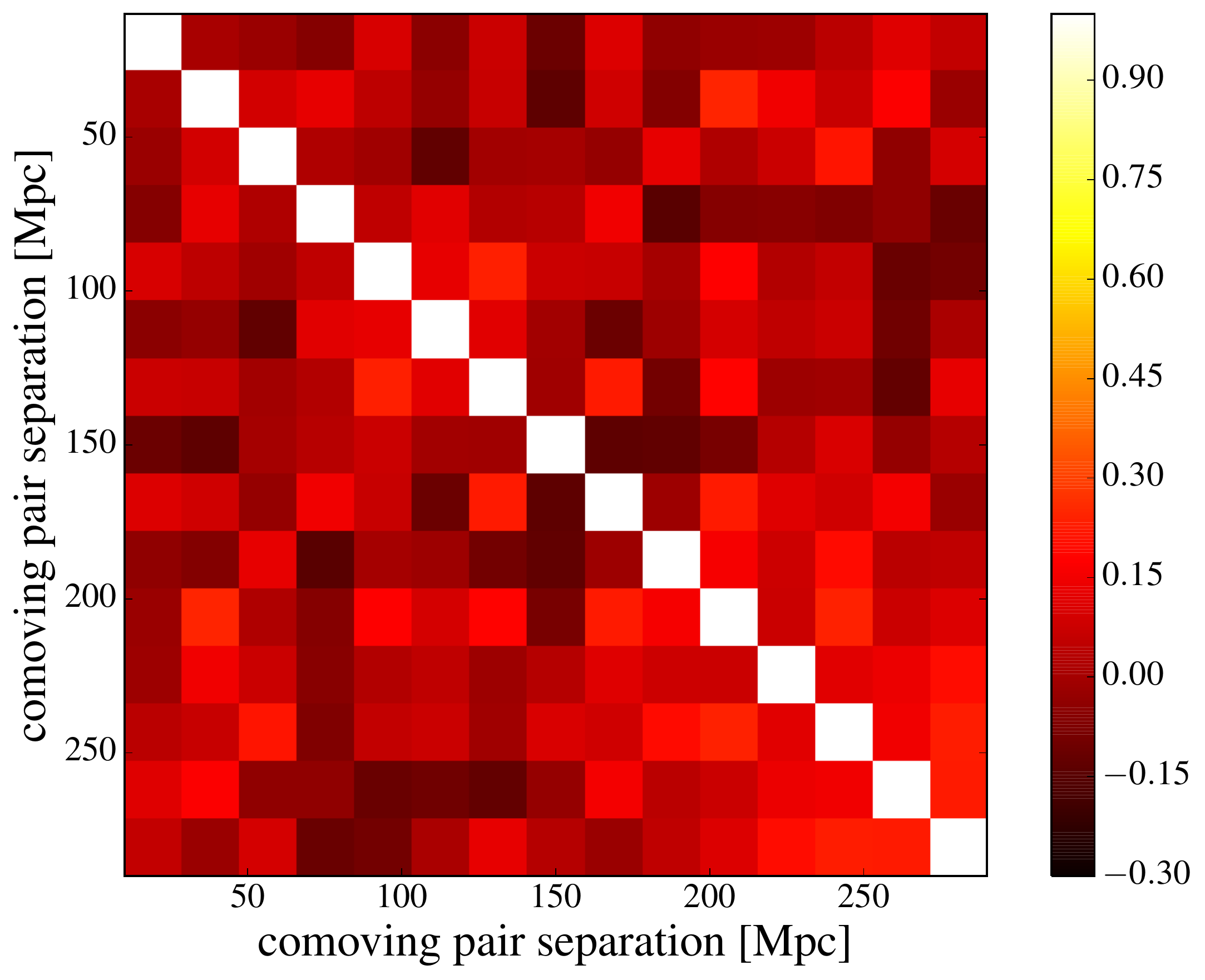}
		\caption{Correlation matrix of the pairwise kSZ measurement of Fig.~\ref{fig:pksz_result} estimated from 120 jack-knife samples drawn from the cluster catalogue.}
		\label{fig:corrmat}
	\end{center}
\end{figure}
\begin{figure}
	\begin{center}
		\includegraphics[width=\columnwidth]{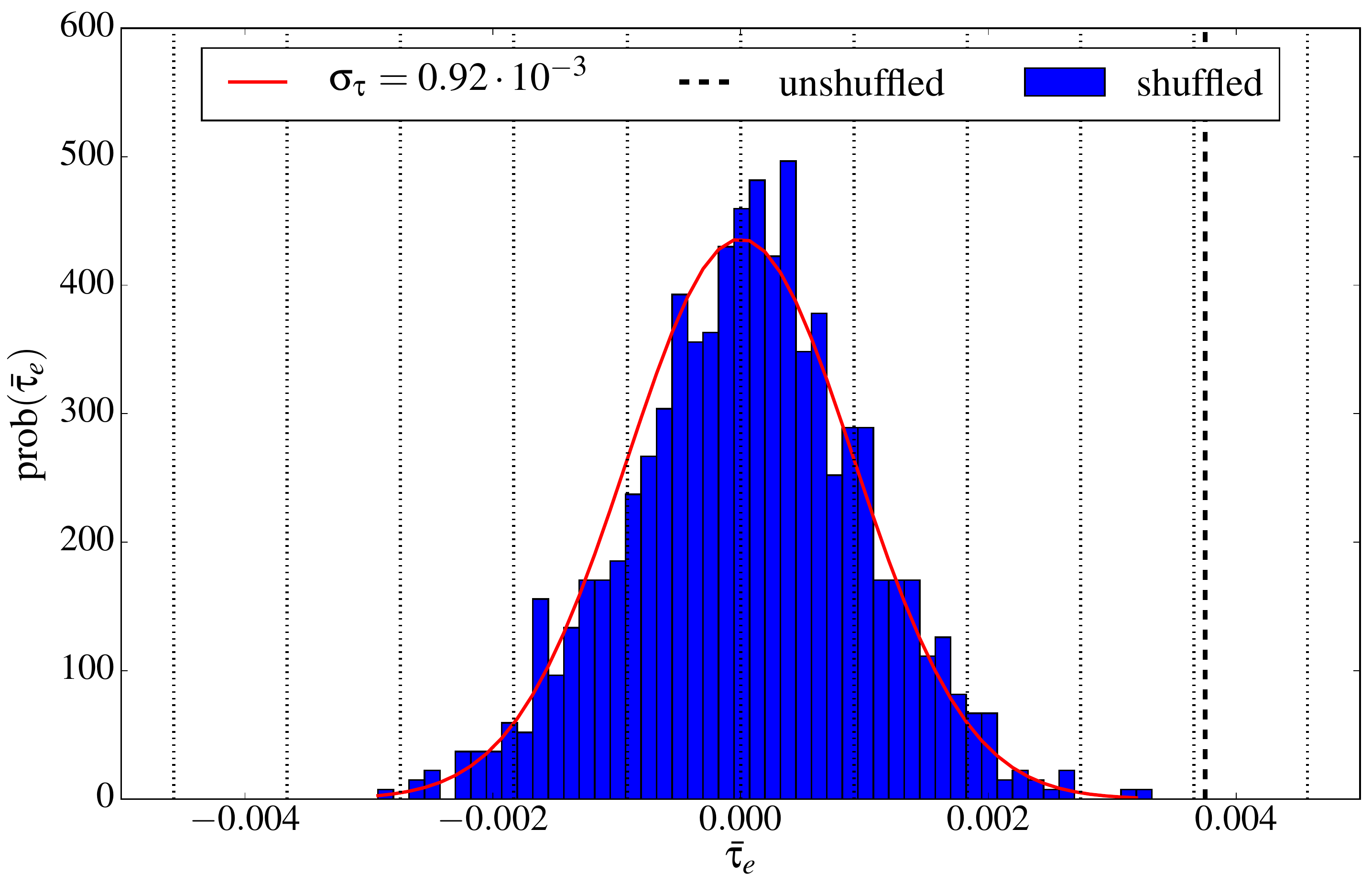}
		\caption{
			Validating the statistical significance of the template fit:
            We show in blue a histogram of the best-fitting $\bar{\tau}_e$ values of 1,000 null-signal realisations obtained by shuffling the cluster pairs. The red curve is a normal distribution fit to the histogram and has a width of $\sigma_{\bar{\tau}_e} = 0.92 \cdot 10^{-3}$. This validates the ${\sim} 4\sigma$ significance of the unshuffled result ($\bar{\tau}_e = 3.75 \cdot 10^{-3}$), which is shown by the vertical thick dashed line.
            The thin dotted lines represent the mean and the $\pm 1\sigma$ to $\pm 5\sigma$ confidence intervals.}
		\label{fig:shuffletest}
	\end{center}
\end{figure}

We show in Fig.~\ref{fig:pksz_result} the pairwise kSZ signal measured from the DES redMaPPer cluster catalogue and the SPT CMB temperature maps using the estimator of equations~\ref{eq:pkszest} and \ref{eq:T_zevolcorr}; this is the main result of our paper.
The uncertainties on the pairwise kSZ amplitude are determined from the JK covariance matrix (equation~\ref{eq:covmat}),
which is estimated as described in Section~\ref{sec:covariance}.
We show the corresponding correlation matrix, which we find to be nearly diagonal, in Fig.~\ref{fig:corrmat}.

The significance of this detection is estimated as described in Section~\ref{sec:significance}.
Fitting the pairwise velocity template of equation~\ref{eq:templ_photoz} to the measured data points yields an average optical depth of
\beq
\bar{\tau}_e = (3.75 \pm 0.89)\cdot 10^{-3} \, ,
\label{eq:mainresult}
\eeq
corresponding to a $4.2 \sigma$ detection of the pairwise kSZ effect.
This measurement represents the first detection of the kSZ using photometric redshift data.
The signal is consistent with that obtained from simulations: see Section~\ref{sec:validation} and Fig.~\ref{fig:sim_full_zerr}.
We discuss the implications of this result for cluster astrophysics in Section~\ref{sec:interpretation} below.

As an additional test of the detection significance, we calculate the $\chi^2$ with respect to the no-signal hypothesis ($\chi^2_0$) as defined in Section~\ref{sec:significance};
we find $\chi^2_0 = 29$ for 15 \textit{d.o.f.}.
This corresponds to a PTE of $1.6\%$ or, assuming Gaussian uncertainties, a rejection of the no-signal hypothesis at the $2.4\sigma$ level.
As expected, this test yields a lower significance than the template fit, which includes additional information about the predicted shape of the pairwise kSZ signal.
However, even with the agnostic approach of the $\chi^2_0$ procedure, the no-signal hypothesis is rejected at a statistically significant level.

To further validate the ${\simeq} 4\sigma$ significance of the template fit,
we create a set of 1,000 null measurements by shuffling the cluster pairs and then estimating the pairwise kSZ amplitude;
a histogram of the measured $\bar{\tau}_e$ from these realisations is shown in Fig.~\ref{fig:shuffletest}.
Their distribution should be statistically consistent with a null signal -- this is one of the null tests discussed in Section~\ref{subsec:nulltests}.
We find a mean of $\langle \bar{\tau}_e \rangle \simeq -10^{-5} $
and a standard deviation of $\sigma_{\bar{\tau}_e} = 0.92 \cdot 10^{-3} $, in excellent agreement with the template fit uncertainty.
This test supports the ${\simeq} 4 \sigma$ significance of the template fit.

\subsection{Dependence on filtering scale and profile}
\label{subsec:filteringscale}

In this section we explore how the detection significance depends on the details of the CMB filtering we introduced in Section~\ref{sec:filter}.
The top panel of Table~\ref{tab:ksz_SPTdat_gold_filtertest} shows the dependence of the best-fitting optical depth on the filter scale, $\theta_c$.
A significant ($>3\sigma$) signal is detected for filter scales up to $\theta_c=2'$;
this is expected since the instrumental beam dominates the filter shape for these small filters.
The recovered amplitude $\bar{\tau}_e$ drops monotonically with increasing $\theta_c$:
in this sense $\bar{\tau}_e$ can be seen as an `effective parameter' that is sensitive to the filtering scale.
A physical interpretation of $\bar{\tau}_e$ is nevertheless possible by using a filtering scale that is matched to the actual angular size of the cluster; in our case, this choice corresponds to the $\theta_c = 0.5'$ filter.\footnote{For $\theta \gtrsim 1'$, the $\beta$-profile with $\theta_c = 0.5'$ matches within ${\lesssim} 30\%$ accuracy a projected NFW profile with concentration $c_{200} \simeq 3.5$ and $\theta_{200} \simeq 3'$, which are the mean NFW-parameters from the clusters in our main sample.}
We have established this possibility of a physical interpretation in Section~\ref{subsec:physical_tau} and will discuss the implications for cluster astrophysics in Section~\ref{sec:interpretation} below.

We further test whether the signal depends on the filter profile shape by replacing the $\beta$-profile from equation~\ref{eq:beta_prof} with a projected Navarro, Frenk, and White (NFW) density profile \citep{navarro96}.
The 3D NFW profile is given by
\beq
\rho(r) = \rho_0 \, \frac{1}{r/r_s \left(1+r/r_s\right)^2}\htwo ,
\eeq
where $r$ is the (3D) radius from the cluster centre and $r_s \equiv R_{500}/c_{500}$ is the scale radius.
Here, $R_{500}$ is the radius inside which the mass density of the halo is equal to 500 times the critical density at the cluster redshift, and $c_{500}$ is the dimensionless concentration parameter. We set the latter to $c_{500} = 3$ based on the typical cluster mass and redshift and the mass-concentration relation of \cite{Bhattacharya2013}.
We use the analytic 2D projection of this profile given by \cite{wright00}, and parametrize the angular scale in terms of $\theta_{500}$, the angular counterpart to $R_{500}$.
For the clusters in our sample, we find the mean $\theta_{500}$ to be around 2'.

In the bottom panel of Table~\ref{tab:ksz_SPTdat_gold_filtertest} we show the optical depth measured
using the NFW-profile filter as a function of filter scale, $\theta_{500}$.
The signal behaves in essentially the same way as for the $\beta$-filter:
the amplitude decreases monotonically with increasing $\theta_{500}$.
We have tested NFW filters with $\theta_{500} \in [0.75',3.5']$, and obtain a significant detection with all of them. The relative size of the optical depths obtained with the $\beta$- and NFW profiles is roughly consistent with the expectation based on $\theta_{500} \sim 5 \theta_c$ (e.g. \citealt{plagge10,Liu2015}).
Additionally, the maximum significance for either filter profile is identical ($4.4 \sigma$).

Finally, we have also investigated using an adaptive filter scale based on the cluster radius and its angular diameter distance, and have detected the signal at a comparable significance. The results of this section
demonstrate that our detection significance is not sensitive to the details of the assumed cluster profile.

\begin{table*}
	\begin{center}
		\caption{Filter profile dependence: We show the best-fitting optical depth and signal-to-noise ratio for different choices of the filter profile and scale.
			\textbf{Top:} We explore various angular core radii $\theta_c$ of the $\beta$-profile. The detection significance is almost independent of the filtering scale for $\theta_c \leq 1'$ and weakly decreases with $\theta_c$ for $\theta_c > 1'$.
			The highlighted scale of $\theta_c = 0.5'$ was used for the main analysis.
			\textbf{Bottom:} Filtering with a projected NFW profile with various angular radii $\theta_{500}$. Both the amplitude and detection significance are comparable to the results with the smallest $\beta$-profile filters.}
		
		\label{tab:ksz_SPTdat_gold_filtertest}
		\begin{tabular}{lccccc}
			\toprule
			Filter type &  & \multicolumn{4}{c}{Filter scale} \\
			\midrule
			$\beta$-profile  & $\theta_c$ & 0.25\arcmin & \textbf{0.5\arcmin} &  1\arcmin & 2\arcmin \\
			& $10^3 \times \bar{\tau}_e$ & $7.63 \pm 1.72\htwo (4.4\sigma)$ & $ \mathbf{3.75 \pm 0.89 \htwo (4.2\sigma)}$  &  $2.15 \pm 0.58\htwo (3.7\sigma$) & $1.68 \pm 0.51\htwo (3.3\sigma)$ \\
			\midrule
			NFW-profile & $\theta_{500}$  & 0.75\arcmin & 1.5\arcmin  & 2.5\arcmin & 3.5\arcmin \\
			& $10^3 \times \bar{\tau}_e$ & $11.26 \pm 2.55\htwo (4.4\sigma)$ & $8.00 \pm 1.82\htwo (4.4\sigma)$ &  $6.27 \pm 1.46\htwo (4.3\sigma)$ & $5.46 \pm 1.32\htwo (4.1\sigma)$\\
			\bottomrule
		\end{tabular}
	\end{center}
\end{table*}

\subsection{Redshift dependence}
\label{subsec:redshiftdep}

At large scales, where linear perturbation theory holds, the pairwise kSZ signal can be written as $T_\mathrm{pkSZ} \propto b \, \bar{\tau}_e \, \xi^{\delta v}$. The redshift evolution of the signal is thus driven by the evolution in bias, average optical depth, and peculiar velocities of the cluster sample.

From the linearised continuity equation~\ref{eq:thetadelta}, it follows that peculiar velocities grow as $v \propto a^{1/2}$ during matter domination.
In the standard cosmological model, this growth slows down and the velocity field decays at late times, where the cosmological constant dominates.
With our analysis, we are probing the redshift range in the transition between these two regimes. To first approximation, we can thus assume that the peculiar velocities do not evolve significantly with redshift.
This simple argument is confirmed by evaluating the full redshift dependence of $\xi^{\delta v}(r,z)$ from equation~\ref{eq:xi_deltav}.
On large scales, and in the redshift range considered here, we find $\xi^{\delta v}(r,z)$ to be within ${\sim}20\%$ from its value at the median redshift of $z_m \simeq 0.5$.

Secondly, the growth of clusters leads to an increased mean optical depth towards low redshifts.
On the other hand, if we apply a constant mass threshold to a sample that is entirely complete and pure,
we would be selecting objects that are more strongly biased at high redshift.
In this somewhat idealistic case, these different effects influence the pairwise kSZ amplitude in opposite directions and would therefore partially cancel, leading to a relatively weak redshift evolution.
However, as explained in Section~\ref{subsec:redmapper}, the high-redshift end of the cluster sample is still partially affected by Malmquist bias (increasing the effective richness threshold) and Eddington bias (decreasing the effective threshold by scattering lower-richness objects into the sample).
Furthermore, the richness-mass relation can also evolve with redshift. A better theoretical understanding of the redshift evolution of the measured pairwise kSZ signal would require a detailed modelling of all of these effects, which is beyond the scope of this work.

\begin{figure}
	\begin{center}
		\includegraphics[width=\columnwidth]{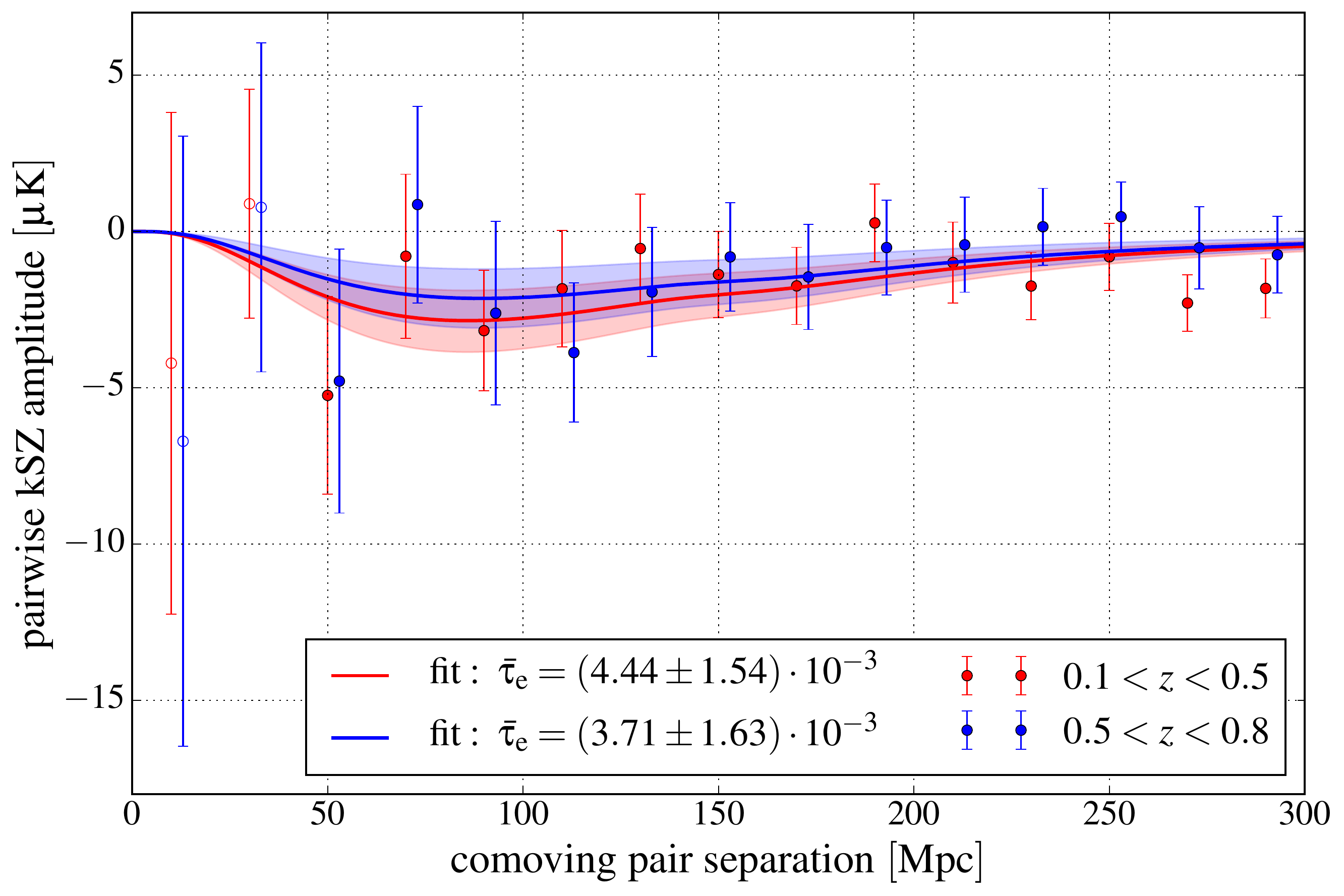}
		\caption{Redshift dependence:
			Pairwise kSZ amplitude from DES Y1 and SPT-SZ data in two redshift bins, below and above the median redshift $z_m \simeq 0.5$. We detect the signal in both redshift bins with comparable amplitudes.}
		\label{fig:ksz_SPTdat_zsplit}
	\end{center}
\end{figure}

We first test our expectation for the redshift evolution with our mock catalogue, i.e.~for the case of a complete and pure sample. For this purpose, we split the sample at the median redshift of $z_m \simeq 0.5$ and recompute the pairwise kSZ amplitude from the two bins. Indeed we find that the two measured amplitudes are comparable.
We then proceed to the measurement on real data, where additionally the sample selection effects mentioned above could influence the redshift dependence.
We show in Fig~\ref{fig:ksz_SPTdat_zsplit} the results for the two redshift bins:
we obtain $\bar{\tau}_e = (4.44 \pm 1.54) \cdot 10^{-3}$ ($2.9 \sigma$ template fit significance) from the low-redshift bin, whereas from the high-redshift bin we find $\bar{\tau}_e = (3.71 \pm 1.63) \cdot 10^{-3}$  ($2.3 \sigma$ template fit significance).
The amplitudes from the two bins are comparable and in good agreement with our main result of equation~\ref{eq:mainresult}.
We further note that the combined significance of the two bins ($3.7\sigma$) is slightly lower than the main result with only one redshift bin because the redshift split removes all pairs with radial separations that cross the redshift boundary.

Given the measurement uncertainties on our data,
there is no evidence that $\bar{\tau}_e$ evolves significantly with redshift.
If future data (see also Section~\ref{sec:conclusions} below) provides
 sufficiently large, volume-limited cluster catalogues, the signal could be measured in multiple redshift bins with higher precision. This would yield significantly tighter constraints on the redshift evolution of the pairwise kSZ, and hence of the peculiar velocity field, than currently possible.

\subsection{Richness limits}
\label{subsec:richnlimits}
In general, the mean optical depth should scale roughly as $\bar{\tau_e} \propto \bar{M} \propto \bar{\lambda}$,
where $\bar{M}$ and $\bar{\lambda}$ are the mean mass and richness of the cluster sample.
Including clusters with lower richness into the sample increases the number counts significantly, but at the price of a higher uncertainty in purity, centring, and photometric redshifts. Here we test the effect of a less stringent richness cut of $10 < \tilde{\lambda} < 60$, which leaves 28,760 (instead of 6,693) clusters in the sample. In this case, the result of the template fit is $\bar{\tau}_e = (1.37 \pm 0.41)\cdot 10^{-3}$.
The best-fitting optical depth has decreased by a factor of ${\simeq}3$ compared to the fiducial $20 < \tilde{\lambda} < 60$ sample,
which is a stronger trend than one would expect based on the simple scaling given above.
This steeper decrease could point to mis-centring or impurities in the cluster catalogue being more pronounced in the low-mass sample.
On the other hand,
as the number of clusters is larger, the error on $\bar{\tau}_e$ is smaller as well,
so that the overall significance ($3.4 \sigma$) is broadly comparable to the higher-mass sample.
Despite the large increase in number counts, the low-$\lambda$ sample does not add to the detection significance of the main sample,
and indeed there is only marginal signal in the $10 < \tilde{\lambda} < 20$ range: here we find $\bar{\tau}_e = (0.77 \pm 0.39) \cdot 10^{-3}$.

To compare the $10 < \tilde{\lambda} < 60$ sample from the DES data to simulations, we select a sample with $ 4 \cdot 10^{13} M_\odot < M_{500c} < 4 \cdot 10^{14} M_\odot$ from our mock catalogue.
It contains ${\sim} 28,000$ clusters, which is comparable to the number in the $10 < \tilde{\lambda} < 60$ sample.
We further simulate photometric redshifts by adding random errors with rms $\sigma_z = 0.02$, which is comparable to the photo-$z$ uncertainty in the real data. From this sample we obtain a best-fitting optical depth of $\bar{\tau}_e = (2.66 \pm 0.68) \cdot 10^{-3}$, corresponding to a ${\sim} 4 \sigma$ detection.
While the detection significance is comparable to the real data, the value of $\bar{\tau}_e$ from the simulations is somewhat higher, and consistent with the expectation based on the mass limits. When analysing the $ 4 \cdot 10^{13} M_\odot < M_{500c} < 9 \cdot 10^{13} M_\odot$ range (corresponding to $10 < \tilde{\lambda} < 20$) separately, we find a similar behaviour.
These observations support the hypothesis that mis-centring and imperfections in the cluster catalogue are non-negligible at the low-mass end.
Additionally, given the increased level of redshift uncertainty in the $10 < \tilde{\lambda} < 60$ sample, it is not surprising that we do not find a more significant detection from these clusters.

We have also explored other lower richness limits, such as $\tilde{\lambda} > 15$, which approximately splits the $10 < \tilde{\lambda} < 60$ sample into a low-richness and high-richness half.
Our findings are consistent with the trend we have observed from the two main samples: including clusters with $\tilde{\lambda} < 20$ significantly reduces the optical depth and does not improve the overall $S/N$ ratio.

Finally, we have explored raising the lower richness threshold and found that this also does not enhance the overall S/N ratio, mainly because of the significant reduction in cluster number.
As an example, we have measured the kSZ signal from the 2,324 clusters with $30 < \tilde{\lambda} < 60$. We find $\bar{\tau}_e = (5.16 \pm 2.20) \cdot 10^{-3}$, corresponding to a $2.3 \sigma$ detection.
The optical depth is about $40\%$ higher than in our main sample; the change is consistent with the expectation based on the increased mean cluster mass.
This indicates that imperfections in the cluster catalogue such as mis-centring do not further decrease significantly when increasing the lower richness threshold beyond $\tilde{\lambda} = 20$.
Also, the photo-$z$ performance does not improve significantly any more; we find only a ${\sim} 10\%$ decrease in the rms distance uncertainty compared to the main sample.
Furthermore, we have studied the $20 < \tilde{\lambda} < 30$ sample containing 4,369 clusters. Again we obtain a relatively weak detection, but the change in the optical depth is consistent with the expectation.
Overall, the results of this section demonstrate that $\tilde{\lambda} = 20$ is as good choice for the lower richness limit for kSZ studies with current cluster data.

\section{Tests for systematics}
\label{sec:systematics}
\begin{table*}
	\begin{center}
		\caption{
			Tests for potential systematic effects:
			we show the measured optical depth for all simulated data scenarios considered in this work.
			The first two rows are estimated from equation~\ref{eq:ksz} using the true cluster velocities from the simulations (see Section~\ref{subsec:physical_tau}).
			The remaining rows are obtained with the pairwise method used in the main analysis,
			i.e.~by fitting the template of equation~\ref{eq:templ_photoz} to the result of the pairwise kSZ estimator (equations~\ref{eq:pkszest}, \ref{eq:T_zevolcorr}).
			The individual columns show the respective data set used,
			the section(s) in which it is analysed or discussed,
			its best-fitting optical depth,
			the reference data combination and row number used for the comparison,
			and the deviation from this reference value in units of the statistical uncertainty of the full mock measurement averaged over multiple realisations of the photo-$z$s,
			$\sigma_{\bar{\tau}_e}^{^\mathrm{sim, full}} = 1.26 \cdot 10^{-3}$.
			This detailed comparison allows us to isolate the effect of each individual potential systematic.
			We find that all cases are consistent within the realistic statistical errors of our measurement,
			demonstrating that the effect of systematics on our measurement is small given the present statistical uncertainties.
		}
		\label{tab:tausim}

		\begin{tabular}[width=\columnwidth]{ccccc}
			\toprule
			Data used & Section(s) & $10^3 \times \bar{\tau}_e$ & reference data [row number] & $(\bar{\tau}_e - \bar{\tau}_e^\mathrm{ref})/ \sigma_{\bar{\tau}_e}^{^\mathrm{sim, full}}$ \\
			\midrule
			\midrule
			\textbf{velocity correlation}: kSZ only (`true') & \ref{subsec:physical_tau} & $3.39 \pm 0.02$ & ---  &  --- \\
			\midrule
			\midrule
			\textbf{velocity correlation:} full CMB, filtered & \ref{subsec:physical_tau} & $3.13 \pm 0.20$ & `true' [1] & $-0.2$ \\
			\midrule
			\textbf{pairwise}: {kSZ only} & \ref{sec:validation} & $3.79 \pm 0.26$  &`true' [1] & $+0.3$  \\
			\midrule
			\midrule
			\textbf{pairwise}: kSZ + primary, noise, foregrounds (`no tSZ') & \ref{sec:validation}  &  $3.65 \pm 0.65$ & pairwise: kSZ only [3] &  $-0.1$  \\
			\midrule
			+ tSZ ({`full CMB')}& \ref{sec:validation}, \ref{subsec:tszcont} & $4.46 \pm 0.86$  & pairwise: kSZ only [3] & $+0.5$ \\
			\midrule
			\midrule
			\textbf{pairwise:} full CMB + photo-$z$ (mult. realis.) & \ref{sec:validation}, \ref{subsec:physical_tau} & $4.07 \pm 1.26$ & full CMB [5] & $-0.3$\\
			\midrule
			full CMB + mis-centring Johnston & \ref{subsec:centring} & $4.03 \pm 0.82$ & full CMB [5] & $-0.3$ \\
			\midrule
			full CMB + mis-centring Saro  & \ref{subsec:centring}  & $3.91 \pm 0.82$ & full CMB [5] & $-0.4$ \\
			\midrule
			full CMB + mass scatter  & \ref{subsec:massscatter}  & $3.56 \pm 0.86$ & full CMB [5] & $-0.7$ \\
			\midrule
			\midrule
			\textbf{pairwise: full CMB + photo-$z$ } & \ref{sec:validation}, \ref{subsec:physical_tau} & $4.07 \pm 1.26$ & `true' [1] &  $+0.5$ \\
			\bottomrule
		\end{tabular}
	\end{center}
\end{table*}

In this section, we perform multiple tests to estimate the effects of potential systematics on our result.
First we carry out a set of null tests using the real data.
Subsequently, we test for contamination by thermal SZ, centring uncertainties and mass scatter using the simulations. We show a compilation of the test results in Table~\ref{tab:tausim}.
Finally, we briefly list and discuss other potential systematics, such as position-dependent foregrounds in the optical catalogue.
Overall, we find that none of the systematic candidates considered has a significant effect on our results, given the current measurement uncertainties.

\subsection{Null tests}
\label{subsec:nulltests}
\begin{figure}
\begin{center}
	\includegraphics[width=\columnwidth]{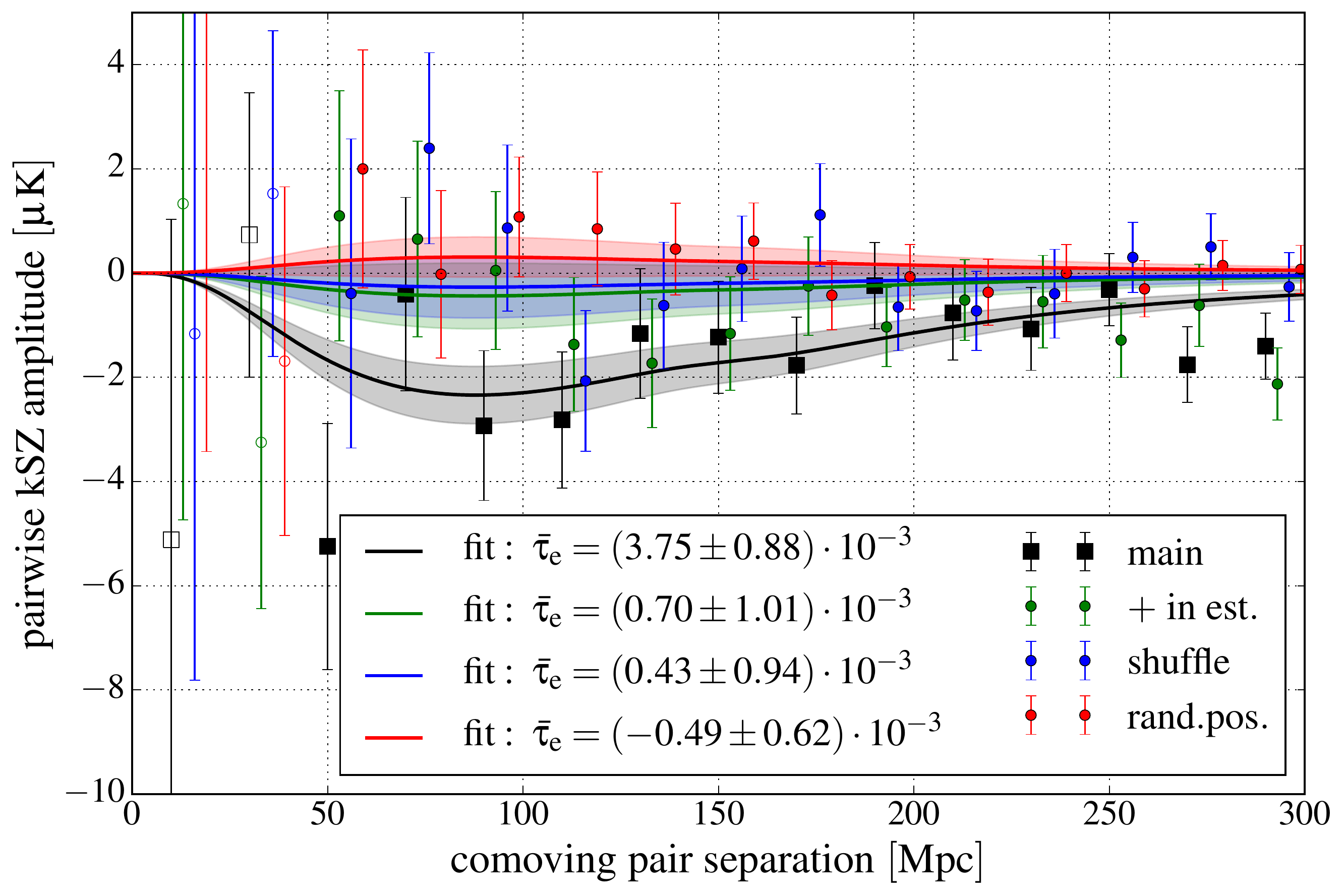}
	\caption{Null tests for the $20 < \tilde{\lambda} < 60$ sample: The large black squares show the actual signal; the green (null test `+ in estimator'), blue (null test `shuffled') and red (null test `random positions') points display the results of the three null tests described in Section~\ref{subsec:nulltests}. All three tests are consistent with the expectation of a null signal.}
	\label{fig:nulltest}
\end{center}
\end{figure}

We first produce null tests by modifying the pairwise estimator (equation~\ref{eq:pkszest}) in three simple ways. The results of these null tests are shown in Fig.~\ref{fig:nulltest}: in all three cases, the best-fitting optical depth is consistent with zero within the $1\sigma$ uncertainties.

(1) We replace the minus sign in the estimator with a plus to remove the sensitivity to the pairwise kSZ signal (H12). The result is consistent with a null signal ($\chi^2_0 \approx 15$ for 15 \textit{d.o.f.}).

(2) We randomly shuffle the clusters before estimating the pairwise kSZ amplitude, but keep the same weights $c_{ij}$ as in the unshuffled case (see H12). We have already shown in Fig.~\ref{fig:shuffletest} a histogram of the best fitting amplitude of 1,000 realisations of this shuffling; the results are consistent with a null signal. Furthermore, Fig.~\ref{fig:nulltest} displays one representative realisation of this shuffling with $\chi^2_0 \approx 11$ (15 \textit{d.o.f.}).

(3) We compute the pairwise kSZ amplitude from a catalogue of random positions.
For this purpose, we use random points produced with the redMaPPer algorithm by creating random $\{z,\lambda\}$ pairs from the real catalogue and assigning random positions on the sky to them,
while ensuring that they cover the same volume and have the same distribution in richness and redshift as the original catalogue \citep{Rykoff2016}.
We then apply the same cuts and process the random points with the same analysis steps as for the real data before estimating the pairwise kSZ amplitude. This test yields $\chi^2_0 \approx 4$ (15 \textit{d.o.f.}), so it is fully consistent with a null signal.
The success of these three null tests demonstrates that there are no obvious systematic contaminants affecting our measurement.

\subsection{Contamination by the tSZ effect}
\label{subsec:tszcont}
On individual clusters the kSZ signal is only a subdominant contribution to the millimetre-wave signal,
which is instead dominated by primary CMB, noise residuals, and the thermal SZ component.
The primary CMB and noise are uncorrelated with cluster positions, and they would therefore already average out in a stacking analysis, whereas the thermal SZ component is only removed by the pairwise estimator (equation~\ref{eq:pkszest}).
This cancellation might however be insufficient
for the most massive clusters, as these are rare objects and have the highest tSZ temperature decrement.
For this reason we have excluded the most massive clusters ($\tilde{\lambda}>60$) from the sample and hence we do not expect a high degree of tSZ contamination.
Nonetheless, we test for this possible systematic by processing a mock CMB data set with all components except for the tSZ component through the same pipeline as before, in order to estimate the impact of the tSZ to our measurement.

The comparison of the result from the `kSZ-only', the `full CMB' and the `no tSZ' mocks were shown
in Fig.~\ref{fig:sim_full_zerr} above, and are summarised  in Table~\ref{tab:tausim} for convenience.
Here we find that the measured optical depth from the `no tSZ' case is in excellent agreement with the `kSZ-only' model, whereas the full mock yields a marginally higher result
($0.8 \sigma$ using the true redshifts, reducing to $0.5\sigma$ when adding redshift uncertainties);
nevertheless all three are consistent within the measurement uncertainties.\footnote{We note that this small, but non-zero degree of tSZ contamination could also be caused by a chance correlation in our particular simulated realisation. This could be studied in more detail only with multiple independent realisations of simulated tSZ and kSZ signals, which is computationally expensive, and beyond the scope of the present work (see also Section~\ref{sec:covariance}).}
We have also tested the impact of including the high-mass clusters on the degree of tSZ contamination. We find that including them reduces the overall signal-to-noise, but still tSZ contamination does not cause our result for the optical depth to be substantially biased.
Therefore we conclude that, with the currently available data, tSZ contamination is not a significant issue for the pairwise kSZ measurement.

\subsection{Centring uncertainties}
\label{subsec:centring}

Offsets between the cluster centre derived from optical data and the centre of the SZ emission have the potential of diluting the pairwise kSZ signal.
This mis-centring can happen for clusters in which the brightest cluster galaxy (BCG) is not at the location of the potential minimum, such as in unrelaxed or merging clusters.
Furthermore, there are cases where the galaxy labelled as the central galaxy
by the redMaPPer algorithm is not truly the BCG.
Here, we test the impact of mis-centring on the measurement of $\bar{\tau}_e$ by introducing random offsets into the mock cluster catalogue.

Following F16, we consider here two different mis-centring models.
Our first mis-centring model follows \citet{Johnston:2007uc}, where we assume that a fraction $f_{\mathrm J}=0.75$ of clusters in the sample have the correct centre, and a fraction $1-f_{\mathrm J}$ of clusters are mis-centred, following a 2D-Gaussian distribution with width $\sigma_{\mathrm{J}}=300~\kpc$ (see F16 for a discussion of the choice of the parameters $f_{\mathrm J}$ and $\sigma_{\mathrm{J}}$).
The second mis-centring model adopts the best-fitting parameters of the model by \citet{Saro2015},
which characterises the offset distribution as a 2-component Gaussian distribution.
We show in Table~\ref{tab:tausim} the best-fitting optical depth measured from simulations where we have introduced random offsets in the cluster positions according to these two mis-centring models.
In agreement with F16, we find that for both models mis-centring reduces the pairwise kSZ amplitude by $\lesssim10\%$, which is not significant given the error bars in our measurement.
We therefore conclude that mis-centring at the level suggested by the models considered here is not a significant problem for our analysis.
We note however that for more precise measurements of the pairwise kSZ amplitude in the future, a better understanding of the mis-centring distribution and potentially a correction of this small bias will be necessary.

\subsection{Mass scatter}
\label{subsec:massscatter}
In order to obtain a good match between the optical cluster catalogue (selected in richness) and the mock catalogue (selected in mass), an accurate knowledge of the mass range and average mass of the cluster sample is required.
Furthermore, the calculation of the theory template also requires a mass range.
If incompleteness and impurities in the catalogue or a large scatter in the mass-richness relation substantially affect the typical cluster mass in the optical catalogue,
the inferred optical depth could be biased, both with respect to the true value and the simulations.
In this work we have selected our sample based on the mass-richness relation calibrated by \cite{Saro2015}, who also accounted for scatter in the relation, finding good agreement between the optical depth from the real data and simulations.

To further test for the impact of scatter in the mass-richness relation, we use an alternative approach for selecting our mock sample:
here we explicitly model the scatter by adding a logarithmic mass error drawn from a normal distribution with width $\sigma_{\ln M} = 0.3$ to the masses in our catalogue and selecting the sample based on this `observed' mass.
We use a mass range of $ 1 < M_{500c} / 10^{14} M_\odot < 3$, where the lower and upper limits corresponds roughly to $\lambda = 20$ and $\lambda = 60$, respectively (e.g. \citealt{Rykoff2012}).
We then recompute the pairwise kSZ amplitude from the alternative sample. To be able to disentangle the effect of the mass scatter from the photometric redshift uncertainties, we do not add additional photo-$z$ errors here.
Averaging over several realisations of the mass scatter, we find $\bar{\tau}_e = (3.56 \pm 0.86)\cdot 10^{-3}$, which is ${\sim} 1 \sigma$ lower than the result with our main sample.
However, when adding redshift errors, the uncertainties on the pairwise kSZ amplitude increase, so that the difference between the two samples reduces to ${\sim} 0.7\sigma$ (see Table~\ref{tab:tausim}).

We further note that our main sample was not selected in the \textit{richness} $\lambda$, but in the \textit{galaxy counts} $\tilde{\lambda}$ (see Section~\ref{subsec:redmapper}). This more conservative cut tends to remove clusters with large richness uncertainties even if they have $\lambda > 20$ (see top right panel of Fig.~\ref{fig:redmapper}).
Therefore it is not surprising that our result for the optical depth slightly decreases if we directly translate a lower richness bound of $\lambda = 20$ into a mass limit.
Finally, our result for $\bar{\tau}_e$ from the data is consistent with both the main and the alternative mock sample within the uncertainties.
Therefore, we conclude that given the current measurement uncertainties mass scatter does not introduce a statistically significant bias into our results.

\subsection{Other potential systematics}
\label{subsec:uncert}
We finally discuss other systematic effects that could potentially affect the pairwise kSZ measurement and bias our estimate of $\bar{\tau}_e$.

(1) The calculation of the pairwise velocity template relies on several assumptions, such as the validity of the model in the quasi-linear regime, a prescription for the halo bias, and a negligible velocity bias. We have shown that our theory template is accurate to within ${\sim} 10\%$ (see Fig.~\ref{fig:veltheory_sim}), significantly below the current measurement uncertainties, so that we discard remaining model inaccuracies as a potential source of significant bias.

(2) Additionally, when relating the mean pairwise velocity to the pairwise kSZ signal in equation~\ref{eq:templ}, we implicitly assume that there are no strong correlations between the velocities and optical depths of individual clusters, such that the average $\langle v \tau_e \rangle$ effectively reduces to $\langle v \rangle \times \langle \tau_e \rangle$.
We have shown in Section~\ref{subsec:physical_tau} that the optical depth measured from the pairwise kSZ is consistent with the true optical depth of the cluster sample.
Therefore this assumption does not bias our results given the current measurement uncertainties.

(3) As described in Section~\ref{sec:photoz}, we have heuristically modelled the effect of photometric redshift uncertainties on the pairwise kSZ signal. We have shown in Section~\ref{sec:validation} that given the current measurement uncertainties this simple model is sufficient to obtain results that are unbiased within the uncertainties.

(4) Finally, position-dependent foregrounds of the optical catalogue, such as Galactic extinction, seeing, airmass and sky brightness, can contaminate the cluster sample, thus potentially affecting the results. We describe in Appendix~\ref{sec:positionsys} a number of tests we have performed on these potential systematics, and we demonstrate that their impact is negligible.

We therefore conclude that at the  significance level reported here, these potential systematics are not significant. With the precision measurements that are likely to be achieved with future data however (see e.g. F16 for a forecast), more careful quantifications and corrections will be required.

\section{Implications for cluster astrophysics}
\label{sec:interpretation}

As described in Section~\ref{subsec:cosmodependence}, the pairwise kSZ amplitude is sensitive to a combination of astrophysical and cosmological parameters, namely $b \bar{\tau}_e \times f \sigma_8^2$, where $b$ is the cluster bias, $\bar{\tau}_e$ is the mean central optical depth of the cluster sample,  $f$ is the growth rate of density perturbations, and $\sigma_8$ is the current rms linear matter fluctuation on scales of $8~h^{-1}\Mpc$.
In principle, it is possible to combine a measurement of the pairwise kSZ amplitude with external constraints on $b$ (e.g.~from a clustering measurement) and $\bar{\tau}_e$ (from the thermal SZ signal or X-ray observations) to derive constraints on $f \sigma_8^2$. The latter could be used as a test of gravity, complementary to other probes such as redshift-space distortions.
At the current detection significance, however, the resulting cosmological constraints would be weak and strongly degenerate with the still relatively uncertain cluster astrophysics.
We have therefore fixed the cosmological parameters to the fiducial \Planck 2015 values of the base \LCDM model\footnote{We have tested that the change of the theory template when varying the cosmology is subdominant compared to the measurement uncertainties. Therefore there is no need to marginalise over the cosmological parameters.}
and have used the measurement of the pairwise kSZ signal to place constraints on the average optical depth of the cluster sample, $\bar{\tau}_e $.
As discussed in Section~\ref{subsec:physical_tau}, physical interpretations of $\bar{\tau}_e $ are only considered at the fiducial filtering scale $\theta_c=0.5'$.

We first compare our results for the optical depth to those obtained by \cite{Planck_KSZ} when extracting the kSZ signal by cross-correlation with reconstructed velocities.
For our fiducial filtering scale of $\theta_c = 0.5'$, we have obtained a value of $\bar{\tau}_e = (3.75 \pm 0.89) \cdot 10^{-3}$.
This is significantly larger than $\bar{\tau}_e = (1.4 \pm 0.5) \cdot 10^{-4}$ found by \cite{Planck_KSZ}.
The difference is however not surprising:
the \Planck analysis uses a catalogue of around 260,000 central galaxies, whose typical host halo masses are significantly below the mass range used in this work.
Furthermore, the authors of \cite{Planck_KSZ} used aperture photometry (as opposed to a matched filter), thus the significantly broader beam of \Planck (FWHM $5'-7'$ for the maps used) dilutes the SZ signal over a much larger area, resulting in a lower effective optical depth.

\begin{figure}
	\begin{center}
		\includegraphics[width=\columnwidth]{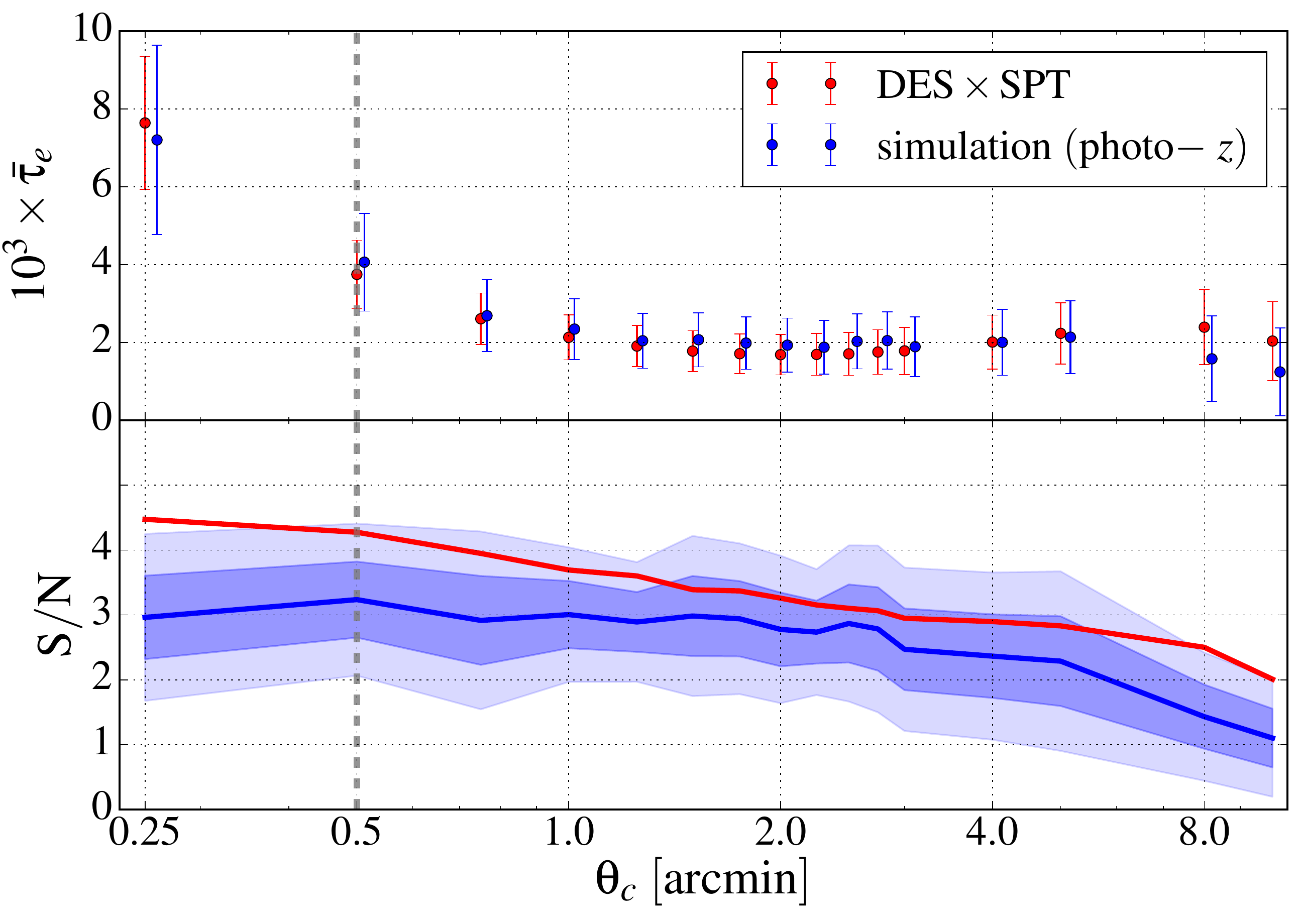}
		\caption{Pairwise kSZ detection as a function of filtering scale $\theta_c$ for the measurement from DES$\times$SPT (red) and our mock catalogues (blue). The dashed vertical line marks our fiducial filtering scale of $\theta_c = 0.5'$.
		\textbf{Top}: We show the best-fitting optical depth, in the case of the simulations averaged over 40 random realisations of the photo-$z$ errors.
		\textbf{Bottom:}
		The lines represent the detection significance $\bar{\tau}_e / \sigma_{\bar{\tau}_e}$ for both data and simulations.
		For the latter we additionally show the scatter caused by the random realisations of the photo-$z$ errors:
		the shaded regions in dark (light) blue are the $68\%$ ($95\%$) confidence regions estimated from the scatter of the realisations.
		}
		\label{fig:ksz_interp}
	\end{center}
\end{figure}

Next, we study the behaviour of $\bar{\tau}_e$ as a function of the matched-filter scale; this probes the spatial extent of the average cluster profile.
Our main result, $\bar{\tau}_e = (3.75 \pm 0.89) \cdot 10^{-3}$, is derived using a matched filter that follows a $\beta$-profile with $\theta_c = 0.5'$, which corresponds roughly to the average cluster profile in our sample.
Here we vary the filter scale and plot the best-fitting $\bar{\tau}_e$ as a function of filter scale in the top panel of Fig.~\ref{fig:ksz_interp}.
As already demonstrated in Section~\ref{subsec:filteringscale}, the pairwise kSZ amplitude decreases with increasing filter scale as the filter essentially averages over a region larger than the cluster.
This trend flattens off at $\theta_c \approx 1'$; for larger $\theta_c$ we measure a roughly constant value of $\bar{\tau}_e \simeq 2 \cdot 10^{-3}$.

The signal-to-noise ratio of the measurement is displayed in the bottom panel of Fig.~\ref{fig:ksz_interp}, where the shaded band indicates the uncertainties caused by the random realisations of the photometric redshift errors.
We detect the signal with $S/N \geq 3$ out to $\theta_c = 3'$; even at $\theta_c=10'$ there is still a marginal detection ($S/N \simeq 2$).
The behaviour of the signal-to-noise as a function of filter scale is consistent with the expectations from the simulations, i.e. there is no evidence in the data for an additional ionized gas component that is not included in the astrophysical model.

It is important to note that the signal found using $\theta_c=10'$ should not be interpreted as a detection of diffuse ionized gas at large distances from the cluster centre;
when a large-aperture matched filter is used, a signal may be detected even from clusters with a significantly smaller spatial extent.
The spatial extent of the gas could be studied with a compensated top-hat filter (i.e., aperture photometry, \citealt{HM2015}), however, this would reduce the significance of the already marginal detection (see, for example, F16).

As shown in the Section~\ref{subsec:physical_tau}, our result for $\bar{\tau}_e$ is a good estimate of the true average optical depth when using our fiducial filtering scale of $\theta_c = 0.5'$.
Therefore we can convert it into the average gas fraction $f_\mathrm{gas}$ as follows:
the matched filter estimates the central amplitude of the temperature profile, hence $\bar{\tau}_e$ corresponds to the optical depth along a line of sight through the centre of the cluster.
The central electron density $n_{e,0}$ is therefore related to $\bar{\tau}_e$ via
\beq
\bar{\tau}_e = \sigma_T \int_{-R}^{R} \mathrm{d}r \htwo n_e(r) \, ,
\label{eq:tau_int}
\eeq
with the 3D $\beta$-profile $n_e(r) = n_{e,0} \, (1+ r^2/r_c^2)^{-3/2}$. The total number of electrons is
\beq
N_e = 4 \pi\int_0^R \mathrm{d}r \, r^2 n_e(r) \, .
\label{eq:Ne_int}
\eeq

As $n_e(r) \propto r^{-3}$ for large $r$, this integral has a logarithmic divergence and needs to be truncated (the same holds true for the NFW profile).
The number of electrons associated with a cluster is of course finite; the divergence merely reflects the fact that for sufficiently large radii the gas profile will deviate from the simple functional form of the $\beta$-profile.
Here we truncate the integrals in equations~\ref{eq:tau_int} and~\ref{eq:Ne_int} at a fixed physical radius given by the mean $R_{500}$ (w.r.t. critical density) of our mock cluster sample, $\langle R_{500} \rangle \simeq 0.65~\Mpc$.
The gas fraction within $R_{500}$ is then computed as $f_\mathrm{gas} = N_e \mu_e m_p / \bar{M}$,
where $m_p$ is the proton mass and the mean particle weight per electron is given by $\mu_e = 1.17$ assuming primordial abundances.
We further use \mbox{$\bar{M} = \langle M_{500} \rangle \simeq 1.4 \cdot 10^{14} M_\odot$},
which is the mean mass of the clusters in our mock catalogue.

Assuming that the average density profile of the cluster sample follows a $\beta$-model profile with $\theta_c = 0.5'$, we obtain
\beq
f_\mathrm{gas}^{500} = 0.080 \pm 0.019 \, \mathrm{(stat.)} \, ,
\eeq
which is in good agreement with the values suggested by X-ray observations of clusters in a similar mass range (e.g.~\citealt{vikhlinin06, Arnaud2007, Sun2009, Giodini2009})
and recent results from hydrodynamical simulations (e.g.~\citealt{McCarthy2016}).
From the simulations by F16 analysed in this work, we infer a gas fraction of
\mbox{$f_\mathrm{gas}^{500} = 0.086 \pm 0.027 \, \mathrm{(stat.)}$}.
In addition to these statistical errors,
systematic uncertainty arises from determining the central optical depth $\bar{\tau}_e$ via the pairwise kSZ measurement, and from converting the latter into the gas fraction.
Concerning the former, Section~\ref{subsec:physical_tau} demonstrated that in simulations, we recover the `true' optical depth to within statistical uncertainties.
Regarding the conversion to gas fraction, the `true' value of $f_\mathrm{gas}$ in the simulations is given by the fiducial model of \citet{Shaw2010}, which was used to generate the kSZ simulations used here;
this model predicts $f_\mathrm{gas}^{500} \simeq 0.09$ for the clusters in the mass range considered here (see Fig.~4 in that work).
Our results from simulations and also from the real data are in good agreement with this value.

Our choice of profile and integration radius are, however, somewhat arbitrary.
Integrating to $ R = \langle R_{200} \rangle \simeq 1~\Mpc$, and using $\bar{M} = \langle M_{200} \rangle \simeq 2.2 \cdot 10^{14} M_\odot$,
we obtain $f_\mathrm{gas}^{200} =  0.070 \pm 0.016~\mathrm{(stat.)}$ from the data and $f_\mathrm{gas}^{200} = 0.076 \pm 0.023~(\mathrm{stat.})$ from the simulations.
These values are slightly smaller than the corresponding results for $f_\mathrm{gas}^{500}$. This is against the expectation of a growing gas fraction with increasing $R$,
and could indicate that the simple $\beta$-profile does not describe the kSZ signal from the cluster outskirts accurately (see also \citealt{vikhlinin06} for similar findings from X-ray data).
Our results for $f_\mathrm{gas}^{200}$ are also in some tension with simulation results by \citet{Battaglia13} and extrapolations from X-ray observations of clusters \citep{Mantz14},
which points to the limitations of our simple calculation of $f_\mathrm{gas}$.
We reiterate here that the assumed profile shape is a significant source of systematic uncertainty in our measurement of  $f_\mathrm{gas}$, and extracted $f_\mathrm{gas}$ values are only valid where the profile is a good match to the actual cluster profile.
An alternative, equally valid choice would be to replace the $\beta$-profile with the appropriate NFW profile matched to our sample (see Section~\ref{subsec:filteringscale});  however, the NFW profile diverges for $r \to 0$, so that the simple integration of equation~\ref{eq:tau_int} would require a further regularisation in the cluster centre.
The same holds true for the universal pressure profile by \citet{Arnaud2010}.

Future kSZ studies with improved CMB and/or cluster data will need to address these limitations, but have the potential to study cluster astrophysics
in significantly more detail.
In particular, high-significance kSZ detections would allow us to place tighter constraints on the gas fraction, both as a function of cluster mass and distance from the cluster centre.

\section{Conclusions}
\label{sec:conclusions}
The pairwise kSZ signal is a probe of the mean relative velocities of galaxy clusters and, as such,
it has the potential to provide important information about both cosmology and cluster astrophysics.
In this work, we have presented a detection of the pairwise kSZ signal with a statistical significance of $4.2 \sigma$ by combining a cluster sample obtained from the first year of DES data with CMB temperature maps at 150 GHz from the SPT-SZ survey.
This is the first detection of the kSZ effect using cluster redshifts derived from photometric data.\footnote{
We note that in the final stages of preparing this manuscript, a kSZ detection with a complementary method using the projected density field by \cite{Hill2016} appeared.}

Our main results are based on a catalogue of galaxy clusters selected with the redMaPPer algorithm on the ${\sim}$1,200 deg$^2$ of DES-Y1 data that overlap with the SPT-SZ survey.
We have used clusters within the richness range $20<\tilde{\lambda}<60$ and redshift range \mbox{$0.1<z<0.8$},
in combination with CMB temperature maps from the SPT that have been match-filtered to extract an estimate for the SZ signal at the position of the clusters.
In parallel to the measurement on these data, we have repeated the analysis on a set of mock cluster catalogues and CMB maps derived from a new suite of high-resolution kSZ simulations, introduced by \cite{Flender2015}.

In both data and simulations, the temperature estimates at the positions of individual clusters are dominated by thermal SZ and primary CMB, foregrounds and noise residuals.
We have removed these contaminants statistically with a differential measurement that isolates the pairwise kSZ signal.
We have then fit the recovered signal with an analytic template that models the mean pairwise velocity on linear and mildly non-linear scales and incorporates the effect of photo-$z$ uncertainties.
Given a set of cosmological parameters and a prescription for the halo bias, this template completely specifies the shape of the pairwise kSZ signal.
Its amplitude, the only free parameter, is given by the average central optical depth of the cluster sample under consideration.
For the main sample used in this work, we have measured $\bar{\tau}_e = (3.75 \pm 0.89) \cdot 10^{-3}$, corresponding to a $4.2\sigma$ detection of the pairwise kSZ signal.
Using the simulations, we have validated this procedure, finding that it recovers the `true' mean optical depth within the statistical uncertainties.

We have tested the robustness of the signal to the details of the analysis:
we have detected the signal at comparable significance over a broad range of filtering scales, and for two different assumptions about the cluster gas profile.
Furthermore, we have split the main sample into two redshift bins and have detected the pairwise kSZ effect in both of them, with no evidence of an evolution of the kSZ amplitude with redshift.
When extending the sample to lower-richness clusters, we have obtained a detection at a slightly lower significance despite the much larger number of clusters. This is likely caused by the higher level of impurities and incompleteness in the optical catalogue, as well as higher centring uncertainties and photometric redshift errors in this sample.
The combination of these effects is currently the main limitation for the statistical significance of our measurement.

Additionally, we have demonstrated with an extensive set of null tests that our result is not significantly affected by systematic uncertainties.
We have explicitly tested for the effect of extinction by Galactic dust and for a range of spatially varying DES observing conditions.
Based on the simulated catalogues, we have also investigated the effect of centring errors, thermal SZ contamination and mass uncertainties, finding that they do not cause any significant bias for our main result.

As an astrophysical application of our measurement, we have studied the optical depth as a function of the scale of the matched filter and found a consistent behaviour between the data and the simulations, indicating that the data are in agreement with the astrophysical model used in the simulations.
At our fiducial filtering scale of $\theta_c= 0.5'$, we have translated the central optical depth into a gas fraction within $R_{500}$ of $f_\mathrm{gas}^{500} = 0.080 \pm 0.019~(\mathrm{stat.})$, in agreement with results from X-ray observations.

Ongoing and upcoming observational campaigns will provide larger and deeper cluster catalogues, as well as CMB maps to higher sensitivity, with which the pairwise kSZ signal will be detectable at high significance:
a cluster catalogue corresponding to our current $10 < \tilde{\lambda} <  60$ sample, but with spectroscopic redshifts,
would result in an ${\sim} 9 \sigma$ detection
of the pairwise kSZ effect with current CMB data.
 A more accurate modelling of $v_{12}(r)$ on smaller scales (see e.g.~\citealt{Sugiyama2015}),
enabling the use of separations \mbox{$r<40~\Mpc$}, would further enhance the detection significance.
Another potential analysis improvement, relevant for future data sets, would be an additional weighting of clusters pairs in the pairwise estimator, based on their expected contribution to the signal given their two masses.
The extra weighting could reduce the scatter in the estimated pairwise signal caused by large variations in cluster masses, which would be particularly relevant for future analyses pushing towards lower mass thresholds and hence spanning a broader range in cluster masses.

Future spectroscopic cluster catalogues with lower mass limits (and thus larger sample sizes) will enable a vast increase in the kSZ detection significance. The possibilities for photo-$z$ catalogues, however,
are currently limited by the redshift uncertainties, but could improve substantially if deeper imaging data (from e.g. the full depth DES survey or, in the longer term, the Large Synoptic Survey Telescope; \citealt{LSSTref}) also leads to improved photo-$z$ performance.
An interesting middle ground between high-resolution spectroscopic and broad-band photometric surveys will be covered by narrow-band photometric surveys such as PAU \citep{PAU2014} and J-PAS \citep{JPAS} and low-resolution spectroscopic surveys such as SPHEREX \citep{SPHEREX}.
However, kSZ analyses from future photometric or low-resolution spectroscopic data might require a more sophisticated modelling of the effect of photo-$z$ uncertainties to ensure that the results stay unbiased within their significantly reduced uncertainties.

Next-generation CMB experiments like SPT-3G (covering the same region as SPT-SZ, but to much lower noise levels; see \citealt{SPT3G}) and AdvACTPol (covering ${\sim} 15,000$ deg$^2$, but not as deeply as SPT-3G; see e.g.~\citealt{Calabrese2014}) should measure the kSZ signal with dramatically increased precision.
In contrast to the single-frequency analysis presented here, the deep multi-frequency data of these experiments could be combined to further isolate the kSZ signal from noise and tSZ contamination.
F16 forecast that by combining AdvACTPol with spectroscopic clusters from the Dark Energy Spectroscopic Instrument (DESI, \citealt{DESI}), the pairwise kSZ could be detected at the $20 \sigma$ level; this number increases to $27-57 \sigma$ in the optimistic scenario where the kSZ can be separated from the other components.
Similar significances are possible when SPT-3G is combined with a cluster catalogue reaching down to $M_{500c} \sim 3 \cdot 10^{13} M_\odot$ and $\sigma_z/(1+z) \simeq 0.003$, which could be achieved by narrow-band photometric surveys (see e.g.~\citealt{PAU2014,JPAS}).

If the potential systematic uncertainties (such as tSZ contamination, mis-centring and mass-scatter) that we have discussed in Sections~\ref{subsec:tszcont} to \ref{subsec:uncert} can be accurately determined and modelled,
these future measurements will allow for an ${\sim} 5\%$ (${\sim} 2\%$ optimistically) constraint on the cosmological and astrophysical parameters that determine the pairwise kSZ amplitude.
A joint analysis of pairwise kSZ measurements with other probes would then be a valuable ingredient for
precision cosmology and cluster astrophysics:
the combination of kSZ with X-ray and tSZ data could provide valuable insight into cluster profiles and the gas fraction, whereas a combination of the kSZ signal with other cosmological observables such as galaxy clustering and redshift-space distortions could measure the growth rate with high precision and hence tighten the constraints on dark energy or modified gravity.

\section*{Acknowledgements}
BS thanks Anthony Challinor, Jens Chluba, Suet-Ying Mak, Emmanuel Schaan, and Fabian Schmidt for helpful discussions. BS further acknowledges support from an Isaac Newton Studentship at the University of Cambridge and from the Science and Technologies Facilities Council (STFC).
KTS acknowledges support from the Kavli Foundation.
TG further thanks David Bacon for useful discussion, and acknowledges support from the Kavli Foundation and STFC grant ST/L000636/1.
CR acknowledges support from the Australian Research Council's Discovery Projects scheme (DP150103208).

This paper has gone through internal review by the DES collaboration.

We are grateful for the extraordinary contributions of our CTIO colleagues and the DECam Construction, Commissioning and Science Verification
teams in achieving the excellent instrument and telescope conditions that have made this work possible.  The success of this project also
relies critically on the expertise and dedication of the DES Data Management group.

Funding for the DES Projects has been provided by the U.S. Department of Energy, the U.S. National Science Foundation, the Ministry of Science and Education of Spain,
the Science and Technology Facilities Council of the United Kingdom, the Higher Education Funding Council for England, the National Center for Supercomputing
Applications at the University of Illinois at Urbana-Champaign, the Kavli Institute of Cosmological Physics at the University of Chicago,
the Center for Cosmology and Astro-Particle Physics at the Ohio State University,
the Mitchell Institute for Fundamental Physics and Astronomy at Texas A\&M University, Financiadora de Estudos e Projetos,
Funda{\c c}{\~a}o Carlos Chagas Filho de Amparo {\`a} Pesquisa do Estado do Rio de Janeiro, Conselho Nacional de Desenvolvimento Cient{\'i}fico e Tecnol{\'o}gico and
the Minist{\'e}rio da Ci{\^e}ncia, Tecnologia e Inova{\c c}{\~a}o, the Deutsche Forschungsgemeinschaft and the Collaborating Institutions in the Dark Energy Survey.
The DES data management system is supported by the National Science Foundation under Grant Number AST-1138766.

The Collaborating Institutions are Argonne National Laboratory, the University of California at Santa Cruz, the University of Cambridge, Centro de Investigaciones En{\'e}rgeticas,
Medioambientales y Tecnol{\'o}gicas-Madrid, the University of Chicago, University College London, the DES-Brazil Consortium, the University of Edinburgh,
the Eidgen{\"o}ssische Technische Hochschule (ETH) Z{\"u}rich,
Fermi National Accelerator Laboratory, the University of Illinois at Urbana-Champaign, the Institut de Ci{\`e}ncies de l'Espai (IEEC/CSIC),
the Institut de F{\'i}sica d'Altes Energies, Lawrence Berkeley National Laboratory, the Ludwig-Maximilians Universit{\"a}t M{\"u}nchen and the associated Excellence Cluster Universe,
the University of Michigan, the National Optical Astronomy Observatory, the University of Nottingham, The Ohio State University, the University of Pennsylvania, the University of Portsmouth,
SLAC National Accelerator Laboratory, Stanford University, the University of Sussex, and Texas A\&M University.

The DES participants from Spanish institutions are partially supported by MINECO under grants AYA2012-39559, ESP2013-48274, FPA2013-47986, and Centro de Excelencia Severo Ochoa SEV-2012-0234.
Research leading to these results has received funding from the European Research Council under the European Union's Seventh Framework Programme (FP7/2007-2013) including ERC grant agreements
240672, 291329, and 306478.

The South Pole Telescope program is supported by the National Science Foundation through grant PLR-1248097. Partial support is also provided by the NSF Physics Frontier Center grant PHY-0114422 to the Kavli Institute of Cosmological Physics at the University of Chicago, the Kavli Foundation, and the Gordon and Betty Moore Foundation through Grant GBMF\#947 to the University of Chicago.

Argonne National Laboratory's work was supported under the U.S. Department of Energy contract DE-AC02-06CH11357. This research used resources of the ALCF, which is supported by DOE/SC under contract DE-AC02-06CH11357.




\bibliographystyle{mnras}
\bibliography{paper_ksz} 

\begin{thebibliography}{}
\makeatletter
\relax
\def\mn@urlcharsother{\let\do\@makeother \do\$\do\&\do\#\do\^\do\_\do\%\do\~}
\def\mn@doi{\begingroup\mn@urlcharsother \@ifnextchar [ {\mn@doi@}
  {\mn@doi@[]}}
\def\mn@doi@[#1]#2{\def\@tempa{#1}\ifx\@tempa\@empty \href
  {http://dx.doi.org/#2} {doi:#2}\else \href {http://dx.doi.org/#2} {#1}\fi
  \endgroup}
\def\mn@eprint#1#2{\mn@eprint@#1:#2::\@nil}
\def\mn@eprint@arXiv#1{\href {http://arxiv.org/abs/#1} {{\tt arXiv:#1}}}
\def\mn@eprint@dblp#1{\href {http://dblp.uni-trier.de/rec/bibtex/#1.xml}
  {dblp:#1}}
\def\mn@eprint@#1:#2:#3:#4\@nil{\def\@tempa {#1}\def\@tempb {#2}\def\@tempc
  {#3}\ifx \@tempc \@empty \let \@tempc \@tempb \let \@tempb \@tempa \fi \ifx
  \@tempb \@empty \def\@tempb {arXiv}\fi \@ifundefined
  {mn@eprint@\@tempb}{\@tempb:\@tempc}{\expandafter \expandafter \csname
  mn@eprint@\@tempb\endcsname \expandafter{\@tempc}}}

\bibitem[\protect\citeauthoryear{Abazajian et~al.}{Abazajian
  et~al.}{2009}]{Abazajian:2008wr}
Abazajian K.~N.,  et~al., 2009, \mn@doi [Astrophys. J. Suppl.]
  {10.1088/0067-0049/182/2/543}, 182, 543

\bibitem[\protect\citeauthoryear{{Ahn} et~al.,}{{Ahn}
  et~al.}{2012}]{2012ApJS..203...21A}
{Ahn} C.~P.,  et~al., 2012, \mn@doi [\apjs] {10.1088/0067-0049/203/2/21}, \href
  {http://adsabs.harvard.edu/abs/2012ApJS..203...21A} {203, 21}

\bibitem[\protect\citeauthoryear{{Arnaud}, {Pointecouteau}  \&
  {Pratt}}{{Arnaud} et~al.}{2007}]{Arnaud2007}
{Arnaud} M.,  {Pointecouteau} E.,   {Pratt} G.~W.,  2007, \mn@doi [\aap]
  {10.1051/0004-6361:20078541}, \href
  {http://adsabs.harvard.edu/abs/2007A%26A...474L..37A} {474, L37}

\bibitem[\protect\citeauthoryear{{Arnaud}, {Pratt}, {Piffaretti},
  {B{\"o}hringer}, {Croston}  \& {Pointecouteau}}{{Arnaud}
  et~al.}{2010}]{Arnaud2010}
{Arnaud} M.,  {Pratt} G.~W.,  {Piffaretti} R.,  {B{\"o}hringer} H.,  {Croston}
  J.~H.,   {Pointecouteau} E.,  2010, \mn@doi [\aap]
  {10.1051/0004-6361/200913416}, \href
  {http://adsabs.harvard.edu/abs/2010A%26A...517A..92A} {517, A92}

\bibitem[\protect\citeauthoryear{{Baldauf}, {Desjacques}  \&
  {Seljak}}{{Baldauf} et~al.}{2015}]{Baldauf2014}
{Baldauf} T.,  {Desjacques} V.,   {Seljak} U.,  2015, \mn@doi [\prd]
  {10.1103/PhysRevD.92.123507}, \href
  {http://adsabs.harvard.edu/abs/2015PhRvD..92l3507B} {92, 123507}

\bibitem[\protect\citeauthoryear{{Battaglia}, {Bond}, {Pfrommer}  \&
  {Sievers}}{{Battaglia} et~al.}{2013}]{Battaglia13}
{Battaglia} N.,  {Bond} J.~R.,  {Pfrommer} C.,   {Sievers} J.~L.,  2013,
  \mn@doi [\apj] {10.1088/0004-637X/777/2/123}, \href
  {http://adsabs.harvard.edu/abs/2013ApJ...777..123B} {777, 123}

\bibitem[\protect\citeauthoryear{{Baxter} et~al.,}{{Baxter}
  et~al.}{2015}]{baxter15}
{Baxter} E.~J.,  et~al., 2015, \mn@doi [\apj] {10.1088/0004-637X/806/2/247},
  \href {http://adsabs.harvard.edu/abs/2015ApJ...806..247B} {806, 247}

\bibitem[\protect\citeauthoryear{{Baxter} et~al.,}{{Baxter}
  et~al.}{2016}]{Baxter2016}
{Baxter} E.~J.,  et~al., 2016, preprint, \href
  {http://adsabs.harvard.edu/abs/2016arXiv160207384B} {} (\mn@eprint {arXiv}
  {1602.07384})

\bibitem[\protect\citeauthoryear{{Benitez} et~al.,}{{Benitez}
  et~al.}{2014}]{JPAS}
{Benitez} N.,  et~al., 2014, preprint, \href
  {http://adsabs.harvard.edu/abs/2014arXiv1403.5237B} {} (\mn@eprint {arXiv}
  {1403.5237})

\bibitem[\protect\citeauthoryear{{Benson} et~al.,}{{Benson}
  et~al.}{2014}]{SPT3G}
{Benson} B.~A.,  et~al., 2014, in Society of Photo-Optical Instrumentation
  Engineers (SPIE) Conference Series. p. 91531P (\mn@eprint {arXiv}
  {1407.2973}), \mn@doi{10.1117/12.2057305}

\bibitem[\protect\citeauthoryear{{Bernardeau}, {Colombi}, {Gazta{\~n}aga}  \&
  {Scoccimarro}}{{Bernardeau} et~al.}{2002}]{Bernardeau2002}
{Bernardeau} F.,  {Colombi} S.,  {Gazta{\~n}aga} E.,   {Scoccimarro} R.,  2002,
  \mn@doi [\physrep] {10.1016/S0370-1573(02)00135-7}, \href
  {http://adsabs.harvard.edu/abs/2002PhR...367....1B} {367, 1}

\bibitem[\protect\citeauthoryear{Bhattacharya \& Kosowsky}{Bhattacharya \&
  Kosowsky}{2007}]{Bhattacharya2006}
Bhattacharya S.,  Kosowsky A.,  2007, \mn@doi [\apj] {10.1086/517523}, 659, L83

\bibitem[\protect\citeauthoryear{Bhattacharya \& Kosowsky}{Bhattacharya \&
  Kosowsky}{2008}]{Bhattacharya2007}
Bhattacharya S.,  Kosowsky A.,  2008, \mn@doi [Phys.Rev.]
  {10.1103/PhysRevD.77.083004}, D77, 083004

\bibitem[\protect\citeauthoryear{{Bhattacharya}, {Habib}, {Heitmann}  \&
  {Vikhlinin}}{{Bhattacharya} et~al.}{2013}]{Bhattacharya2013}
{Bhattacharya} S.,  {Habib} S.,  {Heitmann} K.,   {Vikhlinin} A.,  2013,
  \mn@doi [\apj] {10.1088/0004-637X/766/1/32}, \href
  {http://adsabs.harvard.edu/abs/2013ApJ...766...32B} {766, 32}

\bibitem[\protect\citeauthoryear{{Bianchini} \& {Silvestri}}{{Bianchini} \&
  {Silvestri}}{2016}]{Bianchini2015}
{Bianchini} F.,  {Silvestri} A.,  2016, \mn@doi [\prd]
  {10.1103/PhysRevD.93.064026}, \href
  {http://adsabs.harvard.edu/abs/2016PhRvD..93f4026B} {93, 064026}

\bibitem[\protect\citeauthoryear{Birkinshaw}{Birkinshaw}{1999}]{Birkinshaw1998}
Birkinshaw M.,  1999, \mn@doi [Phys.Rept.] {10.1016/S0370-1573(98)00080-5},
  310, 97

\bibitem[\protect\citeauthoryear{{Bleem} et~al.,}{{Bleem}
  et~al.}{2015}]{bleem15}
{Bleem} L.~E.,  et~al., 2015, \mn@doi [\apjs] {10.1088/0067-0049/216/2/27},
  \href {http://adsabs.harvard.edu/abs/2015ApJS..216...27B} {216, 27}

\bibitem[\protect\citeauthoryear{{Bonamente} et~al.,}{{Bonamente}
  et~al.}{2012}]{Bonamente12}
{Bonamente} M.,  et~al., 2012, \mn@doi [New Journal of Physics]
  {10.1088/1367-2630/14/2/025010}, \href
  {http://adsabs.harvard.edu/abs/2012NJPh...14b5010B} {14, 025010}

\bibitem[\protect\citeauthoryear{{Calabrese} et~al.,}{{Calabrese}
  et~al.}{2014}]{Calabrese2014}
{Calabrese} E.,  et~al., 2014, \mn@doi [\jcap] {10.1088/1475-7516/2014/08/010},
  \href {http://adsabs.harvard.edu/abs/2014JCAP...08..010C} {8, 010}

\bibitem[\protect\citeauthoryear{{Calabretta} \& {Greisen}}{{Calabretta} \&
  {Greisen}}{2002}]{calabretta02}
{Calabretta} M.~R.,  {Greisen} E.~W.,  2002, \mn@doi [\aap]
  {10.1051/0004-6361:20021327}, \href
  {http://ads.ari.uni-heidelberg.de/abs/2002A%26A...395.1077C} {395, 1077}

\bibitem[\protect\citeauthoryear{Carlstrom, Holder  \& Reese}{Carlstrom
  et~al.}{2002}]{Carlstrom2002}
Carlstrom J.~E.,  Holder G.~P.,   Reese E.~D.,  2002, \mn@doi [Ann.Rev.\aap]
  {10.1146/annurev.astro.40.060401.093803}, 40, 643

\bibitem[\protect\citeauthoryear{{Cavaliere} \& {Fusco-Femiano}}{{Cavaliere} \&
  {Fusco-Femiano}}{1976}]{cavaliere76}
{Cavaliere} A.,  {Fusco-Femiano} R.,  1976, \aap, \href
  {http://adsabs.harvard.edu/abs/1976A%26A....49..137C} {49, 137}

\bibitem[\protect\citeauthoryear{{Dark Energy Survey Collaboration}
  et~al.,}{{Dark Energy Survey Collaboration} et~al.}{2016}]{DESnonDEoverview}
{Dark Energy Survey Collaboration} et~al., 2016, \mn@doi [\mnras]
  {10.1093/mnras/stw641}, \href
  {http://adsabs.harvard.edu/abs/2016MNRAS.460.1270D} {460, 1270}

\bibitem[\protect\citeauthoryear{Das, Louis, Nolta, Addison, Battistelli
  et~al.}{Das et~al.}{2014}]{Das2013}
Das S.,  Louis T.,  Nolta M.~R.,  Addison G.~E.,  Battistelli E.~S.,   et~al.,
  2014, \mn@doi [J. Cosmology Astropart. Phys.]
  {10.1088/1475-7516/2014/04/014}, 1404, 014

\bibitem[\protect\citeauthoryear{{Davis} \& {Peebles}}{{Davis} \&
  {Peebles}}{1977}]{Davis1977}
{Davis} M.,  {Peebles} P.~J.~E.,  1977, \mn@doi [\apjs] {10.1086/190456}, \href
  {http://adsabs.harvard.edu/abs/1977ApJS...34..425D} {34, 425}

\bibitem[\protect\citeauthoryear{{Diaferio}, {Sunyaev}  \& {Nusser}}{{Diaferio}
  et~al.}{2000}]{Diaferio2000}
{Diaferio} A.,  {Sunyaev} R.~A.,   {Nusser} A.,  2000, \mn@doi [\apjl]
  {10.1086/312627}, \href {http://adsabs.harvard.edu/abs/2000ApJ...533L..71D}
  {533, L71}

\bibitem[\protect\citeauthoryear{{Diaferio} et~al.,}{{Diaferio}
  et~al.}{2005}]{Diaferio2004}
{Diaferio} A.,  et~al., 2005, \mn@doi [\mnras]
  {10.1111/j.1365-2966.2004.08586.x}, \href
  {http://adsabs.harvard.edu/abs/2005MNRAS.356.1477D} {356, 1477}

\bibitem[\protect\citeauthoryear{{Dolag} \& {Sunyaev}}{{Dolag} \&
  {Sunyaev}}{2013}]{Dolag2013}
{Dolag} K.,  {Sunyaev} R.,  2013, \mn@doi [\mnras] {10.1093/mnras/stt579},
  \href {http://adsabs.harvard.edu/abs/2013MNRAS.432.1600D} {432, 1600}

\bibitem[\protect\citeauthoryear{{Dolag}, {Komatsu}  \& {Sunyaev}}{{Dolag}
  et~al.}{2015}]{Dolag2015}
{Dolag} K.,  {Komatsu} E.,   {Sunyaev} R.,  2015, preprint, \href
  {http://adsabs.harvard.edu/abs/2015arXiv150905134D} {} (\mn@eprint {arXiv}
  {1509.05134})

\bibitem[\protect\citeauthoryear{{Dor{\'e}} et~al.,}{{Dor{\'e}}
  et~al.}{2014}]{SPHEREX}
{Dor{\'e}} O.,  et~al., 2014, preprint, \href
  {http://adsabs.harvard.edu/abs/2014arXiv1412.4872D} {} (\mn@eprint {arXiv}
  {1412.4872})

\bibitem[\protect\citeauthoryear{Ferreira, Juszkiewicz, Feldman, Davis  \&
  Jaffe}{Ferreira et~al.}{1999}]{Ferreira1998}
Ferreira P.,  Juszkiewicz R.,  Feldman H.,  Davis M.,   Jaffe A.~H.,  1999,
  \mn@doi [\apj] {10.1086/311959}, 515, L1

\bibitem[\protect\citeauthoryear{{Flaugher} et~al.,}{{Flaugher}
  et~al.}{2015}]{Flaugher2015}
{Flaugher} B.,  et~al., 2015, \mn@doi [\aj] {10.1088/0004-6256/150/5/150},
  \href {http://adsabs.harvard.edu/abs/2015AJ....150..150F} {150, 150}

\bibitem[\protect\citeauthoryear{{Flender}, {Bleem}, {Finkel}, {Habib},
  {Heitmann}  \& {Holder}}{{Flender} et~al.}{2016}]{Flender2015}
{Flender} S.,  {Bleem} L.,  {Finkel} H.,  {Habib} S.,  {Heitmann} K.,
  {Holder} G.,  2016, \mn@doi [\apj] {10.3847/0004-637X/823/2/98}, \href
  {http://adsabs.harvard.edu/abs/2016ApJ...823...98F} {823, 98}

\bibitem[\protect\citeauthoryear{{Fry} \& {Gaztanaga}}{{Fry} \&
  {Gaztanaga}}{1993}]{Fry1993}
{Fry} J.~N.,  {Gaztanaga} E.,  1993, \mn@doi [\apj] {10.1086/173015}, \href
  {http://adsabs.harvard.edu/abs/1993ApJ...413..447F} {413, 447}

\bibitem[\protect\citeauthoryear{George, Reichardt, Aird, Benson, Bleem
  et~al.}{George et~al.}{2015}]{George2014}
George E.,  Reichardt C.,  Aird K.,  Benson B.,  Bleem L.,   et~al., 2015,
  \mn@doi [\apj] {10.1088/0004-637X/799/2/177}, 799, 177

\bibitem[\protect\citeauthoryear{{Giannantonio} et~al.,}{{Giannantonio}
  et~al.}{2016}]{Giannantonio2015}
{Giannantonio} T.,  et~al., 2016, \mn@doi [\mnras] {10.1093/mnras/stv2678},
  \href {http://adsabs.harvard.edu/abs/2016MNRAS.456.3213G} {456, 3213}

\bibitem[\protect\citeauthoryear{{Giodini} et~al.,}{{Giodini}
  et~al.}{2009}]{Giodini2009}
{Giodini} S.,  et~al., 2009, \mn@doi [\apj] {10.1088/0004-637X/703/1/982},
  \href {http://adsabs.harvard.edu/abs/2009ApJ...703..982G} {703, 982}

\bibitem[\protect\citeauthoryear{{Habib} et~al.,}{{Habib}
  et~al.}{2016}]{Habib:2014uxa}
{Habib} S.,  et~al., 2016, \mn@doi [\na] {10.1016/j.newast.2015.06.003}, \href
  {http://adsabs.harvard.edu/abs/2016NewA...42...49H} {42, 49}

\bibitem[\protect\citeauthoryear{Haehnelt \& Tegmark}{Haehnelt \&
  Tegmark}{1996}]{Haehnelt1996}
Haehnelt M.~G.,  Tegmark M.,  1996, \mn@doi [\mnras] {10.1093/mnras/279.2.545},
  279, 545

\bibitem[\protect\citeauthoryear{Hand, Addison, Aubourg, Battaglia, Battistelli
   et~al.}{Hand et~al.}{2012}]{Hand2012}
Hand N.,  Addison G.~E.,  Aubourg E.,  Battaglia N.,  Battistelli E.~S.,
  et~al., 2012, \mn@doi [Phys.Rev.Lett.] {10.1103/PhysRevLett.109.041101}, 109,
  041101

\bibitem[\protect\citeauthoryear{Hartlap, Simon  \& Schneider}{Hartlap
  et~al.}{2007}]{Hartlap2006}
Hartlap J.,  Simon P.,   Schneider P.,  2007, \mn@doi [\aap]
  {10.1051/0004-6361:20066170}, 464, 399

\bibitem[\protect\citeauthoryear{{Heitmann} et~al.,}{{Heitmann}
  et~al.}{2016}]{Heitmann:2015xma}
{Heitmann} K.,  et~al., 2016, \mn@doi [\apj] {10.3847/0004-637X/820/2/108},
  \href {http://adsabs.harvard.edu/abs/2016ApJ...820..108H} {820, 108}

\bibitem[\protect\citeauthoryear{{Hern{\'a}ndez-Monteagudo}, {Ma}, {Kitaura},
  {Wang}, {G{\'e}nova-Santos}, {Mac{\'{\i}}as-P{\'e}rez}  \&
  {Herranz}}{{Hern{\'a}ndez-Monteagudo} et~al.}{2015}]{HM2015}
{Hern{\'a}ndez-Monteagudo} C.,  {Ma} Y.-Z.,  {Kitaura} F.~S.,  {Wang} W.,
  {G{\'e}nova-Santos} R.,  {Mac{\'{\i}}as-P{\'e}rez} J.,   {Herranz} D.,  2015,
  \mn@doi [Physical Review Letters] {10.1103/PhysRevLett.115.191301}, \href
  {http://adsabs.harvard.edu/abs/2015PhRvL.115s1301H} {115, 191301}

\bibitem[\protect\citeauthoryear{{Hill}, {Ferraro}, {Battaglia}, {Liu}  \&
  {Spergel}}{{Hill} et~al.}{2016}]{Hill2016}
{Hill} J.~C.,  {Ferraro} S.,  {Battaglia} N.,  {Liu} J.,   {Spergel} D.~N.,
  2016, preprint, \href {http://adsabs.harvard.edu/abs/2016arXiv160301608H} {}
  (\mn@eprint {arXiv} {1603.01608})

\bibitem[\protect\citeauthoryear{{Johnston} et~al.,}{{Johnston}
  et~al.}{2007}]{Johnston:2007uc}
{Johnston} D.~E.,  et~al., 2007, preprint, \href
  {http://adsabs.harvard.edu/abs/2007arXiv0709.1159J} {} (\mn@eprint {arXiv}
  {0709.1159})

\bibitem[\protect\citeauthoryear{Keisler \& Schmidt}{Keisler \&
  Schmidt}{2013}]{Keisler2012}
Keisler R.,  Schmidt F.,  2013, \mn@doi [\apj] {10.1088/2041-8205/765/2/L32},
  765, L32

\bibitem[\protect\citeauthoryear{{Keisler} et~al.,}{{Keisler}
  et~al.}{2011}]{keisler11}
{Keisler} R.,  et~al., 2011, \mn@doi [\apj] {10.1088/0004-637X/743/1/28}, \href
  {http://adsabs.harvard.edu/abs/2011ApJ...743...28K} {743, 28}

\bibitem[\protect\citeauthoryear{{Kirk} et~al.,}{{Kirk}
  et~al.}{2016}]{Kirk2015}
{Kirk} D.,  et~al., 2016, \mn@doi [\mnras] {10.1093/mnras/stw570}, \href
  {http://adsabs.harvard.edu/abs/2016MNRAS.459...21K} {459, 21}

\bibitem[\protect\citeauthoryear{{Komatsu} et~al.,}{{Komatsu}
  et~al.}{2011}]{WMAP7params}
{Komatsu} E.,  et~al., 2011, \mn@doi [\apjs] {10.1088/0067-0049/192/2/18},
  \href {http://adsabs.harvard.edu/abs/2011ApJS..192...18K} {192, 18}

\bibitem[\protect\citeauthoryear{{LSST Science Collaboration} et~al.,}{{LSST
  Science Collaboration} et~al.}{2009}]{LSSTref}
{LSST Science Collaboration} et~al., 2009, preprint, \href
  {http://adsabs.harvard.edu/abs/2009arXiv0912.0201L} {} (\mn@eprint {arXiv}
  {0912.0201})

\bibitem[\protect\citeauthoryear{{Leistedt} et~al.,}{{Leistedt}
  et~al.}{2015}]{Leistedt2015}
{Leistedt} B.,  et~al., 2015, preprint, \href
  {http://adsabs.harvard.edu/abs/2015arXiv150705647L} {} (\mn@eprint {arXiv}
  {1507.05647})

\bibitem[\protect\citeauthoryear{{Levi} et~al.,}{{Levi} et~al.}{2013}]{DESI}
{Levi} M.,  et~al., 2013, preprint, \href
  {http://adsabs.harvard.edu/abs/2013arXiv1308.0847L} {} (\mn@eprint {arXiv}
  {1308.0847})

\bibitem[\protect\citeauthoryear{{Lewis}, {Challinor}  \& {Lasenby}}{{Lewis}
  et~al.}{2000}]{CAMB}
{Lewis} A.,  {Challinor} A.,   {Lasenby} A.,  2000, \mn@doi [\apj]
  {10.1086/309179}, \href {http://adsabs.harvard.edu/abs/2000ApJ...538..473L}
  {538, 473}

\bibitem[\protect\citeauthoryear{{Li}, {Angulo}, {White}  \& {Jasche}}{{Li}
  et~al.}{2014}]{Li2014}
{Li} M.,  {Angulo} R.~E.,  {White} S.~D.~M.,   {Jasche} J.,  2014, \mn@doi
  [\mnras] {10.1093/mnras/stu1224}, \href
  {http://adsabs.harvard.edu/abs/2014MNRAS.443.2311L} {443, 2311}

\bibitem[\protect\citeauthoryear{{Liu} et~al.,}{{Liu} et~al.}{2015}]{Liu2015}
{Liu} J.,  et~al., 2015, \mn@doi [\mnras] {10.1093/mnras/stv080}, \href
  {http://adsabs.harvard.edu/abs/2015MNRAS.448.2085L} {448, 2085}

\bibitem[\protect\citeauthoryear{{Ma} \& {Zhao}}{{Ma} \&
  {Zhao}}{2014}]{MaZhao2014}
{Ma} Y.-Z.,  {Zhao} G.-B.,  2014, \mn@doi [Physics Letters B]
  {10.1016/j.physletb.2014.06.066}, \href
  {http://adsabs.harvard.edu/abs/2014PhLB..735..402M} {735, 402}

\bibitem[\protect\citeauthoryear{{Ma}, {Li}  \& {He}}{{Ma}
  et~al.}{2015}]{Ma2015}
{Ma} Y.-Z.,  {Li} M.,   {He} P.,  2015, \mn@doi [\aap]
  {10.1051/0004-6361/201526051}, \href
  {http://adsabs.harvard.edu/abs/2015A%26A...583A..52M} {583, A52}

\bibitem[\protect\citeauthoryear{{Mantz}, {Allen}, {Morris}, {Rapetti},
  {Applegate}, {Kelly}, {von der Linden}  \& {Schmidt}}{{Mantz}
  et~al.}{2014}]{Mantz14}
{Mantz} A.~B.,  {Allen} S.~W.,  {Morris} R.~G.,  {Rapetti} D.~A.,  {Applegate}
  D.~E.,  {Kelly} P.~L.,  {von der Linden} A.,   {Schmidt} R.~W.,  2014,
  \mn@doi [\mnras] {10.1093/mnras/stu368}, \href
  {http://adsabs.harvard.edu/abs/2014MNRAS.440.2077M} {440, 2077}

\bibitem[\protect\citeauthoryear{{Mart{\'{\i}}}, {Miquel}, {Castander},
  {Gazta{\~n}aga}, {Eriksen}  \& {S{\'a}nchez}}{{Mart{\'{\i}}}
  et~al.}{2014}]{PAU2014}
{Mart{\'{\i}}} P.,  {Miquel} R.,  {Castander} F.~J.,  {Gazta{\~n}aga} E.,
  {Eriksen} M.,   {S{\'a}nchez} C.,  2014, \mn@doi [\mnras]
  {10.1093/mnras/stu801}, \href
  {http://adsabs.harvard.edu/abs/2014MNRAS.442...92M} {442, 92}

\bibitem[\protect\citeauthoryear{{McCarthy}, {Schaye}, {Bird}  \& {Le
  Brun}}{{McCarthy} et~al.}{2016}]{McCarthy2016}
{McCarthy} I.~G.,  {Schaye} J.,  {Bird} S.,   {Le Brun} A.~M.~C.,  2016,
  preprint, \href {http://adsabs.harvard.edu/abs/2016arXiv160302702M} {}
  (\mn@eprint {arXiv} {1603.02702})

\bibitem[\protect\citeauthoryear{{McGaugh}}{{McGaugh}}{2008}]{McGaugh2008}
{McGaugh} S.~S.,  2008, in {Davies} J.~I.,  {Disney} M.~J.,  eds,  IAU
  Symposium Vol. 244, IAU Symposium. pp 136--145 (\mn@eprint {arXiv}
  {0707.3795}), \mn@doi{10.1017/S1743921307013920}

\bibitem[\protect\citeauthoryear{{Melin}, {Bartlett}  \&
  {Delabrouille}}{{Melin} et~al.}{2006}]{Melin2006}
{Melin} J.-B.,  {Bartlett} J.~G.,   {Delabrouille} J.,  2006, \mn@doi [\aap]
  {10.1051/0004-6361:20065034}, \href
  {http://adsabs.harvard.edu/abs/2006A%26A...459..341M} {459, 341}

\bibitem[\protect\citeauthoryear{{Mocanu} et~al.,}{{Mocanu}
  et~al.}{2013}]{Mocanu13}
{Mocanu} L.~M.,  et~al., 2013, \mn@doi [\apj] {10.1088/0004-637X/779/1/61},
  \href {http://adsabs.harvard.edu/abs/2013ApJ...779...61M} {779, 61}

\bibitem[\protect\citeauthoryear{{Mueller}, {de Bernardis}, {Bean}  \&
  {Niemack}}{{Mueller} et~al.}{2015a}]{Mueller2014b}
{Mueller} E.-M.,  {de Bernardis} F.,  {Bean} R.,   {Niemack} M.~D.,  2015a,
  \mn@doi [\prd] {10.1103/PhysRevD.92.063501}, \href
  {http://adsabs.harvard.edu/abs/2015PhRvD..92f3501M} {92, 063501}

\bibitem[\protect\citeauthoryear{{Mueller}, {de Bernardis}, {Bean}  \&
  {Niemack}}{{Mueller} et~al.}{2015b}]{Mueller2014a}
{Mueller} E.-M.,  {de Bernardis} F.,  {Bean} R.,   {Niemack} M.~D.,  2015b,
  \mn@doi [\apj] {10.1088/0004-637X/808/1/47}, \href
  {http://adsabs.harvard.edu/abs/2015ApJ...808...47M} {808, 47}

\bibitem[\protect\citeauthoryear{{Murray}, {Power}  \& {Robotham}}{{Murray}
  et~al.}{2013}]{Murray2013}
{Murray} S.~G.,  {Power} C.,   {Robotham} A.~S.~G.,  2013, \mn@doi [Astronomy
  and Computing] {10.1016/j.ascom.2013.11.001}, \href
  {http://adsabs.harvard.edu/abs/2013A%26C.....3...23M} {3, 23}

\bibitem[\protect\citeauthoryear{{Navarro}, {Frenk}  \& {White}}{{Navarro}
  et~al.}{1996}]{navarro96}
{Navarro} J.~F.,  {Frenk} C.~S.,   {White} S.~D.~M.,  1996, \mn@doi [\apj]
  {10.1086/177173}, \href {http://adsabs.harvard.edu/abs/1996ApJ...462..563N}
  {462, 563}

\bibitem[\protect\citeauthoryear{{Peebles}}{{Peebles}}{1980}]{Peebles1980}
{Peebles} P.~J.~E.,  1980, {The large-scale structure of the universe}.
Princeton University Press

\bibitem[\protect\citeauthoryear{Percival \& White}{Percival \&
  White}{2009}]{Percival2008}
Percival W.~J.,  White M.,  2009, \mn@doi [\mnras]
  {10.1111/j.1365-2966.2008.14211.x}, 393, 297

\bibitem[\protect\citeauthoryear{{Plagge} et~al.,}{{Plagge}
  et~al.}{2010}]{plagge10}
{Plagge} T.,  et~al., 2010, \mn@doi [\apj] {10.1088/0004-637X/716/2/1118},
  \href {http://adsabs.harvard.edu/abs/2010ApJ...716.1118P} {716, 1118}

\bibitem[\protect\citeauthoryear{{Planck Collaboration}}{{Planck
  Collaboration}}{2013}]{Planck12}
{Planck Collaboration} 2013, \mn@doi [\aap] {10.1051/0004-6361/201220040},
  \href {http://adsabs.harvard.edu/abs/2013A%26A...550A.131P} {550, A131}

\bibitem[\protect\citeauthoryear{{Planck Collaboration}}{{Planck
  Collaboration}}{2014}]{Planck2013dust}
{Planck Collaboration} 2014, \mn@doi [\aap] {10.1051/0004-6361/201323195},
  \href {http://adsabs.harvard.edu/abs/2014A%26A...571A..11P} {571, A11}

\bibitem[\protect\citeauthoryear{{Planck Collaboration}}{{Planck
  Collaboration}}{2015b}]{Planck15Like}
{Planck Collaboration} 2015b, preprint, \href
  {http://adsabs.harvard.edu/abs/2015arXiv150702704P} {} (\mn@eprint {arXiv}
  {1507.02704})

\bibitem[\protect\citeauthoryear{{Planck Collaboration}}{{Planck
  Collaboration}}{2015a}]{Planck2015params}
{Planck Collaboration} 2015a, preprint, \href
  {http://adsabs.harvard.edu/abs/2015arXiv150201589P} {} (\mn@eprint {arXiv}
  {1502.01589})

\bibitem[\protect\citeauthoryear{{Planck Collaboration}}{{Planck
  Collaboration}}{2016}]{Planck_KSZ}
{Planck Collaboration} 2016, \mn@doi [\aap] {10.1051/0004-6361/201526328},
  \href {http://adsabs.harvard.edu/abs/2016A%26A...586A.140P} {586, A140}

\bibitem[\protect\citeauthoryear{{Rephaeli} \& {Lahav}}{{Rephaeli} \&
  {Lahav}}{1991}]{Rephaeli1991}
{Rephaeli} Y.,  {Lahav} O.,  1991, \mn@doi [\apj] {10.1086/169950}, \href
  {http://adsabs.harvard.edu/abs/1991ApJ...372...21R} {372, 21}

\bibitem[\protect\citeauthoryear{{Rozo} \& {Rykoff}}{{Rozo} \&
  {Rykoff}}{2014}]{Rozo2013}
{Rozo} E.,  {Rykoff} E.~S.,  2014, \mn@doi [\apj] {10.1088/0004-637X/783/2/80},
  \href {http://adsabs.harvard.edu/abs/2014ApJ...783...80R} {783, 80}

\bibitem[\protect\citeauthoryear{Rozo, Rykoff, Bartlett  \& Melin}{Rozo
  et~al.}{2015}]{Rozo2014a}
Rozo E.,  Rykoff E.~S.,  Bartlett J.~G.,   Melin J.~B.,  2015, \mn@doi [\mnras]
  {10.1093/mnras/stv605}, 450, 592

\bibitem[\protect\citeauthoryear{{Rykoff} et~al.,}{{Rykoff}
  et~al.}{2012}]{Rykoff2012}
{Rykoff} E.~S.,  et~al., 2012, \mn@doi [\apj] {10.1088/0004-637X/746/2/178},
  \href {http://adsabs.harvard.edu/abs/2012ApJ...746..178R} {746, 178}

\bibitem[\protect\citeauthoryear{Rykoff et~al.}{Rykoff
  et~al.}{2014}]{Rykoff2013}
Rykoff E.,  et~al., 2014, \mn@doi [\apj] {10.1088/0004-637X/785/2/104}, 785,
  104

\bibitem[\protect\citeauthoryear{{Rykoff} et~al.,}{{Rykoff}
  et~al.}{2016}]{Rykoff2016}
{Rykoff} E.~S.,  et~al., 2016, \mn@doi [\apjs] {10.3847/0067-0049/224/1/1},
  \href {http://adsabs.harvard.edu/abs/2016ApJS..224....1R} {224, 1}

\bibitem[\protect\citeauthoryear{{Saro} et~al.,}{{Saro}
  et~al.}{2015}]{Saro2015}
{Saro} A.,  et~al., 2015, \mn@doi [\mnras] {10.1093/mnras/stv2141}, \href
  {http://adsabs.harvard.edu/abs/2015MNRAS.454.2305S} {454, 2305}

\bibitem[\protect\citeauthoryear{{Sayers} et~al.,}{{Sayers}
  et~al.}{2013a}]{Sayers13}
{Sayers} J.,  et~al., 2013a, \mn@doi [\apj] {10.1088/0004-637X/768/2/177},
  \href {http://adsabs.harvard.edu/abs/2013ApJ...768..177S} {768, 177}

\bibitem[\protect\citeauthoryear{Sayers, Mroczkowski, Zemcov, Korngut, Bock
  et~al.}{Sayers et~al.}{2013b}]{Sayers2013}
Sayers J.,  Mroczkowski T.,  Zemcov M.,  Korngut P.,  Bock J.,   et~al., 2013b,
  \mn@doi [\apj] {10.1088/0004-637X/778/1/52}, 778, 52

\bibitem[\protect\citeauthoryear{{Schaan} et~al.,}{{Schaan}
  et~al.}{2016}]{Schaan2015}
{Schaan} E.,  et~al., 2016, \mn@doi [\prd] {10.1103/PhysRevD.93.082002}, \href
  {http://adsabs.harvard.edu/abs/2016PhRvD..93h2002S} {93, 082002}

\bibitem[\protect\citeauthoryear{{Schaffer} et~al.,}{{Schaffer}
  et~al.}{2011}]{schaffer11}
{Schaffer} K.~K.,  et~al., 2011, \mn@doi [\apj] {10.1088/0004-637X/743/1/90},
  \href {http://adsabs.harvard.edu/abs/2011ApJ...743...90S} {743, 90}

\bibitem[\protect\citeauthoryear{Schmidt}{Schmidt}{2010}]{Schmidt2010}
Schmidt F.,  2010, \mn@doi [Phys.Rev.] {10.1103/PhysRevD.82.063001}, D82,
  063001

\bibitem[\protect\citeauthoryear{{Shaw}, {Nagai}, {Bhattacharya}  \&
  {Lau}}{{Shaw} et~al.}{2010}]{Shaw2010}
{Shaw} L.~D.,  {Nagai} D.,  {Bhattacharya} S.,   {Lau} E.~T.,  2010, \mn@doi
  [\apj] {10.1088/0004-637X/725/2/1452}, \href
  {http://adsabs.harvard.edu/abs/2010ApJ...725.1452S} {725, 1452}

\bibitem[\protect\citeauthoryear{Sheth, Diaferio, Hui  \& Scoccimarro}{Sheth
  et~al.}{2001}]{Sheth2000}
Sheth R.~K.,  Diaferio A.,  Hui L.,   Scoccimarro R.,  2001, \mn@doi [\mnras]
  {10.1046/j.1365-8711.2001.04457.x}, 326, 463

\bibitem[\protect\citeauthoryear{{Shirokoff} et~al.,}{{Shirokoff}
  et~al.}{2011}]{shirokoff11}
{Shirokoff} E.,  et~al., 2011, \mn@doi [\apj] {10.1088/0004-637X/736/1/61},
  \href {http://adsabs.harvard.edu/abs/2011ApJ...736...61S} {736, 61}

\bibitem[\protect\citeauthoryear{{Story} et~al.,}{{Story}
  et~al.}{2013}]{story13}
{Story} K.~T.,  et~al., 2013, \mn@doi [\apj] {10.1088/0004-637X/779/1/86},
  \href {http://adsabs.harvard.edu/abs/2013ApJ...779...86S} {779, 86}

\bibitem[\protect\citeauthoryear{{Sugiyama}, {Okumura}  \&
  {Spergel}}{{Sugiyama} et~al.}{2016}]{Sugiyama2015}
{Sugiyama} N.~S.,  {Okumura} T.,   {Spergel} D.~N.,  2016, \mn@doi [\jcap]
  {10.1088/1475-7516/2016/07/001}, \href
  {http://adsabs.harvard.edu/abs/2016JCAP...07..001S} {7, 001}

\bibitem[\protect\citeauthoryear{{Sun}, {Voit}, {Donahue}, {Jones}, {Forman}
  \& {Vikhlinin}}{{Sun} et~al.}{2009}]{Sun2009}
{Sun} M.,  {Voit} G.~M.,  {Donahue} M.,  {Jones} C.,  {Forman} W.,
  {Vikhlinin} A.,  2009, \mn@doi [\apj] {10.1088/0004-637X/693/2/1142}, \href
  {http://adsabs.harvard.edu/abs/2009ApJ...693.1142S} {693, 1142}

\bibitem[\protect\citeauthoryear{{Sunyaev} \& {Zeldovich}}{{Sunyaev} \&
  {Zeldovich}}{1970}]{SZ1970}
{Sunyaev} R.~A.,  {Zeldovich} Y.~B.,  1970, \mn@doi [\apss]
  {10.1007/BF00653471}, \href
  {http://adsabs.harvard.edu/abs/1970Ap%26SS...7....3S} {7, 3}

\bibitem[\protect\citeauthoryear{{Sunyaev} \& {Zeldovich}}{{Sunyaev} \&
  {Zeldovich}}{1972}]{SZ1972}
{Sunyaev} R.~A.,  {Zeldovich} Y.~B.,  1972, Comments on Astrophysics and Space
  Physics, \href {http://adsabs.harvard.edu/abs/1972CoASP...4..173S} {4, 173}

\bibitem[\protect\citeauthoryear{{Sunyaev} \& {Zeldovich}}{{Sunyaev} \&
  {Zeldovich}}{1980}]{SZ1980}
{Sunyaev} R.~A.,  {Zeldovich} I.~B.,  1980, \mnras, \href
  {http://adsabs.harvard.edu/abs/1980MNRAS.190..413S} {190, 413}

\bibitem[\protect\citeauthoryear{Swetz, Ade, Amiri, Appel, Battistelli
  et~al.}{Swetz et~al.}{2011}]{ACT}
Swetz D.,  Ade P.,  Amiri M.,  Appel J.,  Battistelli E.,   et~al., 2011,
  \mn@doi [\apjs] {10.1088/0067-0049/194/2/41}, 194, 41

\bibitem[\protect\citeauthoryear{{The Dark Energy Survey Collaboration}}{{The
  Dark Energy Survey Collaboration}}{2005}]{DES}
{The Dark Energy Survey Collaboration} 2005, preprint, \href
  {http://adsabs.harvard.edu/abs/2005astro.ph.10346T} {} (\mn@eprint {arXiv}
  {astro-ph/0510346})

\bibitem[\protect\citeauthoryear{{Tinker}, {Kravtsov}, {Klypin}, {Abazajian},
  {Warren}, {Yepes}, {Gottl{\"o}ber}  \& {Holz}}{{Tinker}
  et~al.}{2008}]{Tinker2008}
{Tinker} J.,  {Kravtsov} A.~V.,  {Klypin} A.,  {Abazajian} K.,  {Warren} M.,
  {Yepes} G.,  {Gottl{\"o}ber} S.,   {Holz} D.~E.,  2008, \mn@doi [\apj]
  {10.1086/591439}, \href {http://adsabs.harvard.edu/abs/2008ApJ...688..709T}
  {688, 709}

\bibitem[\protect\citeauthoryear{{Tinker}, {Robertson}, {Kravtsov}, {Klypin},
  {Warren}, {Yepes}  \& {Gottl{\"o}ber}}{{Tinker} et~al.}{2010}]{Tinker2010}
{Tinker} J.~L.,  {Robertson} B.~E.,  {Kravtsov} A.~V.,  {Klypin} A.,  {Warren}
  M.~S.,  {Yepes} G.,   {Gottl{\"o}ber} S.,  2010, \mn@doi [\apj]
  {10.1088/0004-637X/724/2/878}, \href
  {http://adsabs.harvard.edu/abs/2010ApJ...724..878T} {724, 878}

\bibitem[\protect\citeauthoryear{{Vikhlinin}, {Kravtsov}, {Forman}, {Jones},
  {Markevitch}, {Murray}  \& {Van Speybroeck}}{{Vikhlinin}
  et~al.}{2006}]{vikhlinin06}
{Vikhlinin} A.,  {Kravtsov} A.,  {Forman} W.,  {Jones} C.,  {Markevitch} M.,
  {Murray} S.~S.,   {Van Speybroeck} L.,  2006, \mn@doi [\apj]
  {10.1086/500288}, \href {http://adsabs.harvard.edu/abs/2006ApJ...640..691V}
  {640, 691}

\bibitem[\protect\citeauthoryear{{Wright} \& {Brainerd}}{{Wright} \&
  {Brainerd}}{2000}]{wright00}
{Wright} C.~O.,  {Brainerd} T.~G.,  2000, \mn@doi [\apj] {10.1086/308744},
  \href {http://adsabs.harvard.edu/abs/2000ApJ...534...34W} {534, 34}

\makeatother
\end{thebibliography}



\appendix
\section{Error budget and covariance estimation}
\label{sec:covariance_tests}
In the main analysis, we estimate the covariance matrix from jack-knife resamples drawn from the cluster catalogue.
Here we perform further tests that illustrate the error budget of our measurement, justify our choice of resampling techniques for the covariance estimation, and demonstrate the stability of our estimate.
First we decompose the error on the optical depth into the contributions of the different components of the simulated mm-sky.
Secondly, we demonstrate the stability of our resampling covariance estimate and explore alternative resampling methods.

\subsection{Decomposition of the error budget}
\label{sec:decompos}

We decompose the error budget making use of the different ingredients of the simulated mm-signal that are described in Section~\ref{sec:simdescrip}, namely primary CMB, foregrounds, instrumental noise and thermal SZ.
To identify the strongest source of uncertainty, we measure $\bar{\tau}_e$ for different combinations of the kSZ signal with the various contaminants. We use the same template fit as in the main analysis (see Section~\ref{sec:significance}) and compare the error on the optical depth, $\sigma_{\bar{\tau}_e}$, for the different simulated data scenarios.
This compresses the contribution of the individual components into a single number per simulated data scenario, allowing an easy comparison.

We show in Table~\ref{tab:errdecomp} the error on $\bar{\tau}_e$ for various combinations of simulated mm-signals.
We find that the strongest increase of $\sigma_{\bar{\tau}_e}$, and hence the largest contribution to the total error budget of our measurement, can be attributed to thermal SZ.
The latter is closely followed by instrumental noise, while the primary CMB and foregrounds contribute significantly less.
This ranking can be understood by recalling that the tSZ signal is always negative at 150 GHz and is spatially correlated with the cluster kSZ signal,
while the primary CMB and instrumental noise can be positive or negative and are uncorrelated with the signal.

Because the tSZ signal is the largest source of uncertainty in the measurement, it is not feasible to compute the covariance matrix from Monte Carlo simulations.
We could simulate many realisations of the CMB and instrumental noise, however, random tSZ realisations would lack spatial correlation with the kSZ signal.
Simulating tSZ signal with the correct spatial correlations requires $N$-body simulations, which is not computationally feasible.
This leaves resampling techniques as the only sensible option to estimate the contribution of the tSZ signal to the covariance of our measurement. 

Additionally, for our next dominant noise contribution --- instrumental noise --- the SPT noise is approximately white at the scales of interest for our matched filter (see e.g. Fig.~7 in \citealt{schaffer11}). As the matched-filter strongly down-weights low-$\ell$ modes (thus precluding large-scale correlations), jackknife resampling techniques are reasonable estimates of the covariance for this contribution as well.

\begin{table}
	\begin{center}
		\caption{
		We show in this table the $1\sigma$ uncertainty on $\bar{\tau}_e$ when combining the simulated kSZ signal with other components of the mm-sky. A comparison of the respective errors provides a ranking of the relative importance of the various uncertainties. We find that thermal SZ is the strongest source of noise in our measurement, closely followed by instrumental noise.}
		\label{tab:errdecomp}
		\begin{tabular}[width=\columnwidth]{cc}
			\toprule
			mm-sky components & $10^3 \times \sigma_{\bar{\tau}_e}$ \\
			\midrule
			kSZ only & 0.26 \\
			kSZ + CMB/foregrounds & 0.39 \\
			kSZ + instr. noise & 0.60 \\
			kSZ + CMB/foregrounds + instr. noise (`no tSZ') & 0.65 \\
			kSZ + tSZ & 0.63  \\
			\midrule
			all (`full CMB') & 0.86 \\
			\bottomrule
		\end{tabular}
	\end{center}
\end{table}

\subsection{Stability of the resampling covariance}
\label{sec:resamplingstab}

\begin{figure}
	\begin{center}
		\includegraphics[width=\columnwidth]{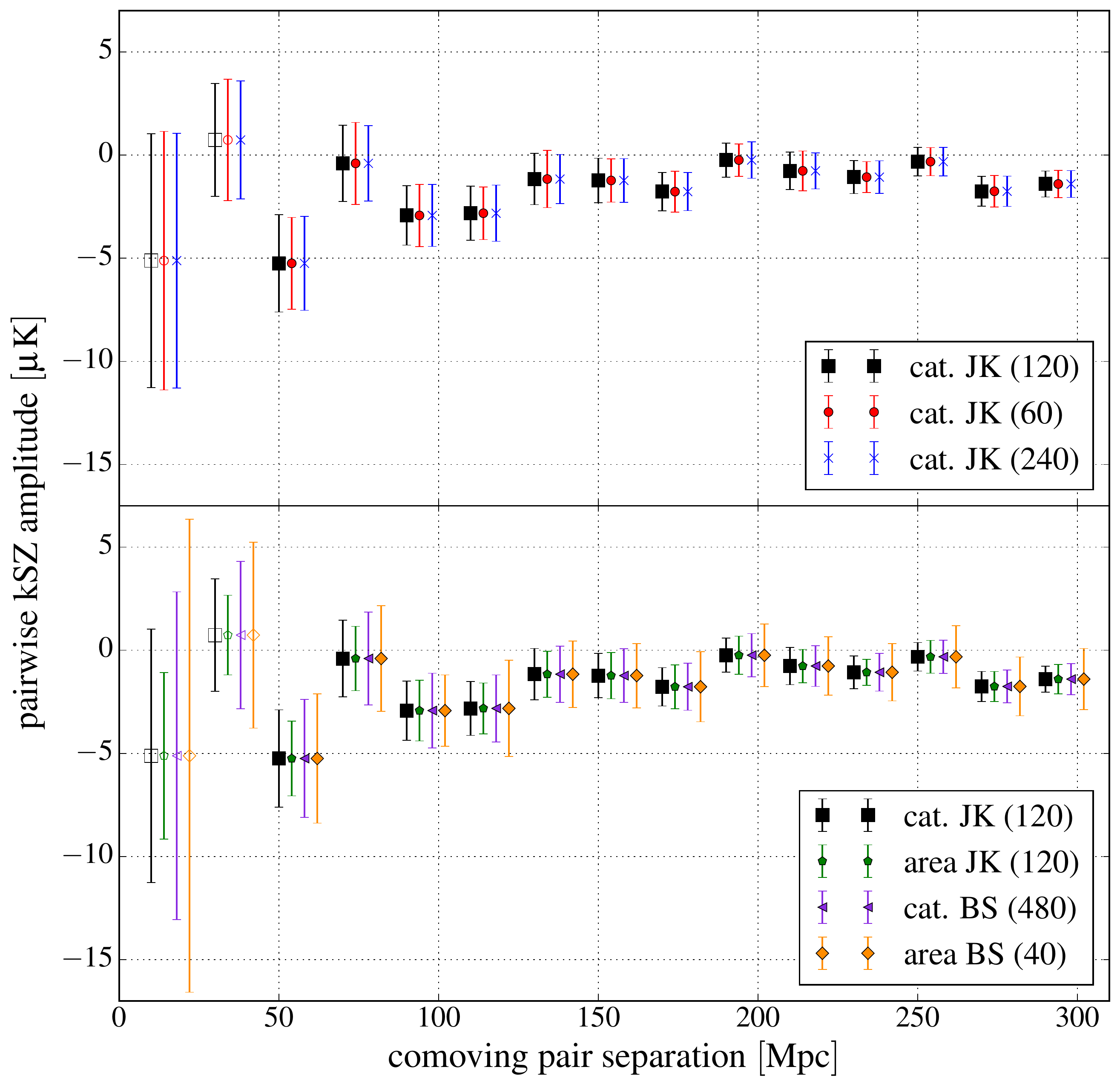}
		\caption{
			Tests for the stability of the error estimate.
			\textbf{Top:} We show here the errors of the pairwise kSZ measurement estimated via jack-knife resampling from the cluster catalogue. The black points and error bars used 120 resamples (as for the main analysis), whereas in the red/blue case we have decreased/increased the number of resamples by a factor of 2.
			\textbf{Bottom:}
			This panel displays error estimates from four different resampling techniques, namely jack-knife and bootstrap, in both cases sampling either directly from the catalogue or by sky area. The numbers in brackets in the legend denote the number of samples used; for a motivation of these numbers and a detailed description of the methods see Appendix~\ref{sec:resamplingstab}.}
		\label{fig:errtest}
	\end{center}
\end{figure}

As described in detail in Section~\ref{sec:covariance}, for the main analysis we create $N_\mathrm{JK} = 120$ jack-knife resamples from the cluster catalogue and then estimate the covariance matrix via equation~\ref{eq:covmat}.
To demonstrate that our estimate is stable with respect to the choice of $N_\mathrm{JK}$, we repeat the covariance estimation using $N_\mathrm{JK} = \{60,240\}$.
We show in the top panel of Fig.~\ref{fig:errtest} the errors on the pairwise kSZ amplitude obtained in this way.
They are in good agreement with our original estimate, demonstrating the stability of our procedure to changes in $N_\mathrm{JK}$.

Secondly, we compare four different strategies of estimating the covariance via JK and bootstrap (BS) resampling, namely:
\begin{itemize}
	\item \textit{Catalogue jack-knife} with $N_\mathrm{JK} = 120$: this is the technique used for the main analysis.

	\item \textit{Area jack-knife} with $N_\mathrm{JK} = 120$: we follow a similar procedure as for the catalogue JK. However, instead of resampling directly from the catalogue, for every resample we remove all clusters located in a given rectangular sky patch covering roughly $1/N_\mathrm{JK}$ of the total survey area.

	\item \textit{Catalogue bootstrap}: we create $N_\mathrm{BS}$ resampled cluster catalogues with the same size as the original catalogue by drawing with replacement from the latter.
	We then calculate the covariance as
	$ \hat{C}_{ij}^\mathrm{BS} = (N_\mathrm{BS}-1)^{-1} \, \sum_{\alpha = 1}^{N_\mathrm{BS}} (\hat{T}_i^\alpha - \bar{T}_i) \, (\hat{T}_j^\alpha - \bar{T}_j)$.
	It is worth noting that the typical BS realisation is more different from the original catalogue than a typical JK realisation.
	Therefore a higher number of resamples is required for the error estimate to converge; here we use $N_\mathrm{BS} = 480$.

	\item \textit{Area bootstrap / sky patches (SP)}: we split the survey area into $N_\mathrm{SP}$ approximately equal-sized subpatches.
	In the presence of an irregularly shaped mask (see Fig.~\ref{fig:skyplot}) it is non-trivial to obtain sub-patches with exactly the same size and similar geometry.
	Here we approximate by restricting the survey area to the largest rectangle oriented along lines of constant RA and Dec.
	The price for this simplification is using a slightly smaller sky area than the main analysis.
	Additionally, cluster pairs across patch boundaries do not contribute to the signal. This limits $N_\mathrm{SP}$ to relatively small numbers; here we use $N_\mathrm{SP}=40$.
	From the pairwise kSZ amplitude computed on the individual patches, we estimate the covariance as for the catalogue bootstrap, but rescale it with a factor of $2/N_\mathrm{SP}$ to account for the smaller size of the individual patch.
	Due to the limitations discussed above, we expect this method to provide slightly larger error estimates than the other three, especially for large pair separations.
	Nevertheless it still provides an instructive cross-check of our error estimate.

\end{itemize}

We show in the bottom panel of Fig.~\ref{fig:errtest} the errors on the pairwise kSZ amplitude estimated with these four techniques.
Both JK methods and the catalogue BS give similar results, whereas as expected the area BS yields slightly larger, but still comparable error estimates.
The slight differences in the uncertainties of the two lowest separation bins do not affect our analysis, as these two bins are excluded from the template fits in the main analysis anyway (see Section~\ref{sec:validation}).
We therefore conclude that our error estimate is robust against the details of the used resampling technique.

~
\section{Tests for position-dependent observational systematics}
\label{sec:positionsys}

\begin{table*}
    \begin{center}
	\caption{Test for observational systematics: \textbf{Left:} The best-fitting optical depth (here shown as $10^3 \times \bar{\tau}_e$) for the given systematic cut. \textbf{Right:} The change in $\bar{\tau}_e$ in units of the expected scatter given the increased uncertainties due to the smaller sample size (equation~\ref{eq:systcheck}).}
	\label{tab:systchecktab}
	\begin{tabular}[width=0.3\columnwidth]{lccccc}
		\toprule
		best & 98\% & 95\% & 90\% & 80\% & 50\% \\
		\midrule
		$E_{B-V}$ & $3.45 \pm 1.00 $ &  $3.42 \pm 0.99 $ &  $3.15 \pm 1.03 $ &  $3.81 \pm 1.11 $ &   $3.18 \pm 1.43 $ \\
		seeing &  $4.04 \pm 1.01 $ &  $3.93 \pm 1.10 $ &  $4.19 \pm 1.11 $ &  $5.07 \pm 1.29 $  & $5.69 \pm 1.86 $\\
		airmass & $3.92 \pm 0.93 $ & $3.73 \pm 0.99 $ &  $3.66 \pm 1.05 $ & $2.59 \pm 1.06 $ & $2.54 \pm 1.65 $\\
		sky br. & $3.72 \pm 0.93 $ & $3.88 \pm 0.93 $ &  $3.84 \pm 0.97 $ & $2.83 \pm 1.03 $ & $2.96 \pm 1.56 $\\
		\bottomrule
	\end{tabular}
	\quad
	\begin{tabular}{lcccr}
		\toprule
		98\% & 95\% & 90\% & 80\% & 50\% \\
		\midrule
		-0.48 & -0.57 & -0.99 & +0.19 & -0.44 \\
		+0.59 & +0.26 & +0.65 & +1.40 & +1.18 \\
		+0.66 & +0.02 & -0.10 & -1.91 & -0.85 \\
		-0.19 & +0.41 & +0.18 & -1.81 & -0.63\\
		\bottomrule
	\end{tabular}
  \end{center}
  \end{table*}
In this appendix, we test for spatially-varying systematics in the optical catalogue that could bias the kSZ measurement.
Although these are taken into account during the production of the cluster catalogue, there could be residual correlation of these systematics with properties of the catalogue, which could cause a contamination of the signal.
Given that the pairwise kSZ estimation is most sensitive to pairs along the line of sight, we expect a possible contamination by these effects to be small.
Nevertheless, here we test the stability of our results with respect to a selection of these contaminants, to ensure the robustness of our signal.
In particular, we consider: extinction by Galactic dust as quantified by the $E_{B-V}$ map produced by \cite{Planck2013dust}, seeing, airmass, and sky brightness. For the latter three we use the $i$-band systematics maps produced within the DES collaboration using the approach described by \cite{Leistedt2015} for the SV area.

\begin{figure}
\begin{center}
	\includegraphics[width=\columnwidth]{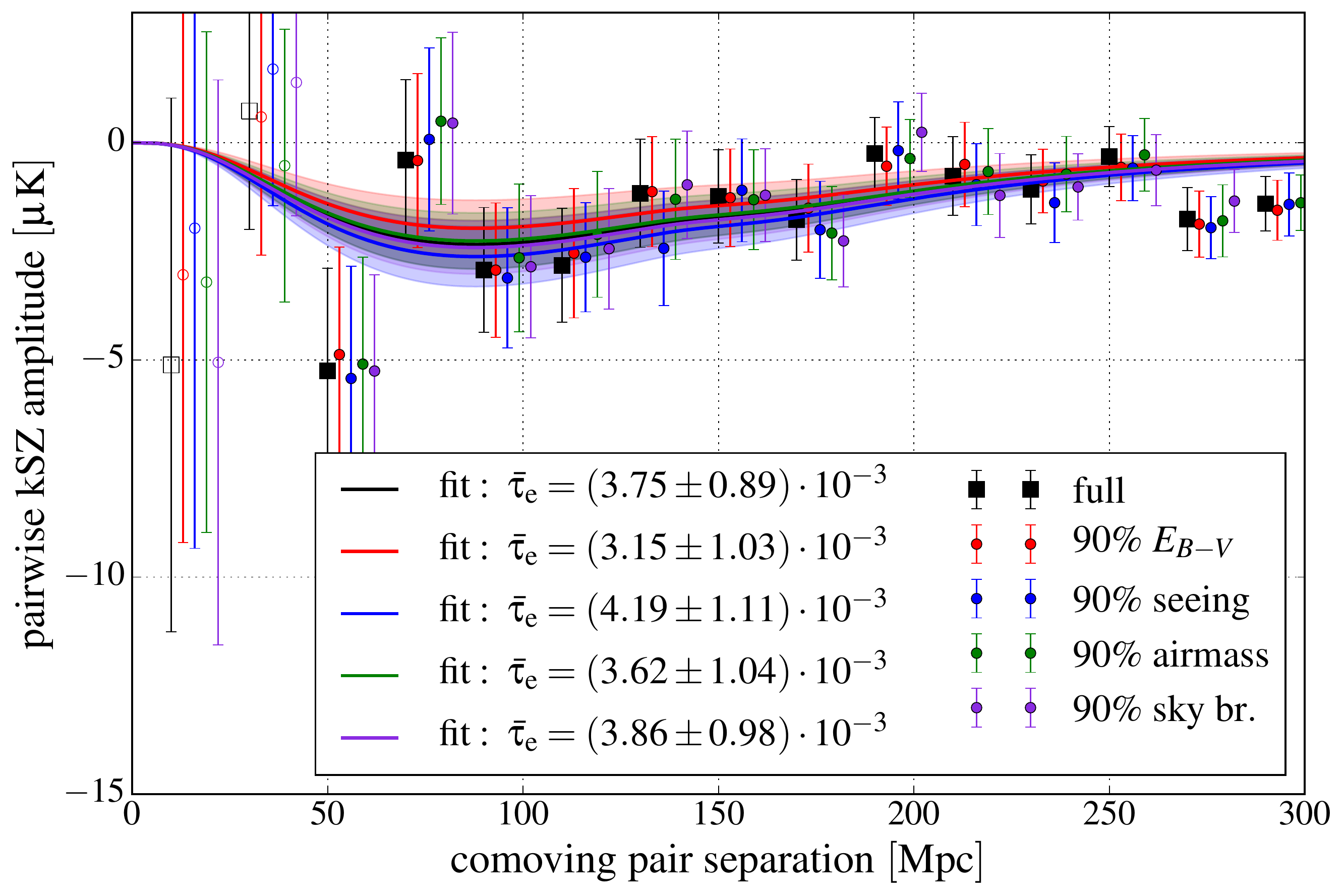}
	\caption{Tests for the impact of spatial variation of DES observing conditions on the pairwise kSZ amplitude: The black points show the original measurement, whereas the colours show the results using only the 90\% best clusters in the four systematic candidates considered here.}
	\label{fig:systchecks}
\end{center}
\end{figure}

To test for a correlation of our results with these systematic candidates, we measure their values at the cluster positions. We then recompute the pairwise kSZ amplitude from the $\{98\%,95\%,90\%,80\%,50\%\}$ of the clusters with the lowest
value in the systematic candidate under consideration. The results are given in the left panel of Table~\ref{tab:systchecktab}. Additionally we show in Fig.~\ref{fig:systchecks} the results for the 90\% cuts in all four systematic candidates.

When using only a subset $S$ of the full data set, we expect the inferred value of $\bar{\tau}_e$ to change on average by
\beq
\langle ( \Delta \bar{\tau}_e )^2 \rangle = \langle ( \bar{\tau}_e^S - \bar{\tau}_e)^2 \rangle = (\sigma_{\bar{\tau}_e}^S) ^2 - (\sigma_{\bar{\tau}_e})^2
\label{eq:systcheck}
\eeq
due to purely statistical scatter, even if there is no effect of the given systematic (e.g.~\citealt{Planck15Like}). We show the results for the change in our estimation of $\bar{\tau}_e$ in units of the expected mean scatter in the right panel of Table~\ref{tab:systchecktab}. We find that the results are mostly consistent with purely statistical scatter. The only exception is a weak trend of increasing $\bar{\tau}_e$ when cutting in seeing full width at half-maximum (FWHM): $\bar{\tau}_e$ changes by $ {\simeq} 1.4 \times \sqrt{ \langle ( \Delta \bar{\tau}_e )^2 \rangle }$ when using the 80\% clusters with the best seeing. Nevertheless the result for this particular cut is still consistent with the main result within the measurement uncertainties. We further note that results from different cuts in the same systematic candidate are correlated due to the overlap in samples, so a weak `trend' should not be overinterpreted.
Furthermore, if testing
multiple different cuts for several systematic candidates, one expects a few of them to show a weak correlation with the data.
In that sense, the $1.8~(1.9) \times \sqrt{ \langle ( \Delta \bar{\tau}_e )^2 \rangle }$ fluctuations from the $80\%$ of sky brightness (airmass) cuts are consistent with the expectation due to statistical scatter.
We therefore conclude that at the level of precision reported here, there is no significant contamination by spatially varying observing conditions or extinction.

 \section*{Affiliations}
\textit{
$^{1}$ Institute of Astronomy, University of Cambridge, Madingley Road, Cambridge CB3 0HA, UK\\
$^{2}$ Kavli Institute for Cosmology, University of Cambridge, Madingley Road, Cambridge CB3 0HA, UK\\
$^{3}$ HEP Division, Argonne National Laboratory, 9700 S. Cass Ave., Lemont, IL 60439, USA\\
$^{4}$ Kavli Institute for Cosmological Physics, University of Chicago, Chicago, IL 60637, USA\\
$^{5}$ Kavli Institute for Particle Astrophysics \& Cosmology, P. O. Box 2450, Stanford University, Stanford, CA 94305, USA\\
$^{6}$ Department of Physics, Stanford University, 382 Via Pueblo Mall, Stanford, CA 94305, USA\\
$^{7}$ Centre for Theoretical Cosmology, DAMTP, University of Cambridge, Wilberforce Road, Cambridge CB3 0WA, UK\\
$^{8}$ SLAC National Accelerator Laboratory, Menlo Park, CA 94025, USA\\
$^{9}$ Fermi National Accelerator Laboratory, P. O. Box 500, Batavia, IL 60510, USA\\
$^{10}$ Department of Astronomy and Astrophysics, University of Chicago, Chicago, IL, 60637, USA\\
$^{11}$ MCS Division, Argonne National Laboratory, 9700 S. Cass Ave., Lemont IL 60439, USA\\
$^{12}$ Department of Physics, McGill University, 3600 Rue University, Montreal, QC, H3A 2T8, Canada\\
$^{13}$ Department of Physics and Astronomy, University of Pennsylvania, Philadelphia, PA 19104, USA\\
$^{14}$ Department of Physics, University of Arizona, 1118 E 4th St, Tucson, AZ 85721, USA\\
$^{15}$ Faculty of Physics, Ludwig-Maximilians-Universit\"at, Scheinerstr. 1, 81679 Munich, Germany\\
$^{16}$ Excellence Cluster Universe, Boltzmannstr.\ 2, 85748 Garching, Germany\\
$^{17}$ Max Planck Institute for Extraterrestrial Physics, Giessenbachstrasse, 85748 Garching, Germany\\
$^{18}$ Universit\"ats-Sternwarte, Fakult\"at f\"ur Physik, Ludwig-Maximilians Universit\"at M\"unchen, Scheinerstr. 1, 81679 M\"unchen, Germany\\
$^{19}$ Department of Physics \& Astronomy, University College London, Gower Street, London, WC1E 6BT, UK\\
$^{20}$ Department of Physics and Electronics, Rhodes University, PO Box 94, Grahamstown, 6140, South Africa\\
$^{21}$ Department of Astrophysical Sciences, Princeton University, Peyton Hall, Princeton, NJ 08544, USA\\
$^{22}$ CNRS, UMR 7095, Institut d'Astrophysique de Paris, F-75014, Paris, France\\
$^{23}$ Sorbonne Universit\'es, UPMC Univ Paris 06, UMR 7095, Institut d'Astrophysique de Paris, F-75014, Paris, France\\
$^{24}$ Department of Physics, University of Chicago, 5640 South Ellis Avenue, Chicago, IL, USA 60637\\
$^{25}$ Laborat\'orio Interinstitucional de e-Astronomia - LIneA, Rua Gal. Jos\'e Cristino 77, Rio de Janeiro, RJ - 20921-400, Brazil\\
$^{26}$ Observat\'orio Nacional, Rua Gal. Jos\'e Cristino 77, Rio de Janeiro, RJ - 20921-400, Brazil\\
$^{27}$ Department of Astronomy, University of Illinois, 1002 W. Green Street, Urbana, IL 61801, USA\\
$^{28}$ National Center for Supercomputing Applications, 1205 West Clark St., Urbana, IL 61801, USA\\
$^{29}$ Institut de Ci\`encies de l'Espai, IEEC-CSIC, Campus UAB, Carrer de Can Magrans, s/n,  08193 Bellaterra, Barcelona, Spain\\
$^{30}$ School of Physics and Astronomy, University of Southampton,  Southampton, SO17 1BJ, UK\\
$^{31}$ Institute of Cosmology \& Gravitation, University of Portsmouth, Portsmouth, PO1 3FX, UK\\
$^{32}$ Department of Physics, University of California, Berkeley, CA 94720\\
$^{33}$ Department of Physics, University of Michigan, Ann Arbor, MI 48109, USA\\
$^{34}$ Department of Astronomy, University of Michigan, Ann Arbor, MI 48109, USA\\
$^{35}$ Department of Physics, The Ohio State University, Columbus, OH 43210, USA\\
$^{36}$ Center for Cosmology and Astro-Particle Physics, The Ohio State University, Columbus, OH 43210, USA\\
$^{37}$ Cerro Tololo Inter-American Observatory, National Optical Astronomy Observatory, Casilla 603, La Serena, Chile\\
$^{38}$ Australian Astronomical Observatory, North Ryde, NSW 2113, Australia\\
$^{39}$ Departamento de F\'{\i}sica Matem\'atica,  Instituto de F\'{\i}sica, Universidade de S\~ao Paulo,  CP 66318, CEP 05314-970, S\~ao Paulo, SP,  Brazil\\
$^{40}$ George P. and Cynthia Woods Mitchell Institute for Fundamental Physics and Astronomy, and Department of Physics and Astronomy, Texas A\&M University, College Station, TX 77843,  USA\\
$^{41}$ Kavli Institute for Astrophysics and Space Research, Massachusetts Institute of Technology, 77 Massachusetts Avenue, Cambridge, MA 02139, USA\\
$^{42}$ Instituci\'o Catalana de Recerca i Estudis Avan\c{c}ats, E-08010 Barcelona, Spain\\
$^{43}$ Institut de F\'{\i}sica d'Altes Energies (IFAE), The Barcelona Institute of Science and Technology, Campus UAB, 08193 Bellaterra (Barcelona) Spain\\
$^{44}$ Jet Propulsion Laboratory, California Institute of Technology, 4800 Oak Grove Dr., Pasadena, CA 91109, USA\\
$^{45}$ School of Physics, University of Melbourne, Parkville, VIC 3010, Australia\\
$^{46}$ Department of Physics and Astronomy, Pevensey Building, University of Sussex, Brighton, BN1 9QH, UK\\
$^{47}$ Astrophysics and Cosmology Research Unit, University of KwaZulu-Natal, Durban, SA\\
$^{48}$ Centro de Investigaciones Energ\'eticas, Medioambientales y Tecnol\'ogicas (CIEMAT), Madrid, Spain\\
$^{49}$ Brookhaven National Laboratory, Bldg 510, Upton, NY 11973, USA\\
$^{50}$ Harvard-Smithsonian Center for Astrophysics, 60 Garden Street, Cambridge, MA, USA 02138\\
$^{51}$ Department of Physics, University of Illinois Urbana-Champaign, 1110 W.\ Green Street, Urbana, IL 61801, USA\\
}


\bsp	
\label{lastpage}
\end{document}